\documentclass[preprint
 amsmath,amssymb,showkeys,nofootinbib,
 aps,
]{revtex4-2}

\usepackage{graphicx}
\usepackage{dcolumn}
\usepackage{bm}
\usepackage{comment}
\usepackage{amsmath}
\usepackage{amsthm}
\usepackage{mathtools}
\usepackage{bbm}
\usepackage{xcolor}
\usepackage{enumitem}
\usepackage{soul}

\DeclareMathOperator{\E}{\mathbb{E}}
\DeclareMathOperator{\U}{\mathcal{U}}

\DeclareMathOperator{\N}{\mathbb{N}}

\DeclareMathOperator{\hrho}{\hat{\rho}}

\newcommand{\Eqref}[1]{Eq.~(\ref{#1})}

\newtheorem{theorem}{Theorem}[section]

\newtheorem{lemma}[theorem]{Lemma}

\newtheorem{definition}[theorem]{Definition}

\begin{document}

\preprint{APS/123-QED}

\title{A Game-Theoretic Principle for Pricing Data as a Commodity}

\author{Pasquale Casaburi}
\author{Giovanni Piccioli}%
\author{Pierpaolo Vivo}
 \email{To whom correspondence should be addressed. E-mail: pierpaolo.vivo@kcl.ac.uk}
\affiliation{%
 Department of Mathematics, King’s College London, The Strand, London WC2R 2LS, UK
}%

\begin{abstract}
Data is the central commodity of the digital economy. Unlike physical goods, data exhibits properties that defy the standard theory of supply and demand: it is non-rival (the same dataset can be sold to multiple buyers without degradation), it is replicable at near-zero cost, and it is traded under heterogeneous licensing rules that restrict lawful use. Determining a new pricing principle to attach a fair price tag to datasets is therefore a difficult but central problem. We propose a game-theoretic framework in which the value of a data string emerges from strategic competition among $N$ players betting on a stochastic process with asymmetric information about past outcomes. A better-informed player may either exploit her advantage or sell part of her dataset to less informed competitors. By analytically deriving the Nash equilibrium, we identify the price range for a mutually beneficial trade. The model reveals market dynamics that depart from textbook intuition: informed players may compete or jointly exploit the least informed; data can be shared even at zero price without reducing the seller’s utility; rivalry among well-informed players can benefit uninformed ones; and trades infeasible in small markets can be viable in larger ones. These findings establish a theoretical foundation for the pricing of intangible goods in interacting digital markets, which are in need of robust valuation principles.
\end{abstract}

                 
\maketitle

\section{Introduction}

The supply and demand equilibrium is a cornerstone of classical economic theory. It allows, at least in principle, to attach a price tag to physical, tangible goods and commodities on the basis of a few elementary principles: scarcity relative to demand, consumers’ willingness to pay, and producers’ marginal costs, with prices adjusting through competition to balance quantity supplied and quantity demanded. In this framework, higher demand or lower supply raises prices, incentivizing greater production and conservation, while lower demand or greater supply pushes prices down, discouraging excess output and restoring equilibrium \cite{back2010asset, skiadas2009asset}. 

In modern societies, \emph{information} in the form of data can be exchanged as a commodity as well. Textual data used to train large language models \cite{commoncrawl}, weather data for predicting renewable energy output \cite{benth2012modeling}, and historical financial data \cite{fin_data} are traded on a daily basis in wide international markets. Quite paradoxically, though, data directly defies most of the  foundational assumptions of classical pricing theory \cite{laney2017infonomics,linde2009pricing,gu2011intangible}.

Data is typically \emph{non-rival}, as the same dataset can be sold to multiple buyers without degradation. It can be reproduced infinitely many times at near-zero marginal costs, and can be licensed under different terms restricting lawful use. This creates a pricing dynamic that traditional supply-demand models, built on the assumption that \emph{scarcity} is one of the main drivers of price, cannot capture. This economic paradox has created a trillion-dollar market \cite{mckinsey,data_marketplaces} built on very shaky foundations and resulting in systematic market inefficiency: tech giants acquire datasets for billions of dollars without rigorous valuation frameworks, while individuals surrender valuable personal data in exchange for free services. 
With business interest expanding rapidly, companies offering \emph{data valuation} services have started proliferating, while the question ``how much is my data worth'' has become a policy priority for many governments. Yet, in the absence of established standards, these problems are usually approached mostly using heuristics or proprietary models, narrowly tailored to specific use cases. Existing approaches to data pricing in the economic literature (reviewed in the Discussion section below) include cost-based models that link data value to production or acquisition costs, privacy-based frameworks that price subjective privacy loss, utility- or quality-based schemes that tie prices to dataset attributes, seller-centric profit-maximization models, and market-design analyses of trading mechanisms. However, most attempts typically isolate specific dimensions of the problem and abstract away from the interconnected, multi-agent strategic complexity through which data value endogenously emerges.

In this paper, we instead develop a \emph{game-theoretic pricing principle} that treats data owners, prospective buyers, and the market in which they interact as components of a complex, interconnected, and adaptive system. From this perspective, the value of data emerges from strategic competition rather than ad hoc valuation methods.
To provide an analytically tractable proof of concept, we study a deliberately stylized model in which $N\geq 3$ players participate in a single-round, zero-sum betting game on a binary outcome arising from an underlying stochastic process whose generating mechanism is unknown to them and about which they possess only partial information. Players interpret their available data as independent and identically distributed observations of a possibly biased coin. The model considers one informed data seller, one prospective buyer, and $N-2$ least-informed players who cannot purchase the data and remain unaware of the transaction. This setting may be interpreted, for instance, as a stylized representation of traders attempting to predict whether the value of an asset will increase or decrease tomorrow.
Within this framework, a more informed player may either exploit her private data exclusively (to make a better model of the
underlying process and bet in a more favorable way) or sell a portion of it to a competitor in exchange for monetary compensation. This trade-off induces an intriguing pricing principle that can be formulated directly in terms of the utilities of the two parties, independently of many details of the particular game. Comparing the seller's utility before and after the data transfer identifies the minimum price at which she is willing to sell, while the corresponding change in the buyer's utility identifies the maximum price she is willing to pay. When these two bounds are compatible, they define an interval of prices within which the transaction is feasible and mutually beneficial.

Applying this pricing principle within our stylized framework reveals a rich landscape of counterintuitive and unconventional market dynamics effects as a function of the data held by the different players and of the beliefs those data induce. For instance, while economic intuition suggests that sharing exclusive information should generally harm the seller by 
strengthening competition, we discover unexpected ``symbiotic'' regimes where data sharing benefit both parties even in the absence of monetary compensation. Moreover, we uncover a paradoxical ``blessing of ignorance'' phenomenon: fierce competition between highly informed players may generate positive utilities for less-informed participants despite their strategic disadvantage, while the most informed player suffers negative returns despite their superior knowledge. 
This demonstrates that information abundance does not guarantee competitive advantage, as strategic market structure can favor less-informed parties through complex interaction effects.
We also find that as the number of players $N$ increases, the competition effects between buyer and seller weaken: identical data trades that prove mutually beneficial in large markets become economically unviable in small ones.  Finally, extending the model to volatility-averse players, we show that in this case the value of a dataset depends not only on its content, but also on its size.

\section{The Model}\label{sec:model_intro}
The data pricing problem can be framed by considering a set of $N\geq 3$ players, who play the following game of chance: there is a concealed stochastic process from which players observe only a binary signal, $w\in\{0,1\}$. No player knows how this signal is actually generated, but each interprets the observed data as a sequence of i.i.d. tosses of a possibly biased coin, with Heads corresponding to $w=1$ and Tails to $w=0$. Each player pays a unit fee to place a bet on the next outcome, and the resulting pot is distributed among those who made a correct prediction. If no player makes a correct guess, all players are refunded.

We now make a further set of assumptions about the information held by each player (see Appendix C for a self-contained construction of the information structure of the game, and Appendix D for the resulting equilibrium concept).
\begin{itemize}
\item {\bf (A1)} All players have access to a public string $\vec s_0$ of $R$ consecutive past outcomes that include $H$ Heads and $R-H$ Tails, immediately preceding the outcome on which they are about to bet. 
\item {\bf (A2)} Player 1 (P1) privately holds additional historical data -- a longer consecutive string with parameters $(H_1, R_1)$, where $R_1>R$ -- extending $\vec s_0$ further back in the past.
\item {\bf (A3)} Each player $i$ makes an estimate $\hat\rho_i$ of the ``coin'' \emph{bias} (the probability that it will land on Heads), based on the information they hold about past outcomes. It is common knowledge that all players use the same method to estimate the coin bias (as described in Section \ref{sec:pricinggame}) - even though the estimate made by P1 may differ from the others', as she holds a larger dataset.
\item {\bf (A4)} Each player $i$ chooses her bet by equilibrium reasoning applied to the game as she perceives it: given her information and her beliefs about the information held by the other players, she computes a Nash equilibrium of that perceived game and adopts the strategy $p_i^\star$ that it assigns to her. Here, $p_i^\star$ denotes the probability with which she bets Heads in the next round\footnote{Note the different meanings of $\hat\rho_i$ and $p_i^\star$: the former encodes player $i$'s belief that the ``coin'' will land Heads, while the latter is her betting strategy. Even if the player is very confident that the coin will land Heads, it may be strategically preferable to bet Tails.}. Within the perceived game, no unilateral deviation from $p_i^\star$ would improve her utility.
\item {\bf (A5)} Players that rely only on the public string $\vec s_0$ are unaware that more informed players exist. They therefore believe that P1, like all others, has access only to $\vec s_0$.
\end{itemize} 

\begin{figure}
\centering
\includegraphics[width=\textwidth]{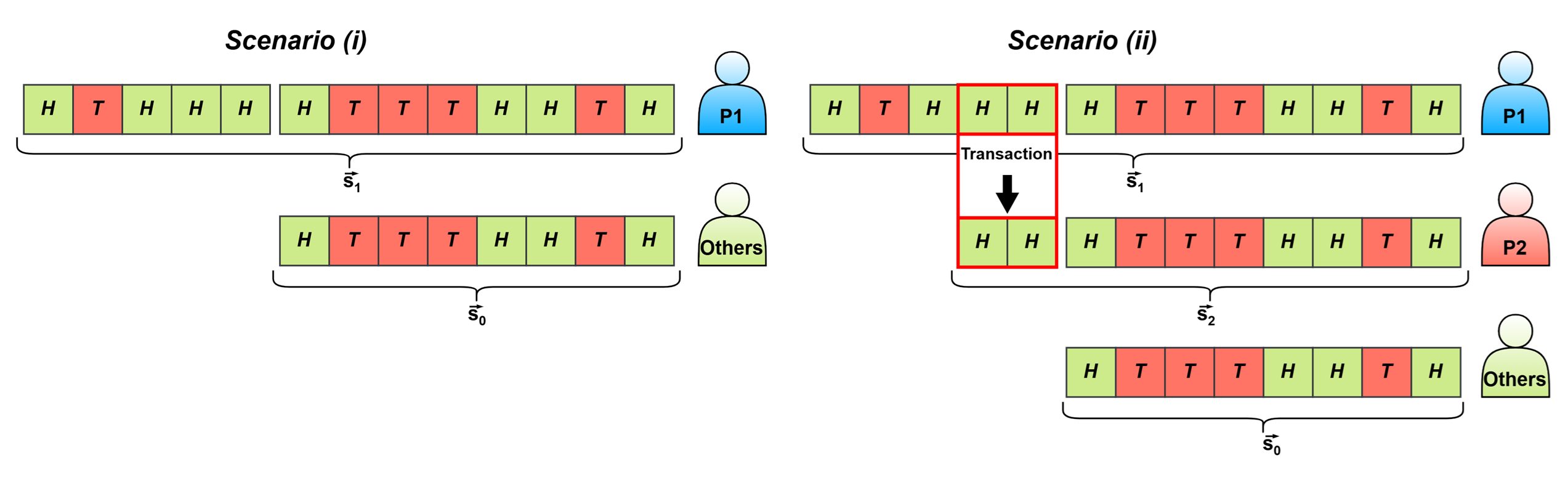}
\caption{Schematic representation of the datasets held by different players in the two scenarios. All players share the publicly available string $\vec s_0$ of past outcomes. P1 is the most informed player and holds a longer string $\vec s_1$. In scenario {\bf (ii)} she sells a portion of her extra string to P2, who therefore can base her betting strategy on the string $\vec s_2$. The length of $\vec s_2$ is intermediate between $\vec s_1$ and the public string $\vec s_0$. Only P1 and P2 can participate in the data transaction. The remaining players $i\geq 3$ cannot purchase the data and remain unaware whether the transaction has occurred.}
\label{fig:infogram_players_strings}
\end{figure}

Assumption~{\bf (A4)} specifies how bets are chosen; the resulting values of $p_i^\star$ are derived in Section~V and in the Appendices. Since players perceive different games, each player computes a \emph{subjective Nash equilibrium} of the game as she perceives it. These equilibria need not agree, and the profile actually implemented need not be a Nash equilibrium of any single game (see Section \ref{sec:Nash_eq} for additional details on subjective Nash equilibria).
This set of assumptions induces a form of \emph{informational asymmetry}: P1 can infer all others' strategies from the public data, while the remaining $N-1$ players cannot compute hers. P1 therefore faces a strategic choice on how to leverage her private information. She can either 
\begin{itemize}
\item {\bf (i)} keep her data private and exploit her informational advantage to benefit from her favorable position in the game relative to less-informed players,
\item {\bf (ii)} choose to sell part of her private data to another player, conventionally denoted as player 2 (P2). 
\end{itemize}
The fundamental principle that sets data apart from tangible goods is indeed that the seller still retains full possession of her dataset even after a possible transaction, leading to competition with the buyer. To complete the description of this possible data transfer, we introduce a further assumption specifying its market structure:
\begin{itemize}
\item {\bf (A6)} We restrict our analysis to a single data transaction. P1 is the sole potential seller, while P2 is the sole potential buyer. Players $i\geq 3$ can neither purchase the data nor observe whether the transaction has occurred. Since they have access to the same data and share the same beliefs about the game, they are equivalent and adopt a common strategy in both scenarios. In scenario {\bf (ii)}, the substring available for sale is specified exogenously, with both its length and content treated as given rather than selected strategically by P1. The transaction gives P2 this ordered substring of P1's extra data, contiguous to the public string, yielding an overall string $\vec s_2$ with parameters $(H_2,R_2)$ ($R_1 \geq R_2 > R$), while P1 still retains her entire dataset, which she uses to devise a new optimal strategy against her new competitor.
\end{itemize}
Figure \ref{fig:infogram_players_strings} provides a schematic representation of the datasets held by the different players in the two scenarios. With this transaction, P1 effectively creates \emph{ex nihilo} a competitor (P2) stronger than any of the other $N-2$ less informed players and potentially gives away some of her edge in the game, which needs to be fairly compensated. The questions we wish to address are therefore:

\begin{enumerate}[label=Q\arabic*.]
\item {\bf Under which conditions is a data transaction mutually beneficial for both P1 and P2?}
\item {\bf If P1 sells part of her private data, what price should she charge P2 for an additional data string of length $R_2 - R$?}
\end{enumerate}

To address these questions, it is first necessary to define how players evaluate outcomes of the game given the information they hold. We therefore begin by introducing the notion of a \emph{utility function}, which encodes players’ preferences and underlies the definition of a fair price for the data set. We then characterize the strategic profiles resulting from Assumptions~(A1)--(A6), and in particular from the subjective Nash equilibrium rule specified in Assumption~(A4), by collecting the strategies actually adopted by the players into a vector $\vec p^{\,\star}$ and determining its form before and after a possible data transaction.

\subsection{Utility functions and Nash equilibrium}\label{sec:Nash}
Given a vector $\vec{p}$ of strategies adopted by all players, and the best estimate $\hat\rho_i$ that player $i$ makes of the ``coin'' bias, we can define the \emph{utility function} of player $i$ as a function $\mathcal{U}_i(\vec{p},\hat\rho_i)$. The utility function encodes player $i$’s preferences over strategy profiles, in the sense that
\begin{equation}
\mathcal{U}_i(\vec{p}_1, \hat{\rho}_i) > \mathcal{U}_i(\vec{p}_2, \hat{\rho}_i)
\quad \Longleftrightarrow \quad
\text{player $i$ prefers $\vec{p}_1$ to $\vec{p}_2$}\ .
\end{equation}
A possible choice in our setting is to let $\mathcal{U}_i(\vec{p},\hat\rho_i)$ represent the expected gain of player $i$ over a single round of the game, if all players bet according to the strategy vector $\vec{p}$ and given player $i$’s belief $\hat{\rho}_i$. In this case, $\mathcal{U}_i$ corresponds to the average monetary outcome for player $i$, defined as the average share of the pot she wins minus the entry fee (see Section \ref{sec:pricinggame} for additional details). This choice corresponds to \emph{volatility-neutral} players and satisfies the preference condition above, but other utility specifications could in principle be considered.

A \emph{Nash equilibrium} of a game is defined as a strategy profile such that no player can improve her utility by unilaterally deviating from her strategy, given the strategies of the others. In our setting, however, players possess different information and hold different beliefs about the information available to others, and therefore perceive different games. Under Assumption~{\bf (A4)}, each player $i$ computes a Nash equilibrium of the game as she perceives it, which we refer to as her \emph{subjective Nash equilibrium}, and adopts the strategy $p_i^\star$ assigned to her by that equilibrium. The vector $\vec p^{\,\star}=(p_1^\star,\ldots,p_N^\star)$ collects the strategies actually implemented, where $0\leq p_i^\star\leq 1$ represents the probability with which player $i$ bets Heads in the next round\footnote{A \emph{pure strategy} corresponds to $p_i^\star \in \{0,1\}$, meaning that player $i$ always bets on the same outcome (Heads for $p_i^\star=1$ and Tails for $p_i^\star=0$). For intermediate values $0<p_i^\star<1$, the player adopts a \emph{mixed strategy}, betting on Heads with probability $p_i^\star$.}. Since players may perceive different games because of their different information and beliefs, the components of $\vec p^{\,\star}$ may originate from distinct subjective Nash equilibria, and the implemented profile need not itself be a Nash equilibrium of any single game (see Section \ref{sec:Nash_eq} for additional details).

The information available to players also differs between the two scenarios introduced in the previous section -- {\bf (i)} no transaction, P1 holds onto the informational advantage exclusively, and {\bf (ii)} a transaction occurs between P1 and P2. Therefore, the equilibrium strategy profile $\vec p^{\,\star}$ defined above must be characterized separately depending on whether a data transaction has occurred or not.

\begin{itemize}
\item Scenario {\bf (i)}: Before any data transaction has occurred, a single player (P1) holds a longer data string than the remaining $N-1$ participants, who all base their strategies on the same public string $\vec{s}_0$. By assumption {\bf (A5)}, these less-informed players are unaware of the existence of any privileged participant and therefore believe that all players rely on the same data set. As a result, they perceive the game as fully symmetric and adopt a common betting strategy, which we denote by $p_{\text{eq}}$. In contrast, P1 knows the data available to the other players and can therefore anticipate the strategy they will adopt. She then determines her own best response to their collective behavior, which we denote by $(q_1)_{\text{eq}}^{\text{pre}}$. Therefore, with all players but P1 operating under the same informational conditions, the equilibrium strategy profile $\vec p^{\,\star}$ will be of the form
\begin{equation}\label{eq:p_pre_eq}
    \vec{p}^{\,\text{pre}} \coloneq ((q_1)_\text{eq}^\text{pre}, p_\text{eq}, \dots, p_\text{eq})\ ,
\end{equation}
where `pre' stands for `before' any transaction has occurred.
\item Scenario {\bf (ii)}: In case a data transaction occurs, the information hierarchy among players is modified. In addition to P1, player P2 now also holds additional information, while the remaining $N-2$ participants are unaware that a transaction has taken place and therefore continue to adopt the same strategy $p_{\text{eq}}$. Since P2 knows the data string on which the less-informed players base their decisions, she can anticipate their behavior and revise her own betting strategy accordingly. We denote the strategy adopted by P2 in this post-transaction setting by $q_{\text{eq}}$. At the same time, P1 can anticipate the strategies adopted by all other players, including the presence of a new, more informed competitor P2. She therefore adapts her behavior and determines a new best response strategy, which we denote by $(q_1)_{\text{eq}}^{\text{post}}$. In the post-transaction setting, the equilibrium profile can therefore be written as
\begin{equation}\label{eq:p_post_eq}
\vec{p}^{\,\text{post}} \coloneq ((q_1)_\text{eq}^\text{post}, q_\text{eq}, p_\text{eq}, \dots, p_\text{eq})\ ,
\end{equation} 
where `post' refers to a scenario where a transaction has occurred. 

\end{itemize}

Under assumptions {\bf (A1–A6)}, the least-informed players adopt the same equilibrium strategy $p_{\text{eq}}$ both before and after a transaction has occurred, and P1 has in both scenarios complete knowledge of the strategies adopted by all other participants. The central technical task is therefore the determination of the equilibrium strategies
$p_{\text{eq}}$, $(q_1)_{\text{eq}}^{\text{pre}}$, $(q_1)_{\text{eq}}^{\text{post}}$, and $q_{\text{eq}}$.
Starting from the utility function defined above, which encodes how players evaluate the outcomes of the game, we derive these strategies by computing the Nash equilibria of the game, following the approach described in Sections~\ref{sec:Nash_eq} and~\ref{sec:procedure}. The specific form of the utility function plays a central role, as it directly shapes the equilibrium strategies that players adopt. Once the equilibrium strategies in Eqs. \eqref{eq:p_pre_eq} and \eqref{eq:p_post_eq} are determined, they can be used to characterize the range of prices for which a data transaction between P1 and P2 is feasible and mutually beneficial. Full analytical derivations are provided in the Appendix.

\subsection{Fair price of data}
We are now ready to define the minimum price $\Psi_{\mathrm{min}}$ that the seller (P1) should demand to sell her data, and the maximum price $\Psi_{\mathrm{max}}$ that the buyer (P2) should be asked to pay for the data, as
\begin{align}
\Psi_\text{min} &\coloneqq \U_1(\vec p^{\,\text{pre}},\hrho_1)-\U_1(\vec p^{\,\text{post}},\hrho_1)\ ,\label{eq:Psi_min}\\
\Psi_\text{max} &\coloneqq \U_2(\vec p^{\,\text{post}},\hrho_1)-\U_2(\vec p^{\,\text{pre}},\hrho_1)\ .\label{eq:Psi_max}
\end{align}
The interpretation of the price bounds above is clear: the act of selling part of her data to P2 may reduce P1's expected earnings due to increased competition, and this potential loss -- measured by the difference of utility functions before and after the transaction -- must be offset by the minimal monetary payment that P2 should be asked to make. At the same time, the extra string purchased by P2 may enhance her ability to make accurate predictions and improve her utility. However, the strategic value of this extra data is bounded: there exists a maximal price $\Psi_\text{max}$ beyond which the cost of purchasing the data exceeds the increase in expected utility for the buyer. 

The price bounds above allow us to answer the questions we posed earlier on: 
\begin{enumerate}[label=Q\arabic*.]
\item {\bf Under which conditions is a data transaction mutually beneficial for both P1 and P2?}\\
\\
A transaction between P1 and P2 is feasible only if $\Psi_{\text{max}} > 0$ and $\Psi_{\text{min}} \leq \Psi_{\text{max}}$. If either of these conditions is violated, the transaction is not advantageous for at least one of the two parties.
\item {\bf If P1 sells part of her private data, what price should she charge P2 for an additional data string of length $R_2 - R_1$?}\\
\\
If the transaction is feasible, any non-negative price $\Psi \geq 0$ within the interval  $[\Psi_\text{min}, \Psi_\text{max}]$ constitutes a mutually beneficial agreement. Therefore any $\Psi$ within the bounds is a \emph{fair price for the data string purchased by P2}.
\end{enumerate}

We emphasize that these prices are defined from the perspective of the seller. More precisely, the terms $\U_2(\vec p^{\,\text{post}},\hrho_1),\U_2(\vec p^{\,\text{pre}},\hrho_1)$ represent P1’s estimate of P2’s utility, as indicated by the use of P1's belief $\hrho_1$. While other choices are possible, defining the price ranges from the point of view of P1 is very natural, as P1 has the most complete information set, enabling her to compute the most accurate estimates of other players' utilities.

In the following, we will always assume that all players model the underlying stochastic process as a ``true'' memory-less coin-tossing (see Assumption {\bf (A3)}). Moreover, we will consider two different choices for the utility functions $\mathcal{U}_i$, corresponding to  \emph{volatility-neutral} and \emph{volatility-averse} players. In the latter case, we include in the utility function an extra term that penalizes large variances in the players' estimate of the coin bias: being less confident about the coin bias is equivalent to forecasting a lower gain on average. In the following section, we report our analytical results in various regimes and for both definitions of the utility function, while in the Methods section below we provide all the technical definitions and mathematical details, including the explicit expressions for the volatility-neutral and volatility-averse utility functions in Eqs.~\eqref{eq:U_capital} and \eqref{eq:U_capital_lambda}, respectively.

\section{Results}
In this section, we characterize the equilibrium strategies, resulting utilities, and the implied price bounds for data exchange before and after a potential transaction. Results are presented separately for volatility-neutral players (Section~\ref{sec:vol_neutral_res}) and volatility-averse players (Section~\ref{sec:vol_averse_res}), highlighting how information asymmetries and attitudes toward uncertainty shape equilibrium behavior and determine the fair price range for data.
The key parameters are the estimates of the probability that the coin lands on Heads, as computed by different players on the basis of the dataset available to them:
\begin{enumerate}
    \item $\hat{\rho}$: the estimate inferred from the public string $\vec{s}_0$, which is common knowledge among all players.
    \item $\hat{\rho}_1$: the estimate inferred from the longer string $\vec{s}_1$, accessible only to the seller P1 and therefore known exclusively to her.
    \item $\hat{\rho}_2$: the estimate inferred by P2 from an intermediate-length string $\vec{s}_2$ following a transaction, and known only to P2 and P1.
\end{enumerate}
Each player interprets the data string available to her as a sequence of independent coin tosses, from which she estimates the coin bias and devises an optimal strategy for the game. Differences in the information available to players therefore lead to different beliefs, equilibrium strategies, utilities, and subjective views of the game. The procedure by which these belief parameters enter the computation of equilibrium strategies and the resulting price bounds is described in Section~\ref{sec:pricinggame} and Section \ref{sec:procedure}, which outlines the hierarchical construction of the equilibria as information asymmetries are introduced. Full technical details and analytical derivations are deferred to the Appendix.

\subsection{Volatility-neutral players}\label{sec:vol_neutral_res}

\begin{figure}[ht]
\centering
\includegraphics[width=\textwidth]{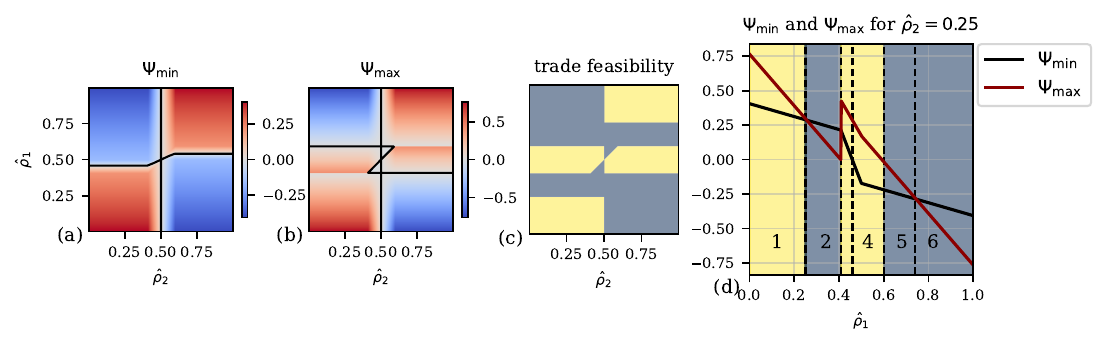}
\caption{Prices and transaction feasibility as a function of $\hat{\rho}_1$ and $\hat{\rho}_2$ for $\hat{\rho}=0.5$ and $N=5$ fixed throughout all panels.
\textbf{(a)} Heatmap of the minimum price $\Psi_\text{min}$.
\textbf{(b)} Heatmap of the maximum price $\Psi_\text{max}$.
\textbf{(c)} Transaction feasibility in the $\hat{\rho}_1,\hat{\rho}_2$ plane: yellow regions indicate where the transaction is possible, dark blue regions where it is not.
\textbf{(d)} $\Psi_\text{min}$ and $\Psi_\text{max}$ as functions of $\hat{\rho}_1$ for fixed $\hat{\rho}_2=0.25$. This corresponds to a vertical slice of the heatmaps. The background color matches panel {\bf (c)}, indicating transaction feasibility. Different strategic regions can be identified: (1) \textit{Cooperative}, $\Psi_{\max}>\Psi_{\min}>0$ --- trade is possible at a strictly positive price;  
(2) \textit{Competitive}, $\Psi_{\min}>\Psi_{\max}>0$ --- trade is not feasible;  
(3) \textit{Cooperative}  --- the possibility of a trade reopens; 
(4) \textit{Symbiotic}, $\Psi_{\min}<0<\Psi_{\max}$ --- trade would be possible also at zero price; 
(5) \textit{Exploitative cooperation}, $\Psi_{\min}<\Psi_{\max}<0$ --- trade would be feasible only if the seller compensates the buyer;  
(6) \textit{No-deal}, $\Psi_{\max}<\Psi_{\min}<0$ --- no mutually beneficial agreement exists.  }
\label{fig:prices}
\end{figure}

\paragraph*{\textbf{Transaction feasibility and strategic regimes.}}
The price bounds $\Psi_{\min}$ and $\Psi_{\max}$ can be computed using \Eqref{eq:Psi_min} and \Eqref{eq:Psi_max}. Figure~\ref{fig:prices} shows these bounds as heatmaps in panels {\bf (a)} and {\bf (b)}, plotted over the $(\hat{\rho}_1,\hat{\rho}_2)$ plane for fixed $\hat{\rho}=0.5$ and $N=5$. The plots exhibit a four-quadrant structure centered at $(\hat{\rho}_1,\hat{\rho}_2)=(\hat{\rho},\hat{\rho})$.

\begin{figure}
\centering
\includegraphics[width=\textwidth]{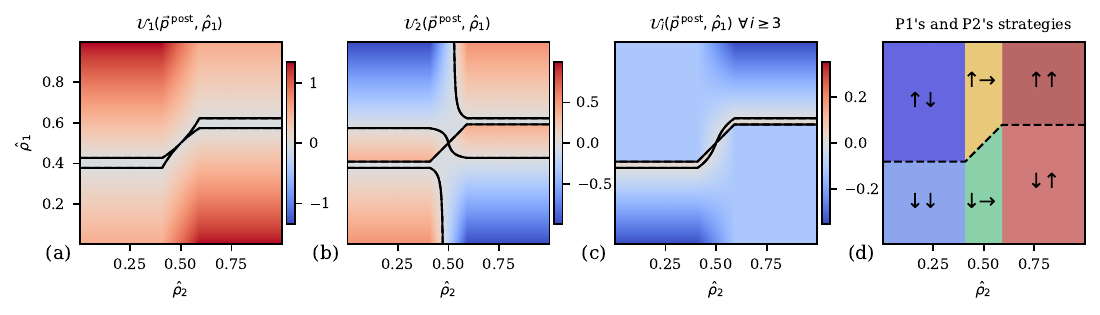}
\caption{Post-transaction equilibrium strategies and utilities as a function of $\hrho_1,\hrho_2$ for $\hrho=0.5$ and $N=5$. In panels \textbf{(a)}, \textbf{(b)}, \textbf{(c)} the black lines separate the region with positive utility from that with negative utility.  \textbf{(a)} P1's utility. \textbf{(b)} P2's utility, computed by P1. \textbf{(c)} Least-informed players' utility, computed by P1. \textbf{(d)} Phase diagram showing the strategies played by P1,P2. Every color corresponds to a different combination of strategies, indicated by the pair of arrows.  The left (right) arrow indicates the strategy of P1 (P2), with $\uparrow,\downarrow, \rightarrow$ corresponding respectively to pure Heads, pure Tails, and mixed strategies. The dashed black line separates the upper region, where P1 plays Heads from the lower region where she plays Tails.}
\label{fig:U_123_post_transaction}
\end{figure}

Qualitatively, both price bounds are positive when $\hat{\rho}_1$ is close to $\hat{\rho}_2$, except for a narrow region near the center. In the other quadrants, both bounds are negative reflecting situations in which P1 would effectively exploit P2 by providing misleading information. This induces P2 to place a bet on the outcome opposite to P1’s, allowing P1 to profit at P2’s expense. Panel {\bf (c)} summarizes the feasibility of the transaction: yellow regions indicate parameter values where a mutually acceptable positive price exists,  while dark blue indicates that no mutually beneficial agreement can be reached. More quantitatively, panel {\bf (d)} displays the price bounds as functions of $\hat{\rho}_1$ for $\hat{\rho}_2=0.25$. The background shares the same color scheme as panel {\bf (c)}, highlighting where a transaction is viable. The plot in panel {\bf (d)} can be divided into six regions:
\begin{enumerate}
    \item Both price bounds are positive and satisfy $\Psi_\text{max}>\Psi_\text{min}>0$, so the transaction is viable provided the seller charges a price above $\Psi_\text{min}$ to offset her expected loss. This defines a \textit{cooperative} regime, where both players benefit from the exchange at a strictly positive price.
    \item  In contrast, in the second region, $\Psi_\text{min}>\Psi_\text{max}>0$, meaning the seller’s expected loss exceeds the buyer’s potential gain, making the transaction unviable. This defines a \textit{competitive} regime, where the buyer's gain is too small to justify the price the seller would ask to cover her loss. 
    \item At the boundary between the second and third region, a discontinuity in $\Psi_\text{max}$ arises due to a shift in P1’s post-transaction strategy, from playing Tails to Heads, as shown in Figure \ref{fig:U_123_post_transaction}{\bf (d)}. This strategic switch increases the buyer’s expected benefit, restoring the possibility of a transaction. The third region is thus again a cooperative regime, analogous to the first.
    \item Entering the fourth region, $\Psi_\text{min}$ becomes negative while $\Psi_\text{max}$ remains positive, indicating that both players benefit from the transaction even if no payment is made. This defines a \textit{symbiotic} regime, where sharing the data alone is sufficient to increase both players’ utilities at the expense of the others. The existence of a symbiotic regime is another paradoxical effect: one would normally expect (and indeed previous literature for data pricing hinged on this \cite{majumdar2025developing,jaisingh2008privacy,ghosh2011selling}) that sharing data necessarily leads to a loss of privacy and hence reduces profits for the sharer. In our case instead, the strategic nature of the game enables both P1 and P2 to profit from the data exchange. This is due to a combination of effects: (i) P2 changes strategy from mixed to pure Tails after the exchange (ii) P1 in response adopts the opposite pure strategy (Heads). Since they play opposite strategies, the two players will never share the pot with each other, in case of victory. This leads to an increase in utility for both. (iii) On the other hand, P2 underestimates the probability of the outcome Heads, so she will win less and less frequently as $\hrho_1$ increases. This last effect offsets the gains when reaching the boundary with region 5.
    \item Moving into the fifth region, $\Psi_\text{max}$ also becomes negative, indicating that the buyer would incur a loss from the transaction. Interestingly, the minimum price remains negative and lies below the maximum price. This creates a hypothetically favorable setting in which player P1 could give the data to P2 at no cost and offer additional compensation, making the transaction mutually beneficial despite the fact that the information would otherwise mislead the buyer and reduce her utility. This defines an \textit{exploitative cooperation} regime, where a transaction remains possible only if the \textit{seller} is willing to pay the buyer to offset the misleading effect of the data sold.
    \item Finally, in the sixth region, both price bounds are negative and satisfy $\Psi_\text{max} < \Psi_\text{min}<0$, confirming that no transaction is beneficial under any price. This defines the \textit{no-deal} regime, where the buyer’s loss cannot be offset by what the seller gains, ruling out any possible agreement.
\end{enumerate}
These regimes illustrate the complex and diverse landscape of outcomes that our model uncovers, highlighting how strategic data trading can generate counterintuitive effects.

\begin{figure}
\centering
\includegraphics[scale=0.96]{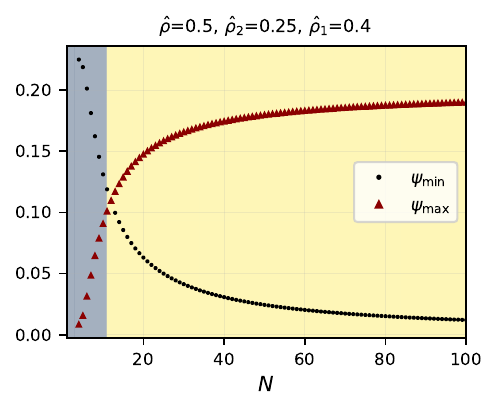}
\caption{Price bounds $\Psi_{\min}$ (black dots) and $\Psi_{\max}$ (red triangles) as functions of the number of players $N$, for fixed beliefs $\hat\rho=0.5$, $\hat\rho_2=0.25$, and $\hat\rho_1=0.4$. The background color matches that of Fig.~\ref{fig:prices}, with yellow indicating a feasible transaction.}
\label{fig:n_dependence}
\end{figure}

\paragraph*{\textbf{Transaction feasibility and market size.}} 
We now focus on how the number $N$ of participants in the game affects the feasibility of a transaction and the price bounds.
Figure~\ref{fig:n_dependence} shows $\Psi_{\min}$  and $\Psi_{\max}$ as functions of $N$ for fixed beliefs $\hat\rho=0.5$, $\hat\rho_2=0.25$, and $\hat\rho_1=0.4$. In this configuration, $\Psi_{\min}$ decreases and $\Psi_{\max}$ increases as the number of players grows. The transaction becomes feasible only for $N \geq 12$, marking a transition from a \emph{competitive} to a \emph{cooperative} regime. 
This behavior reflects the reduced competition between the two informed players as the number of least informed participants increases. As $N$ grows, $\Psi_{\min}$ tends to zero, showing that the privileged player becomes indifferent to P2’s presence when many uninformed players remain to exploit, while $\Psi_{\max}$ increases, since P2 can leverage the newly acquired information to exploit more of the uninformed opponents.

\paragraph*{\textbf{The ``blessing of ignorance'' effect.}} 
In the post-transaction scenario, a data exchange has occurred and an additional informed player is present. The buyer P2 acquires extra information and forms a belief $\hat{\rho}_2$ that generally differs from both the public belief $\hat{\rho}$ and the seller’s belief $\hat{\rho}_1$. The corresponding equilibrium strategy profile is given by $\vec{p}^{\,\text{post}}$, as defined in \Eqref{eq:p_post_eq}. As the most informed player, P1 can compute the equilibrium utilities of all players both before and after the transaction (the corresponding procedure is described in Step~5 of Section~\ref{sec:procedure}). In the post-transaction scenario, these quantities are shown in Figure~\ref{fig:U_123_post_transaction}, which displays the equilibrium utilities evaluated from the perspective of P1, together with the equilibrium strategies adopted by the seller and the buyer as functions of their respective beliefs $\hat{\rho}_1$ and $\hat{\rho}_2$. Panel~\textbf{(d)} illustrates the strategies adopted by P1 and P2: above the dashed line, P1 consistently plays the pure strategy Heads, while below it she switches to the pure strategy Tails. Moving from left to right (increasing $\hrho_2$) P2 starts with a pure Tails strategy, then enters a mixed strategy region and finally plays pure Heads. Panels \textbf{(a)}, \textbf{(b)} and \textbf{(c)} depict respectively the utilities of the seller, buyer, and least-informed players. We remark that the buyer’s utility tends to be negative when $\hat{\rho}_1$ differs significantly from $\hat{\rho}_2$, and correspondingly P1's utility increases. 
Panel~\textbf{(c)} reveals a nontrivial ``blessing of ignorance'' paradoxical effect: there exists a region where the strong competition between P1 and P2  benefits the least-informed players, leading to positive utilities for them at the expense of P1 and P2. 
In Panel~\textbf{(a)} we can observe the same paradoxical effect: in a small region P1's utility is negative, despite her being the most informed player. This effect shows that having more information does not guarantee better utilities than others. The competitive structure of the game instead creates complex interactions that can actually favor less informed parties.
Conversely, when $\hat{\rho}_1 \approx \hat{\rho}_2$ and both are far from the public estimate $\hat{\rho}$, the least-informed players suffer substantial losses. This is the region where the more informed players jointly extract the most value from the least-informed ones.

\subsection{Volatility-averse players}\label{sec:vol_averse_res}
We now extend the analysis to volatility-averse players. The underlying game structure remains unchanged, but players are assumed to take into account not only their expected gain, but also the uncertainty associated with their estimate of the coin bias. To capture this behavior, the utility function (whose precise functional form is given in Eq.~\eqref{eq:U_capital_lambda} of Section~\ref{sec:vol_aversion}) is modified by adding a term that penalizes large variances in the players’ subjective estimates: lower confidence in the coin bias is treated as forecasting a lower gain on average. The strength of this penalty is controlled by a common parameter $\lambda \geq 0$, which quantifies players’ aversion to uncertainty, with larger values corresponding to stronger volatility aversion. When $\lambda = 0$, the penalty vanishes and the model reduces to the volatility-neutral case discussed above. As a result, equilibrium strategies reflect not only expected profitability, but also the confidence players have in their estimates.
As in the volatility-neutral setting, we determine the equilibrium strategies of all players both before and after the transaction:
$\vec{p}^{\,\text{pre}} = \left((q_1)_\text{eq}^\text{pre},\, p_\text{eq},\, \dots,\, p_\text{eq}\right)$ and 
$\vec{p}^{\,\text{post}} = \left((q_1)_\text{eq}^\text{post},\, q_\text{eq},\, p_\text{eq},\, \dots,\, p_\text{eq}\right)$. These strategies are then used to compute the corresponding utilities and price bounds, following the procedure outlined in Section~\ref{sec:procedure}.
A key difference with respect to the volatility-neutral case is that, under volatility aversion, a player’s equilibrium strategy depends not only on the beliefs  $\hat{\rho}_i$, but also explicitly on the length $R_i$ of the data string they hold. Players with the same estimate $\hat{\rho}_i$ but different $R_i$ may behave differently, as longer strings reduce uncertainty and can lead to higher utilities. A direct consequence of this is that the transaction price bounds also become sensitive to the length of the exchanged string. These bounds can still be computed using \Eqref{eq:Psi_min} and \Eqref{eq:Psi_max}, with the only difference that the utilities are now evaluated using the risk-averse utility function defined in Section \ref{sec:vol_aversion}.

\begin{figure}[tbhp]
\centering
\includegraphics[width=8.7cm]{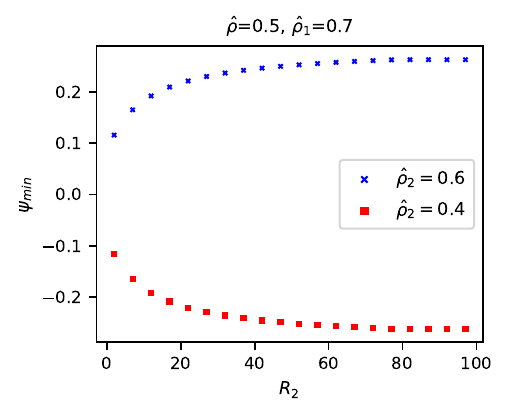}
\caption{Transaction price lower bound $\Psi_\text{min}$ computed by P1 as a function of P2’s total string length $R_2$ after the transaction, for $\lambda = 1$, $N = 5$, with public data $(H = 1, R = 2)$ yielding $\hat{\rho} = 0.5$, and data held by P1 $(R_1 = 98,H_1 = 69)$ yielding $\hat{\rho}_1 = 0.7$. The blue and red curves correspond to $\hat{\rho}_2 = 0.6$ and $\hat{\rho}_2 = 0.4$, respectively. Only feasible values of $R_2$ are shown, corresponding to cases where P1 can sell a private substring of her extra data (not sampled from the public string $\vec s_0$) that, combined with the public data, yields the specified $\hrho_2$.}
\label{fig:psi_lambda_c}
\end{figure}

In Figure~\ref{fig:psi_lambda_c} we plot the lower bound $\Psi_\text{min}$ computed by P1 as a function of the resulting string length $R_2$ that P2 holds after the transaction. The plot is generated for $\lambda = 1$ and $N = 5$. The public estimate is fixed at $\hat{\rho} = 0.5$, corresponding to a public string with $R = 2$ and $H = 1$, while P1’s additional data yields $\hat{\rho}_1 = 0.7$ with $(R_1 = 98, H_1 = 69)$. This means P1 can sell any substring of up to $R_1-R=96$ past outcomes containing at most $H_1-H=68$ heads, since the public portion known to all players is excluded from the sale. 

Given that $\hat{\rho}_1 = 0.7$, the seller believes that the coin is Heads-biased. Along the blue curve, where P2’s posterior shifts from $\hat{\rho} = 0.5$ to $\hat{\rho}_2 = 0.6$, the additional data brings her closer to what P1 considers the correct belief. As $R_2$ increases, the estimate becomes more precise, the variance decreases, and the information becomes more beneficial for P2. This makes the transaction more valuable in a volatility-averse setting, resulting in a price rising with the length of the traded string. In contrast, the red curve corresponds to $\hat{\rho}_2 = 0.4$, where P2 is drawn further away from P1’s belief. In this case, longer strings lead her to an increasingly misleading estimate, making P1 willing to sell the data at a lower price. Consequently, the price is negative and decreases with $R_2$. Overall, these results demonstrate that volatility aversion modifies the equilibrium configuration of the game and the resulting price bounds, while the underlying pricing principle of the model is preserved.

\newpage

\section{Discussion}
In this work we proposed a game-theoretic principle to define and compute the price of data in a competitive setting, where players bet on the outcome of an underlying stochastic process on the basis of asymmetric information about past outcomes. By modeling the potential exchange of data between an informed seller and a less informed buyer, we showed that the value of information naturally emerges from the trade-off between the seller’s expected loss of exclusivity and the buyer’s potential gain in predictive accuracy. Our analysis identified explicit price bounds, $\Psi_{\mathrm{\min}}$ and $\Psi_{\mathrm{\max}}$, which determine whether a mutually beneficial transaction can take place. We characterized different strategic regimes, depending on the beliefs the players hold (or may acquire) about the stochastic process before and after the transaction. 

In some cases, the most informed player (P1) may \emph{cooperate} with a prospective buyer (P2) by sharing valuable information to jointly exploit the least-informed competitors. In a \emph{competitive} regime, instead, the potential revenue from selling the data is not sufficient to cover the seller's losses. 
In a  \emph{symbiotic} regime, the seller can share her data with the buyer at no cost and still improve the utilities of both players involved in the exchange, contradicting the usual intuition that giving up exclusivity necessarily reduces the seller’s profit. Moreover, our analysis shows that for certain combinations of  parameters, the most informed player can paradoxically lose money to the advantage of the least informed players, who instead experience a positive utility. This is due to the fierce competition that can arise between buyer and seller after the transaction has occurred, and further highlights the rich and unusual landscape of interesting and even exotic market effects displayed by intangible assets and foreign to classical textbook economics.

In addition, we found that the feasibility and profitability of data transactions are strongly influenced by market size: as the number of players $N$ increases, competition between the informed parties weakens, and transactions that are unviable in “small” games can become mutually beneficial in “large” ones.

Extending the model to volatility-averse players a new feature arises: equilibrium strategies now explicitly depend on the length of the data strings. This dependence on sample size also makes the transaction price bounds sensitive to the \emph{amount} of data exchanged, not just its  bias, a feature absent in the volatility-neutral case.

Our framework takes inspiration from several approaches proposed in the economic and game theory literature (see \cite{zhang2023survey,pei2020survey,majumdar2025developing} for an overview), but crucially adds the ``complexity'' dimension by treating seller, buyer, and the whole market as an interactive and adaptive system.

Among the existing philosophical approaches to the problem that however did not adopt an interconnected and complex-system perspective, we mention:
\begin{itemize}
    \item \textbf{Cost-based pricing}. Cost-based pricing defines the value of data in terms of the expense incurred to generate information. Classic works in statistical decision theory such as \cite{arrow1949bayes,wald1992sequential} consider the trade-off between experimental cost and informational accuracy, while \cite{bloedel2020cost} extends this to optimal information acquisition strategies. These approaches provide firm economic grounding for how information is produced, but they do not model the dynamics of buyer–seller trade or account for the strategic resale value of data in open markets.
    \item  \textbf{Privacy based pricing}. Under this framework, one aims to assign a price to personal or private data. In   \cite{jaisingh2008privacy}, the authors price personal data based on the perceived subjective loss of privacy. \cite{ghosh2011selling} further develop a formal theory of pricing private data in a differential privacy framework, tying prices to the accuracy of queried data and to the privacy loss individuals experience. The approach has then been expanded in \cite{shen2016pricing,li2014theory}, which focused respectively on large dataset of personal data and on reward mechanisms. While this captures the idea that sharing one's data leads to a loss, the loss is subjective rather than economic, and hence harder to quantify.
    \item \textbf{Economic pricing of exclusivity.}  
    In \cite{bonatti2024selling}, the authors study the sale of information to multiple competing buyers by a monopolist seller who does not itself operate in the market. In their setting, information provision creates negative externalities among buyers, which shape the optimal allocation and pricing of information. In contrast, in our model the seller is an active downstream competitor, so selling data directly reduces the seller’s own strategic advantage. As a result, the cost of information provision in our setting arises from seller–buyer competition rather than exclusively from buyer–buyer externalities.
    \item \textbf{Data utility based pricing}: under this method, the price is determined by the utility of the dataset for the buyer. For example, in \cite{yu2017data} prices account for the data's quality and in \cite{wang1996beyond} several additional attributes are considered, such as data completeness, recentness, and seller trustworthiness. Quality metrics are however subjective and depend on buyer context, making it hard to define universal pricing rules. 
    \item \textbf{Seller profit maximization} 
    A related body of work examines optimal pricing from the seller’s perspective, treating the data owner as a monopolist or dominant provider optimizing a revenue function over buyer segments. Admati and Pfleiderer’s model of a monopolistic seller of information examines how a single seller might price informational goods under demand uncertainty \cite{admati1986monopolistic}, and a more recent work \cite{mehta2021sell} formulates pricing policies for data monetization to maximize seller revenue across multiple buyers. While these models provide insight into seller incentives and product versioning (e.g., tiered data offerings), they typically do not account for strategic competition among buyers or for how sellers’ own use of data affects post-transaction outcomes. 
    \item \textbf{Market‑based pricing}, where data prices emerge from competitive and structural features of data markets. Surveys categorise pricing methods by market structure (sell‑side, buy‑side, two‑sided) and highlight diverse mechanisms and incentives in data marketplaces \cite{zhang2023survey}. In this direction \cite{fernandez2020data} discusses data market platforms where prices and trades result from interactions among traders. These approaches typically analyze specific market designs rather than deriving general equilibrium price bounds under asymmetric information. Our approach fills this gap by analytically characterizing price ranges that emerge from strategic interaction in markets.
\end{itemize}
In summary, while cost-based, privacy-based, quality-based, and seller-centric approaches each contribute valuable perspectives on how information could be valued, they tend to focus on isolated components of the pricing problem and frequently abstract away from strategic interaction, competitive dynamics, and asymmetric information. Market-based pricing models take important steps toward integrating strategy, but often do not analyze rich multi-agent equilibrium structures. Our framework unifies elements of these approaches within a rigorous game-theoretic setting that yields explicit price bounds and uncovers non-trivial market effects emerging from competition and information heterogeneity.

In particular, our framework combines elements from all approaches except for cost-based pricing and privacy-based pricing. Indeed, we price according to both the buyer's utility and the seller's losses, in a competitive setting where utility emerges from exclusivity.
Several market architectures for the exchange of data have been considered. Our basic setting borrows the idea of a \emph{monopolistic seller} from \cite{admati1986monopolistic}. In that case, a single data owner could sell potentially corrupted information to several players. However, in stark contrast with our setting, the data owner was unable to exploit the data herself. 
Our setting is inspired by the statistical mechanics framework introduced in \cite{gamberi2024price}, but departs from it in the 
use of a game-theoretic approach to compute utilities.
Transactions under information asymmetry between the buyer and the seller have been studied in \cite{gradwohl2023selling, bergemann2018design}.
In our study, the information asymmetry is modeled as an information hierarchy, where players can be ranked by the amount of information they possess about the game and about the actions of other players. An analogous concept, the \emph{cognitive hierarchy}, was studied in \cite{stahl1995players, camerer2004cognitive, keynes1936general}.
Finally, our study makes heavy use of the Bayesian games framework, \cite{harsanyi1967games,harsanyi1968games,harsanyi1968games3,zamir2013bayesian}, which is particularly suited to model games with incomplete information.

The simple game-theoretic framework developed here opens up a wealth of opportunities for implementing even more realistic scenarios by relaxing or extending any of the foundational assumptions without compromising the analytical tractability of the model. First, it would be natural to consider a multi-round setting, where the game is played repeatedly rather than in a single round, so that beliefs, strategies, and transaction prices evolve and are updated over time. Another natural extension would be to relax Assumption~{\bf (A5)} by allowing the least-informed players to anticipate the presence of informed agents, as in classical adverse-selection models of financial markets~\cite{glosten1985bid,kyle1985continuous}, and examine how their beliefs about those agents affect the resulting equilibrium strategies and price bounds. Furthermore, allowing players to adopt more sophisticated subjective models of the observed data would also be an interesting direction to explore. If players accounted for temporal correlations, for example through a Markovian model, the ordering of the observations within each string would become relevant and could modify their beliefs, strategies, and the numerical values of the price bounds. The underlying pricing construction would nevertheless remain unchanged, since the price interval would still be obtained by comparing the seller's and buyer's utilities before and after the data transfer. Embedding the game into a complex network of sellers and buyers is likely to generate further feedback loops and even more interesting market effects, as well as going beyond binary outcomes, or allowing for different volatility aversion levels between players. Another extension would be to model the data-selection and disclosure process as a strategic stage, allowing a selection-aware buyer to condition her posterior not only on the transferred string but also on P1's incentives for selecting it. This would require specifying the seller's disclosure rule and the buyer's beliefs about P1's private dataset, and could modify the resulting equilibrium strategies and price bounds. The versatility of this framework and its analytical tractability make it an excellent candidate to explore \emph{in silico} foundational questions about data valuation in interactive markets characterized by asymmetric information.

\section{Methods}
In this section, we provide a formal definition of the game (Section~\ref{sec:pricinggame}), followed by a detailed definition of the (subjective) Nash equilibrium and a description of how it is constructed in scenarios {\bf (i)} (no transaction) and {\bf (ii)} (feasible transaction) (Section~\ref{sec:Nash_eq}). Section~\ref{sec:vol_aversion} introduces the utility function for volatility-averse players. Finally, Section~\ref{sec:procedure} summarizes the computational procedure used to derive equilibrium strategies, utilities, and the associated price bounds.

\subsection{The pricing game}\label{sec:pricinggame}
We consider $N \geq 3$ players who engage in a game of chance based on the outcomes of an underlying stochastic process. Although this process may be arbitrarily complex and correlated, its structure is unknown to the players, who only observe a binary signal $w \in \{0,1\}$ derived from it -- for instance, whether the value of an asset has gone up or down. 
Players pay a unitary entry fee to cast their bets. Once all bets are placed, the outcome $w$ is revealed, and those who correctly predict it share the total pot equally. If no player guesses correctly, the entry fees are refunded. Letting $\vec{a} \in \{0,1\}^N$ represent the action profile (i.e. the bets of all players), the payoff of player $i$, given $w$, is

\begin{equation}\label{eq:round_payoff}
u_i(\vec{a} \mid w) =
\begin{cases}
0\ , & \text{if } \sum_j \delta_{w, a_j} = 0\ , \\[1em]
N  \dfrac{\delta_{w, a_i}}{\sum_j \delta_{w, a_j}} - 1\ , & \text{otherwise}\ ,
\end{cases}
\end{equation}
where $\delta_{w, a_j}$ is the Kronecker delta, equal to 1 if player $j$ guesses correctly (i.e., $a_j = w$), and 0 otherwise. With this payoff form, the game is zero sum, meaning that the sum of all players' payoffs vanishes.

We assume that each player has access to a string $\vec{s}_i \in \{0,1\}^{R_i}$ containing $R_i$ samples from the same process that generates $w$. Each player $i$ further interprets  $\vec{s}_i$  as a sequence of i.i.d. tosses of a possibly biased ``coin'' (Heads $=1$ and Tails $=0$), each coming from a Bernoulli distribution with unknown parameter $\rho \in [0,1]$. By means of Bayes' formula, a player $i$ can form a posterior over the bias $\rho\in[0,1]$ of the coin

\begin{align}
     P_i(\rho|\vec{s}_i) = \frac{P(\vec{s}_i|\rho)\pi_i(\rho)}{\int P(\vec{s}_i|\rho')\pi_i(\rho') d\rho'} = \frac{\rho^{H_i} (1-\rho)^{R_i-H_i}}{Z(R_i,H_i)}\ ,
    \label{eq:posterior}
\end{align}  
where $Z(R,H)=(H! (R-H)!)/(R+1)!$, $H_i \coloneqq \sum_{r=1}^{R_i} (\vec{s_i})_r$ is the number of Heads in $\vec{s}_i$, $P(\vec{s}_i|\rho)=\rho^{H_i}(1-\rho)^{R_i-H_i}$ is the probability of observing the string $\vec{s}_i$, and $\pi_i(\rho)$ is the prior distribution assumed to be uniform. The posterior in \Eqref{eq:posterior} corresponds to a Beta distribution $\mathcal{B}(H_i+1,R_i-H_i+1)$ \cite{berger2013statistical,johnson1995continuous}. 
Accordingly, the subjective probability that player $i$ assigns to the coin landing on Heads in the next toss is given by
\begin{equation}
    \hrho_i \coloneqq P_i(w=1|\vec{s}_i)=\int \rho  P_i(\rho|\vec{s}_i) d\rho =\frac{1+H_i}{2+R_i}\ .
    \label{eq:coin_bias_estimate}
\end{equation}
A further ingredient in the theory is the mixed strategy profile $\vec{p} = (p_1,..., p_N)\in [0,1]^N$, where $p_i$ denotes the probability that player $i$ bets on the outcome $w = 1$ (i.e., the probability that $a_i=1$). Then, the expected payoff for player $i$ must be computed by averaging $u_i(\vec a|w)$ over all possible action profiles as well as over the possible outcomes. Denoting this expectation value as $\U_i(\Vec{p},\hrho_i)$ we have

\begin{equation}\label{eq:U_capital}
\begin{aligned}
    \U_i(\Vec{p},\hrho_i)
    &\coloneqq \E_{\vec{a}}[\E_w[u_i(\Vec{a}|w)]] \\
    &=\hrho_i  \E_{\vec{a}}[u_i(\Vec{a}|w=1)]  +(1-\hrho_i)  \E_{\vec{a}}[u_i(\Vec{a}|w=0)]\ ,
\end{aligned}
\end{equation}
where $w$ is a Bernoulli variable with parameter $\hrho_i$ and $a_j$ is a Bernoulli variable with parameter $p_j$, $\forall j \in \{1,\dots,N\}.$ 
This quantity, linear in $\hrho_i$, represents the utility function that a \emph{volatility-neutral} player $i$ would like to maximize, with respect to her own strategy $p_i$. 
The information available to each player, which includes  their data string $\vec s_i$, therefore plays a central role in shaping strategic behavior. In what follows, we consider a setting in which players may hold different information about the underlying stochastic process, so that the betting strategy each player adopts is directly determined by the data she observes, as well as by her beliefs about the information held by other players. In the next section, we make these informational assumptions explicit and describe how private data may be retained or exchanged among players.

\subsection{Nash equilibrium with asymmetric information}\label{sec:Nash_eq}
A \emph{Nash equilibrium} is a strategy profile such that no player can improve their utility by unilaterally deviating from their strategy, given the strategies of the others. In our setting this notion cannot be applied directly, because strategies and utilities are not common knowledge: private data lead players to different estimates of the bias $\rho$, and -- by Assumption {\bf (A5)} -- the least-informed players are not even aware that such differences exist. Each player therefore reasons about a \emph{perceived game} of her own and computes an equilibrium of that game; following Assumption {\bf (A4)}, she then plays the strategy that this perceived equilibrium assigns to her. We refer to it as a \emph{subjective} Nash equilibrium, to stress that the game being solved is player-dependent.

To make this precise we must first specify what each player believes about the others. We denote by $\hat{\rho}_j^{(i)}$ the bias that player $i$ believes player $j$ attributes to the coin. Under the informational hierarchy induced by Assumptions {\bf (A1)}--{\bf (A2)} and {\bf (A5)}, this quantity takes a very simple form (derived in Appendix C):
\begin{equation}\label{eq:rho_ij}
    \hat{\rho}^{(i)}_j=\hat{\rho}_{\max(i,j)}\ ,
\end{equation}
namely the belief of the \emph{less informed} of the two players. Indeed, if $j$ is less informed than $i$ (i.e. $j>i$), player $i$ knows the data $j$ holds and therefore knows $j$'s belief exactly, $\hat\rho^{(i)}_j=\hat\rho_j$. If instead $j$ is better informed ($j<i$), player $i$ cannot conceive of an informational advantage that she does not possess, and treats $j$ as a copy of herself, $\hat\rho^{(i)}_j=\hat\rho_i$.

Equation \eqref{eq:rho_ij} has a direct consequence for how the perceived equilibrium must be constructed. Player $i$ can correctly anticipate the behaviour of every less-informed player $j>i$; but she also knows that $j$ chooses that behaviour by solving $j$'s \emph{own}, poorer, perceived game, not $i$'s. In other words, the more informed can predict the less informed, but not vice versa, and this asymmetry has to be built into the equilibrium itself. Accordingly, the subjective Nash equilibrium perceived by player $i$
\begin{equation}\label{eq:subjectve_nash_eq}
    \vec{p}^{\,\star}(i) \coloneq (p^\star_1(i),p^\star_2(i),\dots,p^\star_N(i))\ ,
\end{equation}
is specified by two conditions. First, the strategies that $i$ attributes to the less-informed players are the ones those players actually adopt
\begin{equation}\label{eq:Nash_eq_tail}
    p^\star_j(i)=p^\star_j(j)\ ,\qquad \forall j>i\ ;
\end{equation}
second, the strategies of the remaining players, who in $i$'s perception all share her own belief $\hat\rho_i$, form a Nash equilibrium among themselves, given the strategies in \Eqref{eq:Nash_eq_tail}:
\begin{equation}\label{eq:Nash_eq}
    \U_j(\vec{p}^{\,\star}(i), \hat{\rho}_j^{(i)})
    \geq \U_j((p_j, \vec{p}^{\,\star}_{-j}(i)), \hat{\rho}_j^{(i)})\ ,
    \qquad \forall j \leq i\ ,\ \forall p_j \in [0,1]\ .
\end{equation}
Here $\vec{p}^{\,\star}_{-j}(i)$ denotes the vector of strategies of all players except $j$, i.e., $\vec{p}^{\,\star}_{-j}(i) \coloneqq (p_1^{\star}(i), \dots, p_{j-1}^{\star}(i), p_{j+1}^{\star}(i), \dots, p_N^{\star}(i))$.\footnote{We denote by $(p_j, \vec{p}^{\,\star}_{-j}(i))$, the vector $(p_1^{\star}(i),..., p_{j-1}^{\star}(i),p_j, p_{j+1}^{\star}(i),..., p_N^{\star}(i))$.} A subjective Nash equilibrium is said to be \emph{symmetric} if all the players that $i$ perceives as identical to one another adopt the same strategy.

Equations \eqref{eq:Nash_eq_tail}--\eqref{eq:Nash_eq} determine $\vec p^{\,\star}(i)$ recursively. One starts from the least-informed player, $i=N$: there \Eqref{eq:Nash_eq_tail} is empty and \Eqref{eq:Nash_eq} reduces to an ordinary Nash equilibrium of a fully symmetric game in which all players share the public belief $\hat\rho$. One then moves up the hierarchy, $i=N-1,\dots,1$, at each step solving a Nash equilibrium problem among the first $i$ players only, while the strategies of the remaining $N-i$ players are already fixed by \Eqref{eq:Nash_eq_tail}. The recursion terminates after $N$ steps and always admits a solution. When more than one Nash equilibrium exists at a given step, we select the symmetric one, which is the natural choice because the players involved perceive one another as identical\footnote{Other Nash equilibria may exist at each step; we assume throughout that all players adopt this same selection rule.}.
Given a strategy profile $\vec{p}_{-j}$ for all players except player $j$, according to player $i$ the \emph{best response} of player $j$ is the strategy $p_{j,(i)}^{\text{br}}(\vec{p}_{-j})$ that maximizes her utility function
\begin{equation}\label{eq:best_resp}
   p_{j,(i)}^{\text{br}}(\vec{p}_{-j}) \coloneqq \mathop{\arg\max}_{p_j \in [0,1]} [\U_j((p_j, \vec{p}_{-j}), \hat{\rho}_j^{(i)})]\ .
\end{equation}
For the game under consideration, a best response always exists; however, it is not necessarily unique: multiple strategies may yield the same maximum utility\footnote{In case of non-uniqueness $p_{j,(i)}^{\text{br}}(\vec{p}_{-j})$ is the set of values of $p_j$ for which the maximum in \Eqref{eq:best_resp} is attained.}. With this definition, condition \eqref{eq:Nash_eq} can be equivalently stated as the requirement that, from the point of view of player $i$, the strategy $p_j^{\star}(i)$ of each player $j\leq i$ be a best response to the strategies of the others
\begin{equation}
p_j^{\star}(i) \in p_{j,(i)}^{\text{br}}(\vec{p}^{\,\star}_{-j}(i)) \quad \forall j \leq i\ .
\end{equation}
This is the form in which the equilibrium conditions are solved in practice, following the procedure of Section~\ref{sec:procedure}.

From the subjective equilibria, we can finally define the actual equilibrium profile as
\begin{equation}\label{eq:p_star}
    \vec p^{\,\star}\coloneqq (p^\star_1(1),p^\star_2(2),\dots,p^\star_N(N))\ ,
\end{equation}
namely the collection of strategies actually adopted by each player based on their own subjective equilibrium $p^\star_i(i)$.\footnote{$\vec p^\star$ is not common knowledge; each player $i$ in general has access only to the component $p^\star_i(i)$, i.e., their own strategy. To ease the notation, in Section~\ref{sec:model_intro} we denoted the quantity $p_i^\star(i)$ simply as $p_i^\star$.} Note that in general $p^{\star}_j(i)\neq p^{\star}_j(k)$ for two players $i \neq k$, as they hold different information and therefore form different beliefs about the others. In particular, $\vec p^{\,\star}$ is not the Nash equilibrium of any single game played by all $N$ participants on a common representation of the information structure: in such a game the least-informed players would best-respond to P1's actual strategy, which Assumption {\bf (A5)} explicitly rules out. What \Eqref{eq:p_star} does describe is the profile of strategies that are simultaneously optimal, each within the game that the corresponding player perceives. This is a standard, if less familiar, way of extending equilibrium reasoning to situations in which players hold mutually incompatible views of the game \cite{mertens1985formulation,zamir2013bayesian}; a self-contained construction of the underlying belief structure and of the resulting equilibrium is given in Appendices C and D.

Because the information available to players differs between the two scenarios introduced in the previous section, the equilibrium strategy profile $\vec p^{\,\star}$ defined above must be characterized separately depending on whether a data transaction has occurred or not. In the pre-transaction case, $\vec p^{\,\star}$ coincides with the profile $\vec p^{\,\text{pre}}$ defined in \Eqref{eq:p_pre_eq}, while if a transaction occurs it coincides with the post-transaction profile $\vec p^{\,\text{post}}$ defined in \Eqref{eq:p_post_eq}. These profiles are introduced and analyzed in Section~\ref{sec:Nash}.

\subsection{Volatility aversion}\label{sec:vol_aversion}
In the setting we just described, the utility function $\U_i(\Vec{p},\hrho_i)$ from \Eqref{eq:U_capital} only depends on the mean $\hrho_i$ of the posterior in \Eqref{eq:coin_bias_estimate}. However, since this estimate $\hat{\rho}_i$ is obtained from a finite data sample, it is inherently subject to statistical fluctuations.
Consequently, even though $\hat{\rho}_i$ serves as a point estimate, it may significantly deviate from the true bias, especially for short strings.

It is therefore natural to assume that rational players might take this uncertainty into account when evaluating their strategies. Specifically, they may consider the variance of their utility with respect to the randomness in the posterior $P_i(\rho|\vec{s}_i)$ as a measure of the uncertainty associated with their estimate $\hat{\rho}_i$. Depending on their preferences, players may be more or less tolerant to it.

We incorporate this idea by considering \textit{volatility-averse} players who adjust their effective utility by penalizing large variance in their subjective estimates. This leads to a modified utility function that reflects not only the expected gain but also the confidence the player has in their estimate of the gain. We denote the utility function of a volatility-averse player as $\U_i(\Vec{p},\hrho_i,R_i;\lambda)$, and define it as

\begin{equation}\label{eq:U_capital_lambda}
\begin{aligned}
    \U_i(\Vec{p},\hrho_i,R_i;\lambda) \coloneqq \U_i(\Vec{p},\hrho_i) - \lambda \operatorname{Var}_{\rho_i}[\U_i(\Vec{p},\rho_i)]\ ,
\end{aligned}
\end{equation}
where $\lambda \ge 0$ is a parameter that quantifies players' aversion to volatility (assumed identical across all players for simplicity), and $ \operatorname{Var}_{\rho_i}[\U_i(\Vec{p},\rho_i)]$ denotes the variance of \Eqref{eq:U_capital} where $\rho_i$ is a random variable following the posterior distribution in \Eqref{eq:posterior}. The choice of a variance penalty is fairly standard in portfolio optimization \cite{e5a1bb8f-41b7-35c6-95cd-8b366d3e99bc,markowitz2000mean,phelps2024user}.
Since $\U_i(\Vec{p},\rho_i)$ is linear in $\rho_i$ (see \Eqref{eq:U_capital}), this variance term can be computed analytically:

\begin{align}
\nonumber \operatorname{Var}_{\rho_i}[\U_i(\Vec{p},\rho_i)] &= (\E_{\vec{a}}[u_i(\Vec{a}|w=1)-u_i(\Vec{a}|w=0)])^2 \operatorname{Var}_{\rho_i}[\rho_i] \\&=  (\E_{\vec{a}}[u_i(\Vec{a}|w=1)-u_i(\Vec{a}|w=0)])^2  \frac{\hrho_i(1-\hrho_i)}{R_i+3}\ ,
\label{eq:U_capital_lambda_complete} 
\end{align}
where we have used the fact that $\rho_i$ follows a Beta distribution $\mathcal{B}(H_i+1, R_i-H_i+1)$. The dependence on $\hrho_i$ and $R_i$ enters solely through the final factor, which decreases with increasing sample size $R_i$.
Higher values of $\lambda$ correspond to stronger aversion to uncertainty. When $\lambda = 0$, the player is volatility-neutral, and the utility function reduces to the expected value in \Eqref{eq:U_capital}. In the case of volatility-averse players, the admissible price range for the transaction can be defined analogously to the volatility-neutral setting, using the updated form of the utility function given in \Eqref{eq:U_capital_lambda}. The definition of the Nash equilibrium also remains formally identical, with the modified utility replacing the original one in all expressions.

\subsection{Summary of computational procedure}\label{sec:procedure}
This section provides a step-by-step roadmap for computing the equilibrium strategies, the corresponding utilities, and the associated price bounds introduced in the previous sections. The methodology proceeds in a hierarchical manner, starting from the least-informed players and progressively incorporating more informed agents. This ordering reflects the information structure of the game and clarifies how equilibrium strategies are constructed as informational asymmetries are introduced. It is precisely the recursion defined by Eqs.~\eqref{eq:Nash_eq_tail}--\eqref{eq:Nash_eq}, run from $i=N$ down to $i=1$.

The same sequence applies to both volatility-neutral and volatility-averse players. The only difference lies in the utility function used: $\U_i(\vec p,\hat{\rho}_i)$ in \Eqref{eq:U_capital} for volatility-neutral players and $\U_i(\vec p,\hat{\rho}_i,R_i;\lambda)$ in \Eqref{eq:U_capital_lambda} for volatility-averse players. All equilibrium definitions and best-response conditions remain formally unchanged.

Analytical derivations of the equilibrium strategies and utilities, together with details of the best-response approach used, are provided in the Appendix (Appendix A for the volatility-neutral case and Appendix B for the volatility-averse case). 

\begin{enumerate}

    \item \textbf{Equilibrium strategy of least-informed players, $p_{\text{eq}}$.} 
    Less informed players, who rely solely on the publicly available string $\vec{s}_0$, assume that all participants have access only to this shared data and base their strategies accordingly. In particular, by Assumption \textbf{(A5)}, they are unaware that some players (notably P1, and potentially P2 if the transaction occurs) may hold additional private information. As a result, they perceive the game as fully symmetric and believe that all participants face identical decision problems.

    Under this \emph{perceived} informational symmetry, it is natural to restrict attention to a symmetric subjective Nash equilibrium in which all players adopt the same strategy\footnote{Other Nash equilibria may exist, our selection of the symmetric equilibrium follows the criterion specified in Section~\ref{sec:Nash_eq}.}. Accordingly, we determine the equilibrium strategy $p_{\text{eq}}$ using a best-response approach in a fully symmetric setting, where all players compute the same belief $\hat{\rho}$ via \Eqref{eq:coin_bias_estimate} and share identical utility functions and strategies. This step fixes the equilibrium strategy $p_{\text{eq}}$ adopted by the least-informed players in both the pre- and post-transaction scenarios. 

    \item \textbf{Pre-transaction equilibrium strategy for P1, $ (q_1)_{\text{eq}}^{\text{pre}}$.}  
    Although the least-informed players perceive the game as fully symmetric, the actual information structure is different. Before any transaction occurs, the prospective seller P1 has access to a longer data string than the other $N-1$ players, allowing her to form a distinct estimate $\hat{\rho}_1$ of the coin bias, which generally differs from the common estimate $\hat{\rho}$.

    Since $\hat{\rho}$ is common knowledge, P1 can anticipate that players $2,\dots,N$ will adopt the symmetric equilibrium strategy $p_{\text{eq}}$ determined in Step~1. P1 can therefore evaluate her utility when all other players use $p_{\text{eq}}$ and determine her best response. This yields P1’s pre-transaction equilibrium strategy $(q_1)_{\text{eq}}^{\text{pre}}$. 

    At this stage, the pre-transaction equilibrium profile $\vec p^{\,\text{pre}}$ defined in \Eqref{eq:p_pre_eq} is fully specified.

    \item \textbf{Post-transaction equilibrium strategy for P2, $q_{\text{eq}}$.}  
    The next step is to characterize the strategy adopted by P2 when P1 sells her a portion of private data. Following the transaction, P2 holds a longer consecutive data string $(H_2,R_2)$ than the least-informed players, which allows her to form a distinct estimate $\hat{\rho}_2$ of the coin bias.

    The transaction also reveals to P2 that P1 possesses a data string of length \emph{at least} $R_2$. In the absence of further information, the only reasonable assumption for P2 is that P1 holds the same information as herself and consequently forms the same estimate $\hat{\rho}_2$. As a result, P2 believes P1 to be strategically identical to herself and expects her to adopt the same betting strategy.

    The remaining $N-2$ players, who are not involved in the transaction and are unaware that more informed players exist, continue to rely on the public string and to adopt the equilibrium strategy $p_{\text{eq}}$ determined in Step~1. Since P2 knows the information on which these players base their decisions, their behavior can be anticipated.

    Under these beliefs, we determine $q_{\text{eq}}$ by looking for a subjective Nash equilibrium that is symmetric between P1 and P2, while the remaining players use $p_{\text{eq}}$. The equilibrium strategy $q_{\text{eq}}$ is obtained by applying the best-response approach in this setting.

    \item \textbf{Post-transaction equilibrium strategy for P1, $(q_1)_{\text{eq}}^{\text{post}}$.}  
    In the previous step, the strategy $q_{\text{eq}}$ was determined under P2’s belief that the seller P1 evaluates utilities using the same estimate $\hat{\rho}_2$. This belief, however, does not reflect the actual information structure of the game. Unless the full dataset is transferred, P1 retains additional private information and continues to base her decisions on her own estimate $\hat{\rho}_1$, which generally differs from $\hat{\rho}_2$.

    Since P1 has complete knowledge of the information held by all other players, she can anticipate both the strategy $q_{\text{eq}}$ adopted by P2 and the strategy $p_{\text{eq}}$ adopted by the remaining players. P1’s post-transaction equilibrium strategy $(q_1)_{\text{eq}}^{\text{post}}$ is then obtained by computing her best response given the strategies $q_{\text{eq}}$ and $p_{\text{eq}}$ adopted by her opponents.

    This completes the derivation of the post-transaction equilibrium profile $\vec p^{\,\text{post}}$ defined in \Eqref{eq:p_post_eq}.

    \item \textbf{Lower price bound: computation of $\Psi_{\text{min}}$.}  
    Having determined the equilibrium strategy profiles both before and after a possible transaction, $\vec p^{\,\text{pre}}$ and $\vec p^{\,\text{post}}$, we evaluate the seller’s utility $\U_1$ at equilibrium in the two scenarios using P1’s own estimate $\hat{\rho}_1$.

    The difference between these two quantities determines the minimum acceptable price $\Psi_{\text{min}}$ that P1 should demand in order to compensate for the potential loss in utility induced by the increased competition following the transaction, as defined in \Eqref{eq:Psi_min}.

    \item \textbf{Upper price bound: computation of $\Psi_{\text{max}}$.}  
    As defined in \Eqref{eq:Psi_max}, the price upper bound $\Psi_{\text{max}}$ corresponds to the increase in P2’s utility induced by the transaction. Importantly, $\Psi_{\text{max}}$ is defined from the point of view of the seller P1, and therefore requires evaluating the buyer’s utility using P1’s information set. A preliminary step is thus to compute P2’s utility before and after the transaction as estimated by P1.
    
    We illustrate this computation in the volatility-neutral setting that underlies the main analysis. Prior to any data exchange, P2 belongs to the group of least-informed players and adopts the common equilibrium strategy $p_{\text{eq}}$. By symmetry, all least-informed players obtain the same utility before the transaction. Since the game is zero-sum, the sum of all players’ utilities must vanish, which implies that P1’s utility is exactly offset by equal and opposite expected utilities among the remaining $N-1$ players. P1 can therefore infer P2’s pre-transaction utility as

    \begin{equation}\label{eq:payoff_least_pre}
        \U_2(\vec{p}^{\,\text{pre}}, \hat{\rho}_1)
    = -\frac{1}{N-1}\,\U_1(\vec{p}^{\,\text{pre}}, \hat{\rho}_1)\ ,
    \end{equation}
    where $\U_1(\vec{p}^{\,\text{pre}}, \hat{\rho}_1)$ has been computed in the previous step. 

    After the transaction, P2’s utility $\U_2(\vec{p}^{\,\text{post}}, \hat{\rho}_1)$ is evaluated by P1 using the same utility expression as $\U_1(\vec{p}^{\,\text{post}}, \hat{\rho}_1)$, computed in the previous step, with the strategies $q_{\text{eq}}$ and $(q_1)_{\text{eq}}^{\text{post}}$ swapped. The difference between the post- and pre-transaction utilities yields the upper bound $\Psi_{\text{max}}$ as defined in \Eqref{eq:Psi_max}.

\end{enumerate}
Once the above steps are completed, the price bounds $\Psi_{\text{min}}$ and $\Psi_{\text{max}}$ defining the fair trading range for the dataset are fully determined. A transaction between P1 and P2 can only occur if $\Psi_{\text{max}} > 0$ and $\Psi_{\text{min}} \leq \Psi_{\text{max}}$, in which case any non-negative price $\Psi \geq 0$ within the interval $[\Psi_{\text{min}}, \Psi_{\text{max}}]$ constitutes a mutually beneficial agreement.

As a final remark, the post-transaction utility of each of the remaining $N-2$ least-informed players can also be evaluated from the point of view of P1. In the volatility neutral case, since these players are symmetric and the game is zero-sum, their post-transaction utility is uniquely determined by the utilities of P1 and P2 as inferred by P1. In particular 
\begin{equation}\label{eq:payoff_least_post_povp1}
\U_i(\vec{p}^{\,\text{post}}, \hat{\rho}_1)
=
-\frac{\U_1(\vec{p}^{\,\text{post}}, \hat{\rho}_1)
+
\U_2(\vec{p}^{\,\text{post}}, \hat{\rho}_1)}{N-2},\quad \forall i \geq 3\ .
\end{equation}

\section*{Acknowledgments}
P.V. acknowledges support from UKRI FLF Scheme (No. MR/X023028/1).

\newpage

\appendix

\section{Volatility-neutral players}\label{sec:Appendix_A}
This Appendix provides the full derivation of the equilibrium strategy profiles $\vec{p}^{\,\text{pre}} =((q_1)_\text{eq}^\text{pre}, p_\text{eq}, \dots, p_\text{eq})$ and $\vec{p}^{\,\text{post}} = ((q_1)_\text{eq}^\text{post}, q_\text{eq}, p_\text{eq}, \dots, p_\text{eq})$,
together with the corresponding equilibrium utilities for the seller and the buyer. The derivation follows the computational procedure outlined in Section V~D of the main text and focuses on the case of volatility-neutral players.

\subsection{Calculation of $p_\mathrm{eq}$}\label{A1_uninformed}
In this section, we determine the equilibrium strategy $p_{\text{eq}}$ adopted by the least-informed players, who rely exclusively on the publicly available string $\vec{s}_0$. These players assume that all participants have access only to this shared information and are unaware that any player may hold additional private data. As a result, they perceive the game as fully symmetric and believe that all players face identical decision problems.
Under this perceived informational symmetry, it is natural to focus on a symmetric subjective Nash equilibrium in which all players adopt the same strategy. Each player uses the public string to estimate the coin bias via Eq.~(8) of the main text, leading to a common belief $\hat{\rho}$ and identical utility functions across players. The equilibrium strategy $p_{\text{eq}}$ is then obtained using a standard best-response approach in this fully symmetric setting.
Given the symmetry of this setting, it is sufficient to focus on a single representative player, say player $i$, as the same analysis applies identically to all others. To find the symmetric Nash equilibrium, we assume that every player except $i$ adopts strategy $p$, and compute the best response $q^{\text{br}}_i(p)$ by evaluating player $i$'s expected payoff under the strategy profile $\vec{p} = (p, \dots, p, q_i, p, \dots, p)$, where $q_i$ appears in the $i$-th position.
Given the action profile $\vec{a} \in \{0,1\}^N$, which represents all players' bets, and the toss outcome $w$, let $u_i(\vec{a} \mid w)$ denote the payoff of player $i$ conditional on $w$. Her expected payoff is then the average over all possible outcomes, weighted by the probabilities she assigns to them:

\begin{equation}
    \E_w[u_i(\Vec{a}|w)]=p_i(w=1)  u_i(\Vec{a}|w=1)  +p_i(w=0)  u_i(\Vec{a}|w=0)\ ,
\end{equation}
where $p_i(w=1)= \hrho$ and is the same for all players. Using Eq.~(6) of the main text we get:

\begin{equation}\label{eq:bets_on_head}
 \begin{aligned}
     \E_w[u_i(a_i=1,\Vec a_{-i}|w)] = \hrho \left[  \frac{N}{1+\sum_{j \neq i} \delta_{1 a_j}}-1 \right] + (1-\hrho)\left[- \mathbbm{1}(\sum_{j \neq i} \delta_{0 a_j}>0)\right]\ ,
     \end{aligned}
\end{equation}
\begin{equation}\label{eq:bets_on_tails}
 \begin{aligned}
     \E_w[u_i(a_i=0,\Vec a_{-i}|w)] =  \hrho \left[- \mathbbm{1}(\sum_{j\neq i} \delta_{1 a_j}>0)\right] + (1-\hrho) \left[  \frac{N}{1+\sum_{j\neq i} \delta_{0 a_j}}-1\right]\ ,
     \end{aligned}
\end{equation}
where we have used the binary indicator function $\mathbbm{1}(P)=1$ if  $P$  is true, and zero otherwise.  In \Eqref{eq:bets_on_head}, player $i$ bets on Heads ($a_i=1$). She wins with probability $\hat{\rho}$, in which case the pot is shared among all players who also guessed correctly (a total of $\sum_{j \neq i} \delta_{1a_j}$ opponents). With probability $(1-\hat{\rho})$, her prediction is wrong, and she loses her entry fee unless no other player predicted the correct outcome (i.e., when $\mathbbm{1}(\sum_{j \neq i} \delta_{0a_j} > 0) = 0$). Analogous considerations apply to the case where she bets on Tails, as given in \Eqref{eq:bets_on_tails}. To compute player $i$'s utility, we must also average over possible actions of all other players. Since in the considered strategy profile $\vec{p}$ each of the remaining $N-1$ players bets on Heads with probability $p$, we can write:

\begin{equation}
 \begin{aligned}
    \U_i(\Vec{p},\hrho) \coloneqq \E_{\Vec{a}}\E_w[u_i(\Vec{a}|w)]= & q_i \sum_{k=0}^{N-1} \binom{N-1}{k} p^k (1-p)^{N-k-1} \E_w[u_i(a_i=1,a_{-i}=[1^k0^{N-k-1}]|w)] + \\ 
    \;& (1-q_i) \sum_{k=0}^{N-1} \binom{N-1}{k} p^k (1-p)^{N-k-1} \E_w[u_i(a_i=0,a_{-i}=[1^k0^{N-k-1}]|w)]\ ,
     \end{aligned}
\end{equation}
where $(a_i=1,a_{-i}=[1^k0^{N-k-1}])$ denotes the case where player $i$ bets on Heads, $k$ of the remaining players also bet on Heads, and the remaining $N-k-1$ players bet on Tails. The analogous notation applies when $a_i = 0$. Using the following identities:
\begin{equation}
 \begin{aligned}
   \sum_{k=0}^{N-1} \binom{N-1}{k} p^k (1-p)^{N-k-1} = 1\ ,
     \end{aligned}
        \label{eq:identities_forsum_start}
\end{equation}
\begin{equation}
 \begin{aligned}
   \sum_{k=0}^{N-1} \frac{(N-1)!}{k!(N-1-k)!} \frac{p^k (1-p)^{N-k-1}}{1+k} = \frac{ 1 - (1-p)^N }{Np}\ ,
     \end{aligned}
\end{equation}
\begin{equation}
 \begin{aligned}
   \sum_{k=0}^{N-1} \binom{N-1}{k} \frac{p^k (1-p)^{N-k-1}}{N-k}=\frac{ 1 - p^N }{N(1-p)}\ ,
     \end{aligned}
             \label{eq:identities_forsum_end}
\end{equation}
we can compute the utility function explicitly:
\begin{equation}
 \begin{aligned}
    \U_i(\Vec{p},\hrho)= &q_i \left\{ \hat{\rho}\frac{1-(1-p)^N}{p} - \hat{\rho} + \hat{\rho}(1-p^{N-1}) - (1-p^{N-1}) \right\} +\\& +(1-q_i) \left\{ \hat{\rho} - \frac{\hat{\rho}[1-p^N]}{1-p}-\hat{\rho}[1-(1-p)^{N-1}] + \frac{1-p^N}{1-p} - 1 \right\} 
    \\&=q_i \left\{\hrho\left[\frac{1-(1-p)^{N-1}}{p}\right] + (1-\hrho)\left[\frac{p^{N-1}-1}{1-p}\right] \right\}+ \frac{1-\hrho}{1-p}(1-p^N)+\hrho(1-p)^{N-1}-1 
    \\&= q_i \left\{ \left[ \hrho - \frac{p-p^N}{1-p^N-(1-p)^N}\right]\frac{1-p^N-(1-p)^N}{p(1-p)}\right\} +\left[  \frac{p-p^N}{1-p^N-(1-p)^N}- \hrho \right]\frac{1-p^N-(1-p)^N}{1-p}\ .
     \end{aligned}
     \label{eq:payoff_uninformed_app}
\end{equation}
Since the utility function is linear in $q_i$, we define its slope $\alpha$ as follows:
\begin{equation}
 \begin{aligned}
    \alpha(\hrho,p,N) \coloneqq \left[ \hrho - \frac{p-p^N}{1-p^N-(1-p)^N}\right]\frac{1-p^N-(1-p)^N}{p(1-p)}\ .
     \end{aligned}
\end{equation}
The sign of $\alpha(\hat{\rho}, p, N)$ determines the best response of player $i$, which maximizes her utility:
\begin{itemize}
    \item If $\alpha(\hat{\rho}, p, N) > 0$, the best response is the pure strategy where the player bets Heads with certainty  ($q^{\text{br}}_i(p) \equiv 1$).
    \item If $\alpha(\hat{\rho}, p, N) < 0$, the best response is the pure strategy where the player bets Tails with certainty ($q^{\text{br}}_i(p) \equiv 0$).
    \item If $\alpha(\hat{\rho}, p, N) = 0$, the utility function is independent of $q_i$, hence any $q_i\in[0,1]$ is a best response.
\end{itemize}
To analyze the sign of  $\alpha(\hat{\rho}, p, N)$, we rewrite it as:
\begin{equation}
 \begin{aligned}
     \alpha(\hat{\rho}, p, N)=[\hrho- K(p,N)]G(p,N)\ ,    
     \end{aligned}
\end{equation}
where we define the auxiliary functions:

\begin{equation}
 \begin{aligned}
    K(p,N) \coloneqq  \frac{p-p^N}{1-p^N-(1-p)^N}\ ,
     \end{aligned}
     \label{eq:Kpn}
\end{equation}
\begin{equation}
 \begin{aligned}
    G(p,N) \coloneqq  \frac{1-p^N-(1-p)^N}{p(1-p)}\ . 
     \end{aligned}
\end{equation}
For all $N \geq 3$ and $p \in (0,1)$, the function $K(p, N)$ is strictly increasing and positive, with limits:
\begin{equation}
\lim_{p \to 0} K(p, N) = \frac{1}{N}\ , \qquad \lim_{p \to 1} K(p, N) = 1 - \frac{1}{N}\ .
\end{equation}
Similarly, $G(p, N)$ is strictly positive for $p \in (0,1)$ and satisfies:
\begin{equation}
\lim_{p \to 0} G(p, N) = \lim_{p \to 1} G(p, N) = N\ .
\end{equation}
These properties imply that the sign of $\alpha(\hat{\rho}, p, N)$ is determined by whether $\hat{\rho}$ is larger or smaller than $K(p, N)$. If the public string is such that $\hrho \in [\frac{1}{N},1-\frac{1}{N}]$, there exists a unique symmetric Nash equilibrium where all players follow the same strategy $p=q_i=p_\mathrm{eq}$ solution of:
\begin{equation}\label{eq:eq_condition_uninformed_app}
 \begin{aligned}
    \hrho =  \frac{p_\mathrm{eq}-p_\mathrm{eq}^N}{1-p_\mathrm{eq}^N-(1-p_\mathrm{eq})^N}=K(p_\mathrm{eq},N)\ ,
     \end{aligned}
\end{equation}
since in this case $\alpha(\hrho,p_\mathrm{eq},N)=0$ and every player utility is fixed by the strategy of the others. This Nash equilibrium corresponds to a pure strategy in the limiting cases. Specifically, when $\hat{\rho} = 1 - \frac{1}{N}$, the unique equilibrium is $p_\mathrm{eq} = 1$, while for $\hat{\rho} = \frac{1}{N}$, it is $p_\mathrm{eq} = 0$. Equation \ref{eq:eq_condition_uninformed_app} is in general not analytically invertible for $N>3$, but can be easily solved numerically to find the equilibrium strategy given $\hrho$.

When $\hat{\rho} > 1 - \frac{1}{N}$, the slope $\alpha$ can only be positive, implying that every player's best response is always to bet on Heads, regardless of what other players are doing. This means that a pure strategy symmetric Nash equilibrium exists with $p=q_i=p_\mathrm{eq}=1$. Conversely, when $\hat{\rho} < \frac{1}{N}$, for the same reasons a pure strategy symmetric Nash equilibrium exists with $p=q_i=p_\mathrm{eq}=0$.

The utility function in \Eqref{eq:payoff_uninformed_app} can be rewritten as:
\begin{equation}
 \begin{aligned}
    \U_i(\Vec{p},\hrho)= (q_i-p) \alpha(\hrho,p,N)\ ,
     \end{aligned}
\end{equation}
from which it directly follows that, in all symmetric equilibrium configurations, the utility of each player is zero.

\subsection{Calculation of $(q_1)_\text{eq}^\text{pre}$}\label{A2_sellerbefore}
In this section, we derive the equilibrium strategy $(q_1)_\text{eq}^\text{pre}$ adopted by the seller in the pre-transaction scenario, where no data exchange has occurred and all of her opponents rely exclusively on the publicly available information.
Accordingly, we now assume that among the $N$ players who share the same information, one of them, the prospective seller, has access to a longer history of past outcomes of the process. This enables the seller to derive a different estimate of the coin’s bias based on a larger data sample.

Denoting this privileged player as Player 1 (P1), we let $\hat{\rho}_1$ represent her estimated bias, which in general differs from the common estimate $\hat{\rho}$ used by the remaining $N - 1$ players.

Since $\hat{\rho}$ is common knowledge among all players, P1 can anticipate the strategy adopted by the remaining $N - 1$ players. Specifically, players $2,\ldots,N$ follow the symmetric equilibrium $p_\mathrm{eq}$ derived in section~\ref{A1_uninformed}, as they are unaware that P1 possesses additional information. Believing that all players, including P1, rely on the same estimate $\hat{\rho}$, they also expect her to act according to this common strategy $p_\mathrm{eq}$.

Therefore, P1 can evaluate her utility assuming the strategy profile $\Vec{p} = (q_1, p_\mathrm{eq}, p_\mathrm{eq}, \dots, p_\mathrm{eq})$, obtaining the same functional form as in \Eqref{eq:payoff_uninformed_app}:
\begin{equation}
 \begin{aligned}
    \U_1(\Vec{p},\hrho_1)= & q_1 \left\{ \left[ \hrho_1 - \frac{p_\mathrm{eq}-p_\mathrm{eq}^N}{1-p_\mathrm{eq}^N-(1-p_\mathrm{eq})^N}\right]\frac{1-p_\mathrm{eq}^N-(1-p_\mathrm{eq})^N}{p_\mathrm{eq}(1-p_\mathrm{eq})}\right\} +\left[  \frac{p_\mathrm{eq}-p_\mathrm{eq}^N}{1-p_\mathrm{eq}^N-(1-p_\mathrm{eq})^N}- \hrho_1 \right]\frac{1-p_\mathrm{eq}^N-(1-p_\mathrm{eq})^N}{1-p_\mathrm{eq}}\ .
     \end{aligned}
\end{equation}
Since the utility function remains linear in $q_1$, and the functional form is the same as in ~\eqref{eq:payoff_uninformed_app}, P1’s best response is again determined by the sign of the slope:
\begin{itemize}
    \item If $\hrho_1 > K(p_\mathrm{eq},N)$, the best response is the pure strategy where the player bets Heads with certainty ($q^{\text{br}}_1(p_\mathrm{eq}) \equiv 1$).
    \item If $\hrho_1 < K(p_\mathrm{eq},N)$, the best response is the pure strategy where the player bets Tails with certainty ($q^{\text{br}}_1(p_\mathrm{eq}) \equiv 0$).
    \item If $\hrho_1 = K(p_\mathrm{eq},N)$, the utility function is independent of $q_1$, hence any $q_1\in[0,1]$ is a best response,
\end{itemize}
where the function $K(p,N)$ is defined in \Eqref{eq:Kpn}.  Since the remaining $N - 1$ players follow the symmetric Nash equilibrium strategy $p_\mathrm{eq}$ determined by the common belief $\hat{\rho}$, P1 can compute $K(p_\mathrm{eq}, N)$ accordingly and compare it with her own estimate $\hat{\rho}_1$:
\begin{align}
    K(p_\mathrm{eq},N)=\begin{cases}
        \frac{1}{N}& \text{ if } \hrho < \frac{1}{N}\\
        1-\frac{1}{N}& \text{ if } \hrho > 1-\frac{1}{N}\\
        \frac{p_\mathrm{eq} - p_\mathrm{eq}^N}{1 - p_\mathrm{eq}^N - (1-p_\mathrm{eq})^N}=\hrho& \text{ if } \hrho \in [\frac{1}{N},1-\frac{1}{N}]
    \end{cases}
\end{align}
It follows that P1’s equilibrium strategy before the transaction, denoted as $(q_1)_\text{eq}^\text{pre}$, can be compactly expressed as:
\begin{align}
\label{eq:P1_strategy_pre}
   (q_1)_\text{eq}^\text{pre} = \begin{cases}
         \mathbbm{1}[\hrho_1>\hrho] & \text{if } \hrho_1\in[1/N,1-1/N]\ , \\
        0 & \text{if } \hrho_1<1/N\ , \\
        1 & \text{if } \hrho_1>1-1/N\ .
    \end{cases}
\end{align}
Note that when $\hat{\rho}_1 = \hat{\rho}$ and both lie within the interval $[1/N, 1 - 1/N]$, P1's utility is completely determined by the strategies of the other players and remains fixed at zero, regardless of her own action. In this special case, any strategy constitutes a best response that yields a zero utility. In \Eqref{eq:P1_strategy_pre}, this specific scenario is handled by setting $(q_1)^\text{pre}_\text{eq} = 0$ by convention, since P1’s choice has no impact on her resulting utility.

\subsection{Calculation of $q_{\mathrm{eq}}$}\label{section:A3_buyer_after}
In this section, we derive the equilibrium strategy $q_{\mathrm{eq}}$ adopted by the buyer when a transaction occurs, in which the seller (P1) shares a portion of her private data with another player (P2). As a result of the transaction, P1 retains her original estimate $\hat{\rho}_1$, which remains unknown to all other players, including the buyer. P2, having acquired partial information from the seller, forms a new estimate $\hat{\rho}_2$, which generally may differ from both $\hat{\rho}_1$ and the public estimate $\hat{\rho}$ used by the remaining $N-2$ players.
In this setting, the following assumptions are made:
\begin{itemize}
    \item The seller P1 has complete information: she knows $\hat{\rho}_1$, $\hat{\rho}_2$, and $\hat{\rho}$.
    \item The buyer P2 knows $\hat{\rho}_2$ and $\hat{\rho}$, and assumes that P1 acts based on $\hat{\rho}_2$. This is the most reasonable assumption because the transaction reveals to P2 that P1 holds a dataset of length \emph{at least} equal to her newly acquired one. Lacking evidence of further private information, the buyer therefore assumes that the seller possesses the same data and hence the same estimate $\hat{\rho}_2$, even though in reality P1 may still have access to additional private information.
    \item Players $3, \dots, N$ are unaware of the information asymmetry and believe that all players, including players 1 and 2, are using the common estimate $\hat{\rho}$. Accordingly, they play the symmetric equilibrium strategy $p_\mathrm{eq}$ derived in section~\ref{A1_uninformed}, which is known to both the buyer and the seller.
\end{itemize}
These assumptions imply that we can compute the utility function of the buyer P2 after the transaction using a strategy profile of the form $\Vec{p}=(q_1,q_2,p_\mathrm{eq},...,p_\mathrm{eq})$ where $p_\mathrm{eq}$ is the strategy used by the remaining $N-2$ players derived in section \ref{A1_uninformed}; known both to P1 and P2 since they know $\hat{\rho}$. This utility function can be written in the compact form:

\begin{equation}
\begin{aligned}
\U_2(\Vec{p},\hrho_2) =& q_1 q_2 \mathbb{E}_{\vec a}\left[ \mathbb{E}_w[u_2(1,1)] \right] + (1-q_1)q_2 \mathbb{E}_{\vec a}\left[ \mathbb{E}_w[u_2(0,1)] \right]+\\&
+q_1(1-q_2) \mathbb{E}_{\vec a}\left[ \mathbb{E}_w[u_2(1,0)] \right] + (1-q_1)(1-q_2) \mathbb{E}_{\vec a}\left[ \mathbb{E}_w[u_2(0,0)] \right]\ ,
\end{aligned}  
\end{equation}
where $\mathbb{E}_{\vec a}\left[\mathbb{E}_w[u_2(a_1,a_2)] \right]$ denotes the expected payoff of P2 when P1 plays $a_1$ and she plays $a_2$, while all other players follow the strategy $p_\mathrm{eq}$. These expectation values can be computed explicitly using identities in Eqs. \ref{eq:identities_forsum_start}-\ref{eq:identities_forsum_end}:

\begin{equation}
\begin{aligned}
\mathbb{E}_{\vec a}\left[ \mathbb{E}_w[u_2(0,0)] \right] &= \sum_{k=0}^{N-2} \binom{N-2}{k} p_\mathrm{eq}^k (1-p_\mathrm{eq})^{N-2-k} \left\{ (1-\hat{\rho}_2) \left[ \frac{N}{2+N-2-k} - 1 \right] - \hat{\rho}_2 \mathbbm{1} \left( \sum_{j>3} \delta_{1a_j} > 0 \right) \right\} \\&=  \frac{(1-\hat{\rho}_2)}{(1-p_\mathrm{eq})^2(N-1)} \left[ N(1-p_\mathrm{eq}) - 1 + p_\mathrm{eq}^N - (N-1)(1-p_\mathrm{eq})^N \right] + (1-p_\mathrm{eq})^{N-2} -1\ ,
\end{aligned}  
\end{equation}
\begin{equation}
\begin{aligned}
\mathbb{E}_{\vec a} \left[\mathbb{E}_w[u_2(0,1)] \right] &= \sum_{k=0}^{N-2} \binom{N-2}{k} p_\mathrm{eq}^k (1-p_\mathrm{eq})^{N-2-k} \left\{ \hat{\rho}_2 \left[ \frac{N}{1+k} - 1 \right] \right\} - (1-\hat{\rho}_2) \\&= \frac{\hat{\rho}_2 N}{(N-1)p_\mathrm{eq}} \left[ 1 - (1-p_\mathrm{eq})^{N-1} \right] - 1\ ,
\end{aligned}  
\end{equation}
\begin{equation}
\begin{aligned}
\mathbb{E}_{\vec a}\left[\mathbb{E}_w[u_2(1,0)] \right] &= \sum_{k=0}^{N-2} \binom{N-2}{k} p_\mathrm{eq}^k (1-p_\mathrm{eq})^{N-k-2} \left\{ (1-\hat{\rho}_2) \left[ \frac{N}{1+N-k-2} - 1 \right] \right\} - \hat{\rho}_2 \\&=  \frac{N(1-\hat{\rho}_2)}{(N-1)(1-p_\mathrm{eq})} \left[ 1 - p_\mathrm{eq}^{N-1} \right] - 1\ ,
\end{aligned}  
\end{equation}
\begin{equation}
\begin{aligned}
\mathbb{E}_{\vec a}\left[\mathbb{E}_w[u_2(1,1)] \right] &= \sum_{k=0}^{N-2} \binom{N-2}{k} p_\mathrm{eq}^k (1-p_\mathrm{eq})^{N-k-2} \left\{ \hat{\rho}_2 \left[ \frac{N}{2+k} - 1 \right] - (1-\hat{\rho}_2) \mathbbm{1} \left[ \sum_{j>3} \delta_{0a_j} >0 \right] \right\} \\&= \frac{\hat{\rho}_2}{(N-1)p_\mathrm{eq}^2} \left[ N p_\mathrm{eq} - 1 + (1-p_\mathrm{eq})^N - (N-1)p_\mathrm{eq}^N \right] + p_\mathrm{eq}^{N-2} - 1\ .
\end{aligned}  
\end{equation}
Since the utility function remains linear in P2's strategy $q_2$, it is convenient to express it in the following form:
\begin{equation} 
\begin{aligned}
&\U_2(\Vec{p},\hrho_2)
=q_2\left[A_N(\hrho_2,p_\mathrm{eq})q_1+B_N(\hrho_2,p_\mathrm{eq})\right]+C_N(\hrho_2,p_\mathrm{eq})q_1+D_N(\hrho_2,p_\mathrm{eq})\ ,
\label{eq:payoff_buyer_after_app}
\end{aligned}
\end{equation}
where the general forms of the coefficients are given by:
\begin{equation}
\begin{aligned}
    A_N(\hrho_2,p) \coloneqq \frac{\hat{\rho}_2 \left(p^N (N (p-1)-2 p+1)+(1-p)^N+p \left((N-2) (1-p)^N+2\right)-1\right)}{(N-1) (p-1)^2 p^2}+\frac{p^N (N (-p)+N+2 p-1)-p^2}{(N-1) (p-1)^2 p^2}\ ,  
\end{aligned}
\label{eq:An}
\end{equation}
\begin{equation}
\begin{aligned}
    B_N(\hrho_2,p) \coloneqq \frac{\hat{\rho}_2 \left(p \left(p^N+(1-p)^N-1\right)-N \left((1-p)^N+p-1\right)\right)}{(N-1) (p-1)^2 p}+\frac{-p^N+N (p-1)+1}{(N-1) (p-1)^2}\ ,  
\end{aligned}
\label{eq:Bn}
\end{equation}
\begin{equation}
\begin{aligned}
     C_N(\hrho_2,p) \coloneqq \frac{\hat{\rho}_2 \left((N (-p)+N+p) p^N+p \left(-\left((N-1) (1-p)^N\right)-1\right)\right)}{(N-1) (p-1)^2 p}+\frac{(N (p-1)-p) p^N+p}{(N-1) (p-1)^2 p}\ ,
\end{aligned}
\label{eq:Cn}
\end{equation}
\begin{equation}
\begin{aligned}
     D_N(\hrho_2,p) \coloneqq \frac{p^N-(N-1) p^2+(N-2) p}{(N-1) (p-1)^2}-\frac{\hat{\rho}_2 \left(p^N-\left((N-1) (1-p)^N\right)-N p+N-1\right)}{(N-1) (p-1)^2}\ ,
\end{aligned}
\label{eq:Dn}
\end{equation}
and the utility function slope can be written as:
\begin{equation}
\begin{aligned}
     \alpha_{2}(q_1,\hrho_2,p_\mathrm{eq},N) \coloneqq A_N(\hrho_2,p_\mathrm{eq})q_1+B_N(\hrho_2,p_\mathrm{eq})\ .
\end{aligned}
\label{eq:alpha_2}
\end{equation}
From linearity it follows that, to maximize her utility function, P2's best response will be:
\begin{equation}
\begin{aligned}
    q_2^{(br)}= \begin{cases}
        1 &\text{ if } \alpha_{2}(q_1,\hrho_2,p_\mathrm{eq},N)>0 \\
        0 &\text{ if } \alpha_{2}(q_1,\hrho_2,p_\mathrm{eq},N)<0 \\
        \text{any $q_2 \in [0,1]$}&\text{ if } \alpha_{2}(q_1,\hrho_2,p_\mathrm{eq},N)=0
    \end{cases}
\end{aligned}
\end{equation}
Since P2 assumes that P1 will use the same bias estimate $\hat{\rho}_2$, she also assumes P1 will play the same strategy as hers, i.e., with ($q_1=q_2$).

This symmetry implies that a strategy with $q_1=q_2=q_{\mathrm{eq}}$, where $q_{\mathrm{eq}}$ solves $\alpha_{2}(q_{\mathrm{eq}},\hrho_2,p_\mathrm{eq}, N)=0$, will be a symmetric Nash equilibrium for the two players (at least from the point of view of P2), since this would correspond to a fixed utility for both, independent of their individual strategy. Therefore, the symmetric equilibrium condition is
\begin{equation}
   \alpha_{2}(q_{\mathrm{eq}},\hrho_2,p_\mathrm{eq},N)=0 \iff q_{\mathrm{eq}}=-\frac{B_N(\hrho_2,p_\mathrm{eq})}{A_N(\hrho_2,p_\mathrm{eq})}\ .
\end{equation}
As long as this equation admits a solution in the interval $[0,1]$, i.e. when the ratio $-B_N(\hrho_2,p_\mathrm{eq})/A_N(\hrho_2,p_\mathrm{eq}) \in [0,1]$, P2 will find a symmetric equilibrium in which the two players' utilities are independent of their respective strategies.
We will now prove that 
even when the ratio $-B_N(\hrho_2,p_\mathrm{eq})/A_N(\hrho_2,p_\mathrm{eq})$ is outside the interval $[0,1]$, a subjective symmetric equilibrium still exists (with $q_1 = q_2 = q_{\mathrm{eq}} \in \{0,1\}$).
We first need a preliminary lemma:
\begin{lemma}\label{lemma:A<0}
     $\forall N\geq3,\forall\hrho_2,p\in[0,1]$, we have  $A_N(\hrho_2,p)<0$, where the coefficient $A_N(\hrho_2,p)$ is defined in \Eqref{eq:An}.
\end{lemma}
\begin{proof}
    We first prove it in the case $p \in (0,1)$. Rewriting $A_N(\hrho_2,p)$ as:
\begin{align}
    &A_N(\hrho_2,p)= \frac{[1+N(p-1)-2p](\hrho_2-1)p^N+\hrho_2[(1-p)^N -1]+p[-p+(2+(N-2)(1-p)^N)\hrho_2]}{p^2 (1-p)^2(N-1)}\ ,
\end{align}
since $p\in(0,1)$ and $N\ge3$ the denominator is always positive and we only need to prove that the numerator 
\begin{align}
    &Q_N(\hrho_2,p)\coloneqq [1+N(p-1)-2p](\hrho_2-1)p^N+\hrho_2[(1-p)^N -1]+p[-p+(2+(N-2)(1-p)^N)\hrho_2],
\end{align}
is always negative. To do so we first notice by direct substitution that $Q_3(\hrho_2,p)=-p^2(1-p)^2 <0$ and we will now prove that this function is decreasing in $N$, i.e. $Q_{N+1}(\hrho_2,p) \le Q_{N}(\hrho_2,p) \ \ \ \forall N \ge3$:
\begin{align}
    &Q_{N+1}(\hrho_2,p) - Q_{N}(\hrho_2,p) = - (N-1) [(1-\hrho_2)(p-1)^2p^N+\hrho_2p^2(1-p)^N]\le0\ .
\end{align}
This concludes the proof for $p \in (0,1)$. To complete the proof we compute the limit:
\begin{equation}
    \begin{aligned}
       A_N(\hrho_2,0)\coloneqq \lim_{p \rightarrow 0} A_N(\hrho_2,p)=-\frac{N(N-3)}{2(N-1)}\hrho_2-\frac{1}{N-1}\ ,
    \end{aligned}
    \label{eq:AN_p0}
\end{equation}
which is clearly negative $\forall N\ge3,\forall\hrho_2\in[0,1]$; and the last case:
\begin{equation}
    \begin{aligned}
               A_N(\hrho_2,1)\coloneqq \lim_{p \rightarrow 1} A_N(\hrho_2,p)=\frac{(N-3) N }{2 (N-1)}\hat{\rho}_2-\frac{N}{2}+1\ ,
    \end{aligned}
     \label{eq:AN_p1}
\end{equation}
that is also negative since $\forall N\ge3$  this coefficient is an increasing function of $\hrho_2$ and $A_N(1,1)=-\frac{1}{N-1}<0$, implying that it is negative $\forall\hrho_2\in[0,1]$.
\end{proof}
Thanks to Lemma \ref{lemma:A<0} we can now prove that (since P2 assumes that P1 is using the same coin bias estimate $\hrho_2$) from her point of view there exists a unique symmetric Nash equilibrium for any value of $N,p_\mathrm{eq}$ and $\hrho_2$.
\begin{lemma}\label{lemma:buyer_eq_post}
    Consider P2's utility function as defined in \Eqref{eq:payoff_buyer_after_app}. If P2 assumes that P1 has the same coin bias estimate $\hrho_2$ (i.e. she assumes the same functional form for P1's utility function), then from the point of view of P2 there exists a unique subjective symmetric equilibrium $\forall N\ge3;\forall \hrho_2,p_\mathrm{eq}\in[0,1]$ with: 
\begin{equation}\label{eq:qeq_buyer_after}
    q_1=q_2=q_{\mathrm{eq}}=\max\left(0,\min\left(1,- \frac{B_N(\hrho_2,p_\mathrm{eq})}{A_N(\hrho_2,p_\mathrm{eq})}\right)\right)\ . 
\end{equation}
\end{lemma}
\begin{proof}
Considering that all less-informed players are following the strategy $p_\mathrm{eq}$ computed in section \ref{A1_uninformed}, the slope of P2's utility function is given by \Eqref{eq:alpha_2}: 
\begin{equation}
\begin{aligned}
     \alpha_{2}(q_1,\hrho_2,p_\mathrm{eq},N)= A_N(\hrho_2,p_\mathrm{eq})q_1+B_N(\hrho_2,p_\mathrm{eq})\ .
\end{aligned}
\end{equation}
First, notice that $\alpha_{2}\left(\frac{-B_N(\hat{\rho}_2,p_\mathrm{eq})}{A_N(\hat{\rho}_2,p_\mathrm{eq})}, \hat{\rho}_2, p_\mathrm{eq}, N\right) = 0$ always holds. Moreover, by Lemma~\ref{lemma:A<0}, $\alpha_2$ is a strictly decreasing function of $q_1$. We now proceed case by case.
\begin{itemize}
        \item Suppose $-\frac{B_N(\hrho_2,p_\mathrm{eq})}{A_N(\hrho_2,p_\mathrm{eq})}\in[0,1]$. In this case the equilibrium is $q_{\mathrm{eq}}=-\frac{B_N(\hrho_2,p_\mathrm{eq})}{A_N(\hrho_2,p_\mathrm{eq})}$, since from P2's perspective this value would correspond to a flat utility function (zero slope) for both informed players, meaning their utility is independent of their individual strategies.
        \item Now suppose that $\frac{-B_N(\hat{\rho}_2, p_\mathrm{eq})}{A_N(\hat{\rho}_2, p_\mathrm{eq})} < 0$. Since $\alpha_2$ is strictly decreasing in $q_1$, it follows that:
         \[
        0 = \alpha_2\left( \frac{-B_N(\hat{\rho}_2, p_\mathrm{eq})}{A_N(\hat{\rho}_2, p_\mathrm{eq})}, \hat{\rho}_2, p_\mathrm{eq}, N \right) > \alpha_2(0, \hat{\rho}_2, p_\mathrm{eq}, N) > \alpha_2(q_1, \hat{\rho}_2, p_\mathrm{eq}, N) \quad \forall q_1 \in [0,1].
         \]
         These inequalities imply that, in this case, P2's best response is to always bet on Tails, regardless of other players' strategies. Therefore, from P2's point of view, there exists a symmetric equilibrium given by $q_{\mathrm{eq}} = 0$.
        \item Now suppose that $-\frac{B_N(\hat{\rho}_2, p_\mathrm{eq})}{A_N(\hat{\rho}_2, p_\mathrm{eq})} > 1$. Since $\alpha_2$ is strictly decreasing in $q_1$, it follows that:
         \[
        0 = \alpha_2\left( \frac{-B_N(\hat{\rho}_2, p_\mathrm{eq})}{A_N(\hat{\rho}_2, p_\mathrm{eq})}, \hat{\rho}_2, p_\mathrm{eq}, N \right) <\alpha_2(1,\hat{\rho}_2, p_\mathrm{eq}, N)<\alpha_2(q_1,\hat{\rho}_2, p_\mathrm{eq}, N) \;\;\; \forall q_1\in[0,1].
        \]
        These inequalities imply that, from P2's perspective, the best response for both herself and P1 is to always bet on Heads, regardless of the other players’ actions. Therefore, there exists a symmetric equilibrium with $q_{\mathrm{eq}}=1$ from the point of view of P2.
\end{itemize}
\end{proof}

\subsubsection{Mixed and pure strategies}\label{subsection:mixed_pure_buyer}
In this subsection, we analyze for which values of $\hat{\rho}_2$ the equilibrium strategy identified in Lemma~\ref{lemma:buyer_eq_post} is a pure strategy. To this end, we first prove the following lemma:

\begin{lemma}\label{lemma:B_incr}
   The function $B_N(\hrho_2,p)$, defined in \Eqref{eq:Bn}, is a strictly increasing function of $\hrho_2$, i.e.:
\begin{align}
        \frac{\partial B_N(\hrho_2,p)}{\partial \hrho_2} > 0 \;\;\;\; \forall p \in [0,1] \;\;\;\;\forall N \ge 3\ .
    \end{align}
\end{lemma}

\begin{proof}
The proof is trivial for the two limit values of $p$:

\begin{equation}
\begin{aligned}
    &B_N(\hrho_2,0) \coloneqq \lim_{p\rightarrow0}B_N(\hrho_2,p)= N\hrho_2-1\ ,
\end{aligned}
 \label{eq:BN_p0}
 \end{equation}
\begin{equation}    
\begin{aligned}
    &B_N(\hrho_2,1)\coloneqq\lim_{p\rightarrow1}B_N(\hrho_2,p)=\frac{N(N+1)}{2(N-1)}\hrho_2-\frac{N}{2}\ .
\end{aligned}
 \label{eq:BN_p1}
\end{equation}
We are only left to consider the case $p \in (0,1)$ in which:
 \begin{align}
 \frac{\partial B_N(\hrho_2,p)}{\partial \hrho_2} = \frac{1}{p(1-p)^2(N-1)}\left\{ p^{N+1}+(N-p)[1-(1-p)^N]-pN\right\}\ .
\end{align}
Since the denominator is always positive we only need to prove that the function:
\begin{align}
    W_N(p) \coloneqq p^{N+1}+(N-p)[1-(1-p)^N]-pN
\end{align}
is also always positive $\forall p \in (0,1) ,\forall N \ge 3$. To this end we notice that $W_3(p)=6(1-p)^2p>0$ and we prove that this function is also increasing in $N$, i.e. $W_{N+1}(p) - W_{N}(p) > 0  \;\;\; \forall N\ge3$. Writing this difference explicitly:
\begin{align}
     W_{N+1}(p)-W_{N}(p) = (1-p)[1-(1-p)^N-p^{N+1}]+p(1-p)^N(N-p)
\end{align}
this term is always positive since the function
\begin{align}
    f_{N}(p) \coloneqq [1-(1-p)^N-p^{N+1}]
\end{align}
is always positive. To prove this we compute again $f_3(p)=(1-p)p(3+p^2) >0$ and $f_{N+1}(p)-f_N(p)=p[(1-p)^N+(1-p)p^N]>0 \;\;\; \forall p\in(0,1)$. 
\end{proof}
By virtue of Lemma~\ref{lemma:B_incr}, we can identify the threshold values of $\hat{\rho}_2$ at which P2 switches from a mixed to a pure equilibrium strategy. Indeed, from Lemmas~\ref{lemma:A<0} and~\ref{lemma:buyer_eq_post}, we know that for P2 at equilibrium:
\begin{align}
    &q_\mathrm{eq}=0 \iff -\frac{B_N(\hrho_2,p_\mathrm{eq})}{A_N(\hrho_2,p_\mathrm{eq})}=\frac{B_N(\hrho_2,p_\mathrm{eq})}{|A_N(\hrho_2,p_\mathrm{eq})|}\le 0\iff B_N(\hrho_2,p_\mathrm{eq}) \le 0\ , \\
    &q_\mathrm{eq}=1 \iff -\frac{B_N(\hrho_2,p_\mathrm{eq})}{A_N(\hrho_2,p_\mathrm{eq})}=\frac{B_N(\hrho_2,p_\mathrm{eq})}{|A_N(\hrho_2,p_\mathrm{eq})|}\ge 1\iff B_N(\hrho_2,p_\mathrm{eq}) \ge |A_N(\hrho_2,p_\mathrm{eq})| = -A_N(\hrho_2,p_\mathrm{eq})\ .
\end{align}
Fixing a value of $p_\mathrm{eq} \in (0,1)$, we define $\hrho_2^{\downarrow}(p_\mathrm{eq},N)$ as the largest value of $\hrho_2$ such that  $B_N(\hrho_2,p_\mathrm{eq}) \le 0$. Due to Lemma \ref{lemma:B_incr} to find this threshold we can solve the equation $B_N(\hrho_2,p_\mathrm{eq})=0$ which yields:
\begin{equation}
\label{eq:rho2_q0}
    \hrho_2^{\downarrow}(p_\mathrm{eq},N)=-\frac{p_\mathrm{eq} \left(-p_\mathrm{eq}^N+N (p_\mathrm{eq}-1)+1\right)}{p_\mathrm{eq} \left(p_\mathrm{eq}^N+(1-p_\mathrm{eq})^N-1\right)-N \left((1-p_\mathrm{eq})^N+p_\mathrm{eq}-1\right)}\ ,
\end{equation}
therefore, by Lemma \ref{lemma:B_incr} we can conclude that
\begin{align}
    q_\mathrm{eq}=0 \iff \hrho_2\le \hrho_2^{\downarrow}(p_\mathrm{eq},N)\ .
\end{align}
Although the coefficient $A_N(\hat{\rho}_2, p)$ does not exhibit the same monotonicity for all values of $p$, it is still linear in $\hat{\rho}_2$, just like $B_N(\hat{\rho}_2, p)$(see  Eqs. \eqref{eq:An},\eqref{eq:Bn}). For this reason we can similarly define $\hrho_2^{\uparrow}(p_\mathrm{eq},N)$ as the smallest value of $\hrho_2$ such that $B_N(\hrho_2,p_\mathrm{eq}) \ge -A_N(\hrho_2,p_\mathrm{eq})$, and this value will be the solution of  $ B_N(\hrho_2,p_\mathrm{eq}) = -A_N(\hrho_2,p_\mathrm{eq})$, which yields:
\begin{equation}
\label{eq:rho2_q1}
    \hrho_2^{\uparrow}(p_\mathrm{eq},N)=\frac{1}{1-\frac{(p_\mathrm{eq}-1) \left((1-p_\mathrm{eq})^N+N p_\mathrm{eq}-1\right)}{N p_\mathrm{eq}^2-p_\mathrm{eq}^N (N+p_\mathrm{eq}-1)}}\ .
\end{equation}
Since we proved that $B_N(\hrho_2,p_\mathrm{eq})$ is an increasing function of $\hrho_2$ we have that for P2:
\begin{align}
    q_\mathrm{eq}=1 \iff \hrho_2\ge \hrho_2^{\uparrow}(p_\mathrm{eq},N)\ .
\end{align}
To determine these thresholds for $p_\mathrm{eq}\in\{0,1\}$ we can either recompute them directly using the limiting values for the coefficients $A_N,B_N$ (given in Eqs.\eqref{eq:AN_p0},\eqref{eq:AN_p1},\eqref{eq:BN_p0},\eqref{eq:BN_p1}), or evaluate the limits of the general expressions as  $p_\mathrm{eq}\to0$ and  $p_\mathrm{eq}\to1$
\begin{align}
   & \hrho_2^{\uparrow}(0,N)\coloneqq \lim_{p_\mathrm{eq}\to 0} \hrho_2^{\uparrow}(p_\mathrm{eq},N)=\frac{2}{N+1}\ ,\\
   &\hrho_2^{\uparrow}(1,N)\coloneqq \lim_{p_\mathrm{eq}\to 1} \hrho_2^{\uparrow}(p_\mathrm{eq},N)=1-\frac{1}{N}\ ,\\
   & \hrho_2^{\downarrow}(0,N)\coloneqq \lim_{p_\mathrm{eq}\to 0} \hrho_2^{\downarrow}(p_\mathrm{eq},N)=\frac{1}{N}\ ,\\
   &\hrho_2^{\downarrow}(1,N)\coloneqq \lim_{p_\mathrm{eq}\to 1} \hrho_2^{\downarrow}(p_\mathrm{eq},N)=\frac{N-1}{N+1}\ .
\end{align}

\subsection{Calculation of $(q_1)_\mathrm{eq}^\mathrm{post}$}\label{A4_seller_post}
In this Section we compute the equilibrium strategy $(q_1)_\mathrm{eq}^\mathrm{post}$ adopted by the seller P1 after the transaction.
In Section~\ref{section:A3_buyer_after}, we fully characterized the behavior of the buyer (P2) after the transaction, under the assumption that she believes the seller (P1) will evaluate utilities using the same coin bias $\hat{\rho}_2$ as hers.
This reflects P2’s belief, not the actual state of the game, though. In reality, unless the full dataset is transferred, P1 retains additional private information and continues to base her strategy on her own estimate $\hat{\rho}_1$, which generally differs from $\hat{\rho}_2$. However, since P1 retains access to the data she sold, she also knows the value $\hat{\rho}_2$ that P2 will estimate from it.
This implies that P1 knows both her own and P2's utility functions.
In fact, the utility computation from P1’s perspective is identical to the one already presented in section~\ref{section:A3_buyer_after}, with the same strategy profile $\Vec{p}=(q_1,q_2,p_\mathrm{eq},...,p_\mathrm{eq})$ and roles of the two players simply interchanged. In formulas, the utilities are
\begin{align}
    \U_2(\Vec{p},\hrho_2) &= q_2 \left( A_N(\hat{\rho}_2, p_\mathrm{eq}) q_1 + B_N(\hat{\rho}_2, p_\mathrm{eq}) \right) + C_N(\hat{\rho}_2, p_\mathrm{eq}) q_1 + D_N(\hat{\rho}_2, p_\mathrm{eq})\ , 
\end{align}
\begin{align}\label{eq:payoff_seller_after_app}
    \U_1(\Vec{p},\hrho_1) &= q_1 \left( A_N(\hat{\rho}_1, p_\mathrm{eq}) q_2 + B_N(\hat{\rho}_1, p_\mathrm{eq}) \right) + C_N(\hat{\rho}_1, p_\mathrm{eq}) q_2 + D_N(\hat{\rho}_1, p_\mathrm{eq})\ ,
\end{align}
where the coefficients are given in Eqs.\eqref{eq:An}-\eqref{eq:Dn}. To compute these expressions at the boundaries $p_\mathrm{eq} \to 0$ or $p_\mathrm{eq} \to 1$,  the appropriate limits should be taken according to Eqs. \eqref{eq:AN_p0},\eqref{eq:AN_p1},\eqref{eq:BN_p0},\eqref{eq:BN_p1}. The value of $p_\mathrm{eq}$ is uniquely determined by the common estimate $\hat{\rho}$, and is given by the equilibrium analysis in section~\ref{A1_uninformed}. Moreover, since P1 knows $\hat{\rho}_2$, she can anticipate that P2 will choose her strategy according to \Eqref{eq:qeq_buyer_after}:
\begin{align}
q_2 = q_\mathrm{eq} = \max\left(0, \min\left(1, -\frac{B_N(\hat{\rho}_2, p_\mathrm{eq})}{A_N(\hat{\rho}_2, p_\mathrm{eq})}\right)\right).
\end{align}
Since P1's utility function in Eq. \ref{eq:payoff_seller_after_app} is linear in her own strategy $q_1$, her best response is fully determined by the sign of the slope evaluated at $q_2 = q_{\text{eq}}$:

\begin{align}
    \alpha_1(\hat{\rho}_1, q_\mathrm{eq}, p_\mathrm{eq}) \coloneqq A_N(\hat{\rho}_1, p_\mathrm{eq}) q_\mathrm{eq} + B_N(\hat{\rho}_1, p_\mathrm{eq})\ .
\end{align}
Accordingly, P1’s best response and equilibrium strategy after the transaction will be:
\begin{align}
    (q_1)_\text{eq}^\text{post} = 
    \begin{cases} 
        1 & \text{if } \alpha_1(\hat{\rho}_1, q_\mathrm{eq}, p_\mathrm{eq}) > 0\ , \\
        0 & \text{if } \alpha_1(\hat{\rho}_1, q_\mathrm{eq}, p_\mathrm{eq}) < 0\ , \\
        \text{any } q_1 \in [0,1] & \text{if } \alpha_1(\hat{\rho}_1, q_\mathrm{eq}, p_\mathrm{eq}) = 0\ .
    \end{cases}
\end{align}
As in the pre-transaction case discussed in section~\ref{A2_sellerbefore}, when $\alpha_1 = 0$ and any strategy is a best response, P1's utility is entirely determined by the strategies of the other players and remains fixed. In this case, we again adopt the convention of setting $(q_1)_{\text{eq}}^{\text{post}} = 0$. With this convention, the equilibrium strategy can be compactly written as:
\begin{equation}
\begin{aligned}
(q_1)_\text{eq}^\text{post}=\mathbbm{1}\left[A_N(\hrho_1,p_\mathrm{eq})q_\mathrm{eq}+B_N(\hrho_1,p_\mathrm{eq})>0\right].
\end{aligned}
\end{equation}
We distinguish three cases for $\alpha_1$ depending on the strategy played by the buyer $q_\mathrm{eq}$:
\begin{align}
    \alpha_1(\hrho_1,q_2,p_\mathrm{eq}) = 
    \begin{cases}
        B_N(\hat{\rho}_1, p_\mathrm{eq}) & \text{if } q_\mathrm{eq} = 0\ , \\[1em]
        A_N(\hat{\rho}_1, p_\mathrm{eq}) + B_N(\hat{\rho}_1, p_\mathrm{eq}) & \text{if } q_\mathrm{eq} = 1\ , \\[1em]
        \dfrac{
            A_N(\hat{\rho}_1, p_\mathrm{eq}) B_N(\hat{\rho}_2, p_\mathrm{eq}) 
            - A_N(\hat{\rho}_2, p_\mathrm{eq}) B_N(\hat{\rho}_1, p_\mathrm{eq})
        }{
            \left| A_N(\hat{\rho}_2, p_\mathrm{eq}) \right|
        } & \text{if } q_\mathrm{eq} \in \,(0,1)\ .
    \end{cases}
\end{align}
Note that since the coefficients $A_N$ and $B_N$ have the same functional form for both players, Lemmas~\ref{lemma:A<0} and~\ref{lemma:B_incr} continue to apply. These lemmas can therefore be used to analyze the sign of $\alpha_1$ in each case, enabling a complete characterization of P1’s best response strategy.

\subsubsection{Case 1: $\hrho\ge 1-\frac{1}{N} \iff p_\mathrm{eq}=1$}
Using the results from subsection \ref{subsection:mixed_pure_buyer} and the coefficients $A_N(\hat{\rho}_1, 1)$ and $B_N(\hat{\rho}_1, 1)$ given in Eqs.~\eqref{eq:AN_p1},\eqref{eq:BN_p1}, we distinguish three cases:
\begin{enumerate}
\item[a.] If $\hrho_2\ge 1-\frac{1}{N}$ then $q_\mathrm{eq}=1$ and:
\begin{align}
    \alpha_1=A_N(\hrho_1,1)+B_N(\hrho_1,1)=1+N(\hrho_1-1)\ ,
\end{align}
meaning that:
\begin{itemize}
    \item $\alpha_1=0 \iff \hrho_1=1-\frac{1}{N}$,
    \item $\alpha_1>0 \iff \hrho_1>1-\frac{1}{N}$,
    \item $\alpha_1<0 \iff \hrho_1<1-\frac{1}{N}$.
\end{itemize}
\item[b.] If $\hrho_2\le \frac{N-1}{N+1}$ then $q_\mathrm{eq}=0$ and:
\begin{align}
    \alpha_1=B_N(\hrho_1,1)=\frac{N}{2(N-1)}[1-N+(1+N)\hrho_1]\ ,
\end{align}
meaning that:
\begin{itemize}
    \item $\alpha_1=0 \iff \hrho_1=\frac{N-1}{N+1}$,
    \item $\alpha_1>0 \iff \hrho_1>\frac{N-1}{N+1}$,
    \item $\alpha_1<0 \iff \hrho_1<\frac{N-1}{N+1}$.
\end{itemize}
\item[c.] If $\hrho_2 \in (\frac{N-1}{N+1}, 1-\frac{1}{N})$ then $q_\mathrm{eq}=-B_N(\hrho_2,1)/A_N(\hrho_2,1)\in (0,1)$ and:
\begin{align}
    \alpha_1=\frac{A_N(\hrho_1,1)B_N(\hrho_2,1)- A_N(\hrho_2,1)B_N(\hrho_1,1)}{|A_N(\hrho_2,1)|}=\frac{N}{2|A_N(\hrho_2,1)|}(\hrho_1-\hrho_2)\ ,
\end{align}
meaning that:
\begin{itemize}
    \item $\alpha_1=0 \iff \hrho_1=\hrho_2$,
    \item $\alpha_1>0 \iff \hrho_1>\hrho_2$,
    \item $\alpha_1<0 \iff \hrho_1<\hrho_2$.
\end{itemize}
\end{enumerate}
\subsubsection{Case 2: $\hrho\le \frac{1}{N} \iff p_\mathrm{eq}=0$}
Using the results from subsection \ref{subsection:mixed_pure_buyer} and the coefficients $A_N(\hat{\rho}_1, 0)$ and $B_N(\hat{\rho}_1, 0)$ given in Eqs.\eqref{eq:AN_p0},\eqref{eq:BN_p0}, we distinguish three cases:
\begin{enumerate}
\item[a.] If $\hrho_2\ge \frac{2}{N+1}$ then $q_\mathrm{eq}=1$ and:
\begin{align}
    \alpha_1=A_N(\hrho_1,0)+B_N(\hrho_1,0)=\frac{N[\hrho_1(N+1)-2]}{2(N-1)}\ ,
\end{align}
meaning that:
\begin{itemize}
    \item $\alpha_1=0 \text{ if } \hrho_1=\frac{2}{N+1}$,
    \item $\alpha_1>0 \text{ if } \hrho_1>\frac{2}{N+1}$,
    \item $\alpha_1<0 \text{ if } \hrho_1<\frac{2}{N+1}$.
\end{itemize}
\item[b.] If $\hrho_2\le \frac{1}{N}$ then $q_\mathrm{eq}=0$ and:
\begin{align}
    \alpha_1=B_N(\hrho_1,0)=-1+N\hrho_1\ ,
\end{align}
meaning that:
\begin{itemize}
    \item $\alpha_1=0 \text{ if } \hrho_1=\frac{1}{N}$,
    \item $\alpha_1>0 \text{ if } \hrho_1>\frac{1}{N}$,
    \item $\alpha_1<0 \text{ if } \hrho_1<\frac{1}{N}$.
\end{itemize}
\item[c.] If $\hrho_2 \in (\frac{1}{N},\frac{2}{N+1})$ then $q_\mathrm{eq}=-B_N(\hrho_2,0)/A_N(\hrho_2,0)\in (0,1)$ and:
\begin{align}
    \alpha_1=\frac{A_N(\hrho_1,0)B_N(\hrho_2,0)- A_N(\hrho_2,0)B_N(\hrho_1,0)}{|A_N(\hrho_2,0)|}=\frac{1}{2|A_N(\hrho_2,0)|}(\hrho_1-\hrho_2)\ ,
\end{align}
meaning that:
\begin{itemize}
    \item $\alpha_1=0 \text{ if } \hrho_1=\hrho_2$,
    \item $\alpha_1>0 \text{ if } \hrho_1>\hrho_2$,
    \item $\alpha_1<0 \text{ if } \hrho_1<\hrho_2$.
\end{itemize}
\end{enumerate}
\subsubsection{Case 3: $\hrho \in (\frac{1}{N},1-\frac{1}{N}) \iff p_\mathrm{eq}\in(0,1)$}
Using the results from subsection \ref{subsection:mixed_pure_buyer} and the general form of the coefficients $A_N(\hat{\rho}_1, p)$ and $B_N(\hat{\rho}_1, p)$ given in Eqs.\eqref{eq:An}, \eqref{eq:Bn} evaluated for $p=p_\mathrm{eq}$, we distinguish three cases:
\begin{enumerate}
\item[a.] If $\hrho_2 \ge \hrho_2^{\uparrow}(p_\mathrm{eq},N)$, then $q_\mathrm{eq} = 1$ and:
\begin{align}
    \alpha_1 = A_N(\hrho_1, p_\mathrm{eq}) + B_N(\hrho_1, p_\mathrm{eq})\ ,
\end{align}
meaning that:
\begin{itemize}
    \item $\alpha_1 = 0 \text{ if } \hrho_1 = \hrho_1^{\uparrow}(p_\mathrm{eq}, N)$,
    \item $\alpha_1 > 0 \text{ if } \hrho_1 > \hrho_1^{\uparrow}(p_\mathrm{eq}, N)$,
    \item $\alpha_1 < 0 \text{ if } \hrho_1 < \hrho_1^{\uparrow}(p_\mathrm{eq}, N)$.
\end{itemize}
Here, $\hat{\rho}_1^{\uparrow}(p_\mathrm{eq}, N) = \hat{\rho}_2^{\uparrow}(p_\mathrm{eq}, N)$, as given in \Eqref{eq:rho2_q1}, since the equation determining the sign of $\alpha_1$ is identical to the one solved for P2 in subsection \ref{subsection:mixed_pure_buyer}, where the sign of $A_N(\hat{\rho}_2, p_\mathrm{eq}) + B_N(\hat{\rho}_2, p_\mathrm{eq})$ was analyzed.
\item[b.] If $\hrho_2 \le \hrho_2^{\downarrow}(p_\mathrm{eq}, N)$, then $q_\mathrm{eq} = 0$ and:
\begin{align}
    \alpha_1 = B_N(\hrho_1, p_\mathrm{eq})\ ,
\end{align}
meaning that:
\begin{itemize}
    \item $\alpha_1 = 0 \text{ if } \hrho_1 = \hrho_1^{\downarrow}(p_\mathrm{eq}, N)$,
    \item $\alpha_1 > 0 \text{ if } \hrho_1 > \hrho_1^{\downarrow}(p_\mathrm{eq}, N)$,
    \item $\alpha_1 < 0 \text{ if } \hrho_1 < \hrho_1^{\downarrow}(p_\mathrm{eq}, N)$.
\end{itemize}
As in the previous case, we have $\hat{\rho}_1^{\downarrow}(p_\mathrm{eq}, N) = \hat{\rho}_2^{\downarrow}(p_\mathrm{eq}, N)$ given by \Eqref{eq:rho2_q0}, since the equation determining the sign of $B_N$ is the same as the one analyzed for P2 in subsection~\ref{subsection:mixed_pure_buyer}.
\item[c.] If $\hrho_2 \in (\hrho_2^{\downarrow}(p_\mathrm{eq}, N), \hrho_2^{\uparrow}(p_\mathrm{eq}, N))$, then $q_\mathrm{eq} = -\frac{B_N(\hrho_2, p_\mathrm{eq})}{A_N(\hrho_2, p_\mathrm{eq})} \in (0,1)$ and:
\begin{align}
    \alpha_1 = \frac{A_N(\hrho_1, p_\mathrm{eq}) B_N(\hrho_2, p_\mathrm{eq}) - A_N(\hrho_2, p_\mathrm{eq}) B_N(\hrho_1, p_\mathrm{eq})}{|A_N(\hrho_2, p_\mathrm{eq})|}\ ,
\end{align}
meaning that now the sign of $\alpha_1$ is given by the sign of the function $F_N(\hrho_1, \hrho_2, p_\mathrm{eq}) = A_N(\hrho_1, p_\mathrm{eq}) B_N(\hrho_2, p_\mathrm{eq}) - A_N(\hrho_2, p_\mathrm{eq}) B_N(\hrho_1, p_\mathrm{eq})$. Writing it explicitly:
\begin{align}
    &F_N(\hrho_1, \hrho_2, p_\mathrm{eq}) \coloneqq \\ 
    &\frac{(1-2 N) p_\mathrm{eq}^{N+1}+(N-1) p_\mathrm{eq}^{N+2}+N p_\mathrm{eq}^N+p_\mathrm{eq}^2 \left((N-1) (1-p_\mathrm{eq})^N+1\right)+p_\mathrm{eq} \left((1-p_\mathrm{eq})^N-1\right)-N (1-p_\mathrm{eq})^N p_\mathrm{eq}^N}{(N-1) (1-p_\mathrm{eq})^3 p_\mathrm{eq}^3} (\hrho_2 - \hrho_1)
\end{align}
We now prove that the function in the numerator: 
\begin{align}
    Z_N(p_\mathrm{eq}) \coloneqq &(1-2 N) p_\mathrm{eq}^{N+1} + (N-1) p_\mathrm{eq}^{N+2} + N p_\mathrm{eq}^N + \\& +p_\mathrm{eq}^2 \left((N-1)(1-p_\mathrm{eq})^N + 1\right) + p_\mathrm{eq} \left((1-p_\mathrm{eq})^N - 1\right) - N (1-p_\mathrm{eq})^N p_\mathrm{eq}^N
\end{align}
is strictly negative $\forall N \ge 3, \forall p_\mathrm{eq} \in (0,1)$. We first notice that $Z_3(p_\mathrm{eq}) = -3(1-p_\mathrm{eq})^3 p_\mathrm{eq}^3 < 0$, and that:
\begin{align}
    Z_{N+1}(p_\mathrm{eq}) - Z_N(p_\mathrm{eq}) =& -N (1-p_\mathrm{eq})^3 p_\mathrm{eq}^N - N p_\mathrm{eq}^3 (1-p_\mathrm{eq})^N + (1-p_\mathrm{eq})^N p_\mathrm{eq}^N [N - N (1-p_\mathrm{eq}) p_\mathrm{eq} - (1-p_\mathrm{eq}) p_\mathrm{eq}] \\
    =& -N (1-p_\mathrm{eq})^N [p_\mathrm{eq}^3 - p_\mathrm{eq}^N] - N (1-p_\mathrm{eq})^3 p_\mathrm{eq}^N - (1-p_\mathrm{eq})^{N+1} p_\mathrm{eq}^{N+1} (N+1) < 0\ ,
\end{align}
proving that $Z_N(p_\mathrm{eq}) < 0$ for all $N \ge 3$ and $p_\mathrm{eq} \in (0,1)$. We can then rewrite the slope:
\begin{align}
    \alpha_1 = -\frac{|Z_N(p_\mathrm{eq})|}{|A_N(\hrho_2, p_\mathrm{eq})|(N-1)(1-p_\mathrm{eq})^3 p_\mathrm{eq}^3} (\hrho_2 - \hrho_1) = \frac{|Z_N(p_\mathrm{eq})|}{|A_N(\hrho_2, p_\mathrm{eq})|(N-1)(1-p_\mathrm{eq})^3 p_\mathrm{eq}^3} (\hrho_1 - \hrho_2)\ ,
\end{align}
which finally allows us to conclude that:
\begin{itemize}
    \item $\alpha_1 = 0 \text{ if } \hrho_1 = \hrho_2$,
    \item $\alpha_1 > 0 \text{ if } \hrho_1 > \hrho_2$, 
    \item $\alpha_1 < 0 \text{ if } \hrho_1 < \hrho_2$.
\end{itemize}
\end{enumerate}
\section{Volatility-averse players, $\lambda>0$}\label{sec:Appendix_B}
In this Appendix, we follow the same analytical steps as in Appendix~A, but consider volatility-averse players who penalize uncertainty in their beliefs in addition to expected payoff. In the main text, we denoted the utility function of a volatility-averse player as $\U_i(\Vec{p},\hrho_i,R_i;\lambda)$, and defined it as:
\begin{equation}
\begin{aligned}
    \U_i(\Vec{p},\hrho_i,R_i;\lambda)\coloneqq\U_i(\Vec{p},\hrho_i) - \lambda \operatorname{Var}_{\rho_i}[\U_i(\Vec{p},\rho_i)]\ ,
\end{aligned}
\end{equation}
where $\lambda \ge 0$ is a parameter that quantifies players' aversion to volatility, and $ Var_{\rho_i}[\cdot]$ denotes the variance computed with respect to the subjective posterior distribution each player has for the coin bias. This posterior is a Beta distribution $\mathcal{B}(H_i+1,R_i-H_i+1)$, where $R_i$ is the length of the string observed by player $i$, and $H_i$ is the number of heads it contains. For computational convenience, we introduce an equivalent parametrization by defining $b_i \coloneqq H_i + 1$ and $c_i \coloneqq R_i + 2$, so that the Beta distribution becomes $\mathcal{B}(b_i, c_i - b_i)$ with first two moments:
\begin{equation}
\begin{aligned}
    \hrho_i = \frac{H_i+1}{R_i+2}=\frac{b_i}{c_i}\ ,
\end{aligned}
\end{equation}
\begin{equation}\label{eq:variance_rho}
\begin{aligned}
    \operatorname{Var}_{\rho_i}[\rho_i]=\frac{(H_i+1)(R_i-H_i+1)}{(R_i+2)^2(R_i+3)}=\frac{b_i(c_i-b_i)}{c_i^2 (c_i+1)}=\frac{\hrho_i(1-\hrho_i)}{c_i+1}\ .
\end{aligned}
\end{equation}
Quantities derived from the public string $\vec{s}_0 \equiv (H,R)$ are denoted without any subscript: $b\coloneqq H+1$ and $c\coloneqq R+2$. This alternative parametrization is fully equivalent to the one used in the main text, and is adopted here solely for notational convenience. 
\subsection{Calculation of $p_\mathrm{eq}$}\label{B1_uninformed}
In this section, we determine the equilibrium strategy $p_{\text{eq}}$ adopted by the least-informed players, who rely exclusively on the publicly available string $\vec{s}_0$.
The setting and assumptions are the same as in Section~\ref{A1_uninformed}, with the key difference that players now exhibit volatility aversion, which modifies their utility computations. As in section \ref{A1_uninformed}, the setting is fully symmetric, so we focus without loss of generality on a generic player, denoted as player $i$. Specifically, the utility function for volatility-neutral players derived in section \ref{A1_uninformed} is linear in $\hat{\rho}$ (see \Eqref{eq:payoff_uninformed_app}) and can thus be rewritten as:
\begin{equation}
\begin{aligned}
\U_i(\Vec{p},\hrho)=S(q_i,p,N) \hat{\rho} + P(q_i,p,N)\ ,
\end{aligned}
\end{equation}
where the strategy profile is $\vec{p} = (p, \dots, p, q_i, p, \dots, p)$, where $q_i$ appears in the $i$-th position, and:
\begin{equation}
\begin{aligned}
S(q_i,p,N) \coloneqq \frac{1 - p^N - (1-p)^N}{(1-p)p} (q_i-p)\ ,
\end{aligned}
\end{equation}
\begin{equation}
\begin{aligned}
P(q_i,p,N) \coloneqq \frac{p^N - p}{(1-p)p} (q_i-p)\ .
\end{aligned}
\end{equation}
For volatility-averse players, the utility function is adjusted by subtracting a penalty proportional to the variance of the posterior distribution, yielding (see \Eqref{eq:variance_rho}):
\begin{equation}\label{eq:payoff_uninformed_lambda}
\begin{aligned}
\U_i(\Vec{p},\hrho,c;\lambda)&=S(q_i,p,N) \hat{\rho} + P(q_i,p,N) - \lambda  \frac{ \hat{\rho}(1-\hat{\rho})}{c + 1} S^2(q_i,p,N) \\
&=Y(p,\hat{\rho},c,\lambda,N)(q_i-p)^2 + X(p,\hat{\rho},N)(q_i-p)\ ,
\end{aligned}
\end{equation}
where,
\begin{equation}\label{eq:Y_uninformed}
\begin{aligned}
Y(p,\hat{\rho},c,\lambda,N) \coloneqq -\frac{\lambda}{c+1} \hat{\rho}(1-\hat{\rho}) \left[\frac{1-p^N-(1-p)^N}{(1-p)p}\right]^2\ ,
\end{aligned}
\end{equation}
\begin{equation}
\begin{aligned}
X(p,\hat{\rho},N) \coloneqq \frac{1-p^N-(1-p)^N}{(1-p)p} \hat{\rho} + \frac{p^N-p}{(1-p)p}\ .
\end{aligned}
\end{equation}
From \Eqref{eq:payoff_uninformed_lambda}, it is clear that the cases $\hat{\rho} \in \{0,1\}$ are equivalent to $\lambda = 0$, meaning the volatility aversion correction vanishes and the results of section \ref{A1_uninformed} are recovered. For this reason, we restrict our analysis to the case $\hat{\rho} \in (0,1)$. It follows that the coefficient $Y(p,\hat{\rho},c,\lambda,N)$ in \Eqref{eq:Y_uninformed} is strictly negative, and the utility function in \Eqref{eq:payoff_uninformed_lambda}, as a function of $q$, is a concave parabola whose vertex corresponds to the best responses of the player:
\begin{equation}
\begin{aligned}
q^{\text{br}}_i(p) = -\frac{X(p,\hat{\rho},N)}{2Y(p,\hat{\rho},c,\lambda,N)} + p\ .
\end{aligned}
\end{equation}
The condition for a symmetric Nash equilibrium is therefore
\begin{equation}
\begin{aligned}
q^{\text{br}}_i(p) = p \iff -\frac{X(p,\hat{\rho},N)}{2Y(p,\hat{\rho},c,\lambda,N)} = 0 \iff X(p,\hat{\rho},N) = 0\ .
\end{aligned}
\end{equation}
This condition gives
\begin{equation}\label{eq:eq_uninformed_lambda}
\begin{aligned}
\hat{\rho} = \frac{p - p^N}{1 - p^N - (1-p)^N} = K(p,N)\ ,
\end{aligned}
\end{equation}
where $K(p,N)$ is given in \Eqref{eq:Kpn}, implying that we recover the same symmetric Nash equilibrium found in section \ref{A1_uninformed} for volatility-neutral players (see Eq. \eqref{eq:eq_condition_uninformed_app}). 
Since $K(p,N)$, for fixed $N$, lies within the interval $\left[ \frac{1}{N}, 1 - \frac{1}{N} \right]$, this equilibrium condition can only hold with $\hrho$ within this range. To investigate potential Nash equilibria outside this interval, we examine the limiting behavior of \Eqref{eq:payoff_uninformed_lambda} as $p \to 0$ and $p \to 1$.
\subsubsection{Limit $p \to 0$}
The limit for $p \to 0$ of \Eqref{eq:payoff_uninformed_lambda} yields:
\begin{equation}
\begin{aligned}
\lim_{p\to 0} \U_i(\Vec{p},\hrho,c;\lambda)=Y(0,\hat{\rho},c,\lambda,N) q_i^2 + X(0,\hat{\rho},N) q_i = -\frac{\lambda}{c+1} \hat{\rho}(1-\hat{\rho}) N^2 q_i^2 + (N\hat{\rho} - 1) q_i\ ,
\end{aligned}
\end{equation}
where:
\begin{equation}
\begin{aligned}
 Y(0,\hat{\rho},c,\lambda,N)\coloneqq \lim_{p\to 0} Y(p,\hat{\rho},c,\lambda,N)=-\frac{\lambda}{c+1} \hat{\rho}(1-\hat{\rho}) N^2\ ,
\end{aligned}
\end{equation}
\begin{equation}
\begin{aligned}
 X(0,\hat{\rho},N)\coloneqq \lim_{p\to 0} X(p,\hat{\rho},N)=(N\hat{\rho} - 1)\ .
\end{aligned}
\end{equation}
Since $Y(0,\hat{\rho},c,\lambda,N)$ is strictly negative, the utility function, as a function of $q$, is again a concave parabola whose vertex corresponds to the players' best response.
\begin{equation}
\begin{aligned}
q^{\text{br}}_i(p=0) = -\frac{X(0,\hat{\rho},N)}{2Y(0,\hat{\rho},c,\lambda,N)}\ ,
\end{aligned}
\end{equation}
In this case, for a symmetric Nash equilibrium, we should require that $q^{\text{br}}_i(p=0)=0$. However, because the parabola is concave and $q_i$ must represent a probability, it is sufficient that the vertex lies to the left of the interval $[0,1]$. In fact, since the parabola is concave, if the vertex lies outside the interval $[0,1]$, the best response will be attained at the nearest boundary.
\begin{equation}
\begin{aligned}
q^{\text{br}}_i(p=0) = 0 \iff -\frac{X(0,\hat{\rho},N)}{2Y(0,\hat{\rho},c,\lambda,N)} \le 0 \iff \frac{X(0,\hat{\rho},N)}{2|Y(0,\hat{\rho},c,\lambda,N)|} \le 0 \iff X(0,\hat{\rho},N) \leq 0 \iff \hat{\rho} \leq \frac{1}{N}\ .
\end{aligned}
\end{equation}
Again, we recover the results of section \ref{A1_uninformed}, finding that if $\hat{\rho} \leq \frac{1}{N}$, the only possible equilibrium is achieved when all players bet on Tails with certainty.

\subsubsection{Limit $p \to 1$}
The limit for $p \to 1$ of \Eqref{eq:payoff_uninformed_lambda} yields:
\begin{equation}
\begin{aligned}
\lim_{p\to 1} \U_i(\Vec{p},\hrho,c;\lambda)&=Y(1,\hat{\rho},c,\lambda,N)(q-1)^2 + X(1,\hat{\rho},N)(q-1) \\&= -\frac{\lambda}{c+1} \hat{\rho}(1-\hat{\rho}) N^2 (q-1)^2 + [N\hat{\rho} + (1-N)] (q-1)\ ,
\end{aligned}
\end{equation}
where:
\begin{equation}
\begin{aligned}
 Y(1,\hat{\rho},c,\lambda,N) \coloneqq\lim_{p\to 1} Y(p,\hat{\rho},c,\lambda,N)=-\frac{\lambda}{c+1} \hat{\rho}(1-\hat{\rho}) N^2\ ,
\end{aligned}
\end{equation}
\begin{equation}
\begin{aligned}
 X(1,\hat{\rho},N) \coloneqq\lim_{p\to 1} X(p,\hat{\rho},N)=N\hat{\rho} + (1-N)\ .
\end{aligned}
\end{equation}
Since $Y(1,\hat{\rho},c,\lambda,N)$ is strictly negative, the utility function, as a function of $q$, is again a concave parabola whose vertex corresponds to the players' best response.
\begin{equation}
\begin{aligned}
q^{\text{br}}_i(p=1) = -\frac{X(1,\hat{\rho},N)}{2Y(1,\hat{\rho},c,\lambda,N)} + 1\ .
\end{aligned}
\end{equation}
By the same reasoning of previous subsection, due to the utility function concavity, we will have a symmetric Nash equilibrium where all players always bet on Heads when $q^{\text{br}}_i(p=1) \ge 1$:
\begin{equation}
\begin{aligned}
q^{\text{br}}_i(p=1) = 1 \iff  \frac{X(1,\hat{\rho},N)}{2|Y(1,\hat{\rho},c,\lambda,N)|} \geq 0 \iff X(1,\hat{\rho},N) \geq 0 \iff \hat{\rho}  \ge 1 - \frac{1}{N}\ .
\end{aligned}
\end{equation}
We can thus conclude that in all cases, even in the presence of volatility aversion, the symmetric equilibrium strategy of the less-informed players remains the same as in the volatility-neutral case, namely $p_\mathrm{eq}$ computed in section \ref{A1_uninformed}.

\subsection{Calculation of $(q_1)_\text{eq}^\text{pre}$}
In this section, we compute the equilibrium strategy $(q_1)_\text{eq}^\text{pre}$ of the volatility-averse seller P1 before any transaction occurs. The seller faces players who rely solely on the public dataset, so she knows their utilities and strategies, which follow the equilibrium conditions derived in section \ref{B1_uninformed}. The setting is the same as in section \ref{A2_sellerbefore}, with the only difference being that the players are volatility averse.
In particular, in the volatility-averse case, the functional form of P1's utility function remains the same as that of the uninformed players (given in \Eqref{eq:payoff_uninformed_lambda}). The only difference from the uninformed players lies in the parameters used, reflecting that P1 has a longer string and additional data. Therefore we can again write:
\begin{equation}\label{eq:payoff_seller_before_lambda}
\begin{aligned}
\U_1(\Vec{p},\hrho_1,c_1;\lambda)=Y(p_\mathrm{eq},\hat{\rho}_1,c_1,\lambda,N)(q_1-p_\mathrm{eq})^2 + X(p_\mathrm{eq},\hat{\rho}_1,N)(q_1-p_\mathrm{eq})\ ,
\end{aligned}
\end{equation}
where the strategy profile is $\Vec{p}=(q_1,p_\mathrm{eq},\dots,p_\mathrm{eq})$ and $p_\mathrm{eq}$ denotes the symmetric equilibrium strategy of the less informed players, as derived in section~\ref{B1_uninformed}, which is known to P1. The coefficients take the following form:
\begin{equation}
\begin{aligned}
Y(p,\hat{\rho}_1,c_1,\lambda,N) = -\frac{\lambda}{c_1+1} \hat{\rho}_1(1-\hat{\rho}_1) \left[\frac{1-p^N-(1-p)^N}{(1-p)p}\right]^2\ ,
\end{aligned}
\end{equation}
\begin{equation}
\begin{aligned}
X(p,\hat{\rho}_1,N) = \frac{1-p^N-(1-p)^N}{(1-p)p} \hat{\rho}_1 + \frac{p^N-p}{(1-p)p}\ .
\end{aligned}
\end{equation}
Note that the functional forms of $Y$ and $X$ are the same as those appearing in the analysis of less informed players in section~\ref{B1_uninformed}, with the only difference being the use of P1’s personalized parameters $(\hat{\rho}_1, c_1)$.
Again, we restrict our analysis to the case $\hat{\rho}_1 \in (0,1)$, since at the boundaries the volatility correction vanishes, recovering the results from section \ref{A2_sellerbefore}. Therefore $Y(p_\mathrm{eq},\hat{\rho}_1,c_1,\lambda,N)$ is strictly negative and the utility function remains a concave parabola in $q_1$. Consequently, P1's optimal strategy corresponds either to the vertex of this parabola or to one of the boundaries of the interval $[0,1]$ if the vertex falls outside it:    
\begin{equation}\label{eq:q1_br_seller_pre}
\begin{aligned}
q_1^{br}(p_\mathrm{eq}) = \min\{\max\{0, -\frac{X(p_\mathrm{eq},\hat{\rho}_1,N)}{2Y(p_\mathrm{eq},\hat{\rho}_1,c_1,\lambda,N)}+p_\mathrm{eq}\}, 1\}\ .
\end{aligned}
\end{equation}
We solve this equation by distinguishing three cases, based on whether the remaining $N-1$ players are playing a pure or mixed strategy.

\subsubsection{$\hrho \in (\frac{1}{N},1-\frac{1}{N}) \iff p_\mathrm{eq} \in (0,1)$}
In this case, we first recall that the symmetric equilibrium condition for the uninformed players is given by \Eqref{eq:eq_uninformed_lambda}:
\begin{equation}\label{eq:condition_K}
\begin{aligned}
K(p_\mathrm{eq},N) = \frac{p_\mathrm{eq} - p_\mathrm{eq}^N}{1 - p_\mathrm{eq}^N - (1-p_\mathrm{eq})^N}  = \hat{\rho}\ ,
\end{aligned}
\end{equation}
that can be solved numerically to find their equilibrium strategy $p_\mathrm{eq}$. We now examine the conditions under which P1’s best response is a mixed strategy, i.e. $q_1^{br}(p_\mathrm{eq})$ lies in $(0,1)$. We have
\begin{equation}\label{eq:condition_zero_seller_pre}
\begin{aligned}
q_1^{br}(p_\mathrm{eq})> 0 &\iff X(p_\mathrm{eq},\hat{\rho}_1,N) + 2p_\mathrm{eq}|Y(p_\mathrm{eq},\hat{\rho}_1,c_1,\lambda,N)| > 0 \iff X(p_\mathrm{eq},\hat{\rho}_1,N) > -2p_\mathrm{eq}|Y(p_\mathrm{eq},\hat{\rho}_1,c_1,\lambda,N)|\\
&\iff \frac{1 - p_\mathrm{eq}^N - (1-p_\mathrm{eq})^N}{(1-p_\mathrm{eq})p_\mathrm{eq}} \hat{\rho}_1 + \frac{p_\mathrm{eq}^N - p_\mathrm{eq}}{(1-p_\mathrm{eq})p_\mathrm{eq}} > -\frac{2p_\mathrm{eq}\lambda}{c_1+1} \hat{\rho}_1(1-\hat{\rho}_1) \left[\frac{1 - p_\mathrm{eq}^N - (1-p_\mathrm{eq})^N}{(1-p_\mathrm{eq})p_\mathrm{eq}}\right]^2\\
&\iff \hat{\rho}_1 - \hat{\rho} > -\frac{2\lambda}{c_1+1} \hat{\rho}_1(1-\hat{\rho}_1) \frac{1 - p_\mathrm{eq}^N - (1-p_\mathrm{eq})^N}{(1-p_\mathrm{eq})}\\
&\iff \hat{\rho}_1 - \hat{\rho} > \frac{2\lambda}{c_1+1} \frac{p_\mathrm{eq} - p_\mathrm{eq}^N}{\hat{\rho}(1-p_\mathrm{eq})} \hat{\rho}_1^2 - \frac{2\lambda}{c_1+1} \frac{p_\mathrm{eq} - p_\mathrm{eq}^N}{\hat{\rho}(1-p_\mathrm{eq})} \hat{\rho}_1 \\
&\iff \frac{2\lambda}{c_1+1} \frac{p_\mathrm{eq} - p_\mathrm{eq}^N}{\hat{\rho}(1-p_\mathrm{eq})} \hat{\rho}_1^2 - \left[1 + \frac{2\lambda}{c_1+1} \frac{p_\mathrm{eq} - p_\mathrm{eq}^N}{\hat{\rho}(1-p_\mathrm{eq})}\right] \hat{\rho}_1 + \hat{\rho} < 0\\
& \iff a \hat{\rho}_1^2 - \left[1 + a\right] \hat{\rho}_1 + \hat{\rho} < 0\ ,
\end{aligned}
\end{equation}
where we have used \Eqref{eq:condition_K} and, for clarity of exposition, we have defined the auxiliary parameter
\begin{equation}
\begin{aligned}
a \coloneqq \frac{2\lambda}{c_1+1} \frac{p_\mathrm{eq} - p_\mathrm{eq}^N}{\hat{\rho}(1-p_\mathrm{eq})} > 0\ .
\end{aligned}
\end{equation}
Since $a > 0$, condition \Eqref{eq:condition_zero_seller_pre} is satisfied when $\hat{\rho}_1$ lies between the two roots of the convex parabola $ a \hat{\rho}_1^2 - \left[1 + a\right] \hat{\rho}_1 + \hat{\rho}$, which are given by
\begin{equation}
\begin{aligned}
\hat{\rho}_1 = \frac{1 + a \pm \sqrt{(1 + a)^2 - 4a\hat{\rho}}}{2a} = \frac{1 + a \pm \sqrt{1 + a^2 + 2a - 4a\hat{\rho}}}{2a}\ .
\end{aligned}
\end{equation}
It is straightforward to prove that the larger root is greater than one
\begin{equation}
\begin{aligned}
 \frac{1 + a + \sqrt{(1+a)^2 - 4a\hat{\rho}}}{2a} &> 1 \iff \sqrt{(1+a)^2 - 4a\hat{\rho}} > a-1\ .\\
\end{aligned}
\end{equation}
This last expression always holds since
\begin{equation}
\begin{aligned}
&\quad a < 1 \Rightarrow \sqrt{(1+a)^2 - 4a\hat{\rho}} > 0 > a-1\\
&\quad a \ge 1 \Rightarrow \sqrt{(1+a)^2 - 4a\hat{\rho}} > a-1 \iff 1 + a^2 + 2a - 4a\hat{\rho} > a^2 + 1 - 2a \iff \hat{\rho} < 1 \quad .\\
\end{aligned}
\end{equation}
We have thus proved that:
\begin{equation}
\begin{aligned}
q_1^{br}(p_\mathrm{eq}) > 0 \iff \hat{\rho}_1 \in \left(\frac{1 + a - \sqrt{(1+a)^2 - 4a\hat{\rho}}}{2a}, 1\right]\ ,
\end{aligned}
\end{equation}
\begin{equation}
\begin{aligned}
q_1^{br}(p_\mathrm{eq}) = 0  \iff \hat{\rho}_1 \in \left[0,\frac{1 + a - \sqrt{(1+a)^2 - 4a\hat{\rho}}}{2a}\right]\ .
\end{aligned}
\end{equation}
In a similar fashion, we can now characterize the condition under which $q_1^{br}(p_\mathrm{eq}) < 1$:
\begin{equation}\label{eq:condition_one_seller_pre}
\begin{aligned}
q_1^{br}(p_\mathrm{eq}) < 1 &\iff \frac{X(p_\mathrm{eq},\hat{\rho}_1,N) + 2p_\mathrm{eq}|Y(p_\mathrm{eq},\hat{\rho}_1,c_1,\lambda,N)|}{2|Y(p_\mathrm{eq},\hat{\rho}_1,c_1,\lambda,N)|} < 1 \iff X(p_\mathrm{eq},\hat{\rho}_1,N) <  2|Y(p_\mathrm{eq},\hat{\rho}_1,c_1,\lambda,N)|(1-p_\mathrm{eq})\\
&\iff \frac{1 - p_\mathrm{eq}^N - (1-p_\mathrm{eq})^N}{(1-p_\mathrm{eq})p_\mathrm{eq}} \hat{\rho}_1 + \frac{p_\mathrm{eq}^N - p_\mathrm{eq}}{(1-p_\mathrm{eq})p_\mathrm{eq}} < \frac{2\lambda}{c_1+1} \hat{\rho}_1(1-\hat{\rho}_1)(1-p_\mathrm{eq})\left[\frac{1-p_\mathrm{eq}^N-(1-p_\mathrm{eq})^N}{p_\mathrm{eq}(1-p_\mathrm{eq})}\right]^2\\
&\iff \hat{\rho}_1 - \hat{\rho} < \frac{2\lambda}{c_1+1} \hat{\rho}_1(1-\hat{\rho}_1) \frac{1-p_\mathrm{eq}^N-(1-p_\mathrm{eq})^N}{p_\mathrm{eq}}\\
&\iff \hat{\rho}_1 - \hat{\rho} < \frac{2\lambda}{c_1+1} \hat{\rho}_1(1-\hat{\rho}_1) \frac{p_\mathrm{eq}-p_\mathrm{eq}^N}{p_\mathrm{eq}\hat{\rho}}\\
&\iff \frac{2\lambda}{c_1+1} \frac{p_\mathrm{eq}-p_\mathrm{eq}^N}{\hat{\rho}p_\mathrm{eq}} \hat{\rho}_1^2 + \left[1 - \frac{2\lambda}{c_1+1} \frac{p_\mathrm{eq}-p_\mathrm{eq}^N}{\hat{\rho}p_\mathrm{eq}}\right] \hat{\rho}_1 - \hat{\rho} < 0 \\
&\iff \widetilde{a}\hat{\rho}_1^2+ [1-\widetilde{a}]\hat{\rho}_1- \hat{\rho} < 0\ ,
\end{aligned}
\end{equation}
where, as before, we apply \Eqref{eq:condition_K} and define the auxiliary parameter
\begin{equation}
\begin{aligned}
\widetilde{a} \coloneqq \frac{2\lambda}{c_1+1} \frac{p_\mathrm{eq} - p_\mathrm{eq}^N}{\hat{\rho}p_\mathrm{eq}}>0\ .
\end{aligned}
\end{equation}
Since $\widetilde{a} > 0$, condition \Eqref{eq:condition_one_seller_pre} is satisfied when $\hat{\rho}_1$ lies between the two roots of the convex parabola $\widetilde{a}\hat{\rho}_1^2+ [1-\widetilde{a}]\hat{\rho}_1- \hat{\rho}$, which are given by
\begin{equation}
\begin{aligned}
\hat{\rho}_1 = \frac{(\widetilde{a} - 1) \pm \sqrt{(\widetilde{a} - 1)^2 + 4\widetilde{a}\hat{\rho}}}{2\widetilde{a}}\ ,
\end{aligned}
\end{equation}
It is immediate to see that the smaller root is negative, and straightforward to verify that the larger root is always strictly less than one:
\begin{equation}
\begin{aligned}
 \frac{(\widetilde{a} - 1) + \sqrt{(\widetilde{a} - 1)^2 + 4\widetilde{a}\hat{\rho}}} {2\widetilde{a}} < 1 & \iff (\widetilde{a} - 1) + \sqrt{(\widetilde{a} - 1)^2 + 4\widetilde{a}\hat{\rho}} < 2\widetilde{a}\\
&\iff \sqrt{(\widetilde{a} - 1)^2 + 4\widetilde{a}\hat{\rho}} < \widetilde{a} + 1 \\
&\iff \widetilde{a}^2 + 1 - 2\widetilde{a} + 4\widetilde{a}\hat{\rho} < \widetilde{a}^2 + 1 + 2\widetilde{a}\\
&\iff 4\widetilde{a} > 4\widetilde{a}\hat{\rho} \iff \hat{\rho} < 1\ .
\end{aligned}
\end{equation}
We have thus proved that:
\begin{equation}
\begin{aligned}
q_1^{br}(p_\mathrm{eq}) < 1 \iff \hat{\rho}_1 \in \left[0, \frac{\widetilde{a} - 1 + \sqrt{(\widetilde{a} - 1)^2 + 4\widetilde{a}\hat{\rho}}}{2\widetilde{a}}\right),
\end{aligned}
\end{equation}
\begin{equation}
\begin{aligned}
q_1^{br}(p_\mathrm{eq}) = 1 \iff \hat{\rho}_1 \in \left[ \frac{\widetilde{a} - 1 + \sqrt{(\widetilde{a} - 1)^2 + 4\widetilde{a}\hat{\rho}}}{2\widetilde{a}},1\right].
\end{aligned}
\end{equation}
To summarize the result of this subsection, when $\hrho \in ]\frac{1}{N},1-\frac{1}{N}[ $ (i.e. when $p_\mathrm{eq} \in (0,1)$) we find that:
\begin{equation}
\begin{aligned}
q_1^{br}(p_\mathrm{eq}) = 0  \iff \hat{\rho}_1 \in \left[0,\frac{1 + a - \sqrt{(1+a)^2 - 4a\hat{\rho}}}{2a}\right],
\end{aligned}
\end{equation}
\begin{equation}
\begin{aligned}
q_1^{br}(p_\mathrm{eq}) = 1 \iff \hat{\rho}_1 \in \left[ \frac{\widetilde{a} - 1 + \sqrt{(\widetilde{a} - 1)^2 + 4\widetilde{a}\hat{\rho}}}{2\widetilde{a}},1\right],
\end{aligned}
\end{equation}
\begin{equation}
\begin{aligned}
q_1^{br}(p_\mathrm{eq}) \in (0,1) \iff \hat{\rho}_1 \in \left(\frac{1 + a - \sqrt{(1+a)^2 - 4a\hat{\rho}}}{2a}, \frac{\widetilde{a} - 1 + \sqrt{(\widetilde{a} - 1)^2 + 4\widetilde{a}\hat{\rho}}}{2\widetilde{a}}\right).
\end{aligned}
\end{equation}
We have verified numerically that this last interval is, in general, a proper non-empty subset of $[0,1]$, indicating that the optimal strategy for P1 may indeed be mixed. This is a novel feature introduced by the volatility aversion of the players, whereas in the setting of section \ref{A2_sellerbefore}, where players were volatility-neutral, only pure strategies emerged as optimal.

\subsubsection{$\hrho \leq \frac{1}{N} \iff p_\mathrm{eq} \to 0$}
The limit for $p_\mathrm{eq} \to 0$ of \Eqref{eq:payoff_seller_before_lambda} yields:
\begin{equation}
\begin{aligned}
\lim_{p_\mathrm{eq} \to 0} \U_1(\Vec{p},\hrho_1,c_1;\lambda)=Y(0,\hat{\rho}_1,c_1,\lambda,N) q_1^2 + X(0,\hat{\rho}_1,N) q_1 = -\frac{\lambda}{c_1+1} \hat{\rho}_1(1-\hat{\rho}_1) N^2 q_1^2 + (N\hat{\rho}_1 - 1) q_1\ ,
\end{aligned}
\end{equation}
where we have defined:
\begin{equation}
\begin{aligned}
 Y(0,\hat{\rho}_1,c_1,\lambda,N) \coloneqq\lim_{p_\mathrm{eq}\to 0} Y(p_\mathrm{eq},\hat{\rho}_1,c_1,\lambda,N)=-\frac{\lambda}{c_1+1} \hat{\rho}_1(1-\hat{\rho}_1) N^2\ ,
\end{aligned}
\end{equation}
\begin{equation}
\begin{aligned}
 X(0,\hat{\rho}_1,N) \coloneqq\lim_{p_\mathrm{eq}\to 0} X(p_\mathrm{eq},\hat{\rho}_1,N)=(N\hat{\rho}_1 - 1)\ .
\end{aligned}
\end{equation}
Since $Y(0,\hat{\rho}_1,c_1,\lambda,N)$ is strictly negative, the utility function remains a concave parabola in $q_1$, allowing us to proceed as in the previous subsection. Following the same approach, we now identify the conditions under which P1's optimal strategy is mixed or pure. In this case, Equation \eqref{eq:q1_br_seller_pre} simplifies to
\begin{equation}
\begin{aligned}
q_1^{br}(p_\mathrm{eq}=0) = \min\{\max\{0,-\frac{X(0,\hat{\rho}_1,N)}{2Y(0,\hat{\rho}_1,c_1,\lambda,N)}\}, 1\}\ .
\end{aligned}
\end{equation}
From this, we deduce that
\begin{equation}
\begin{aligned}
q_1^{br}(p_\mathrm{eq}=0) > 0 \iff X(0,\hat{\rho}_1,N) > 0 \iff \hat{\rho}_1 > \frac{1}{N}\ ,
\end{aligned}
\end{equation}
\begin{equation}
\begin{aligned}
q_1^{br}(p_\mathrm{eq}=0) = 0 \iff X(0,\hat{\rho}_1,N) \le 0 \iff \hat{\rho}_1 \le \frac{1}{N}\ ,
\end{aligned}
\end{equation}
\begin{equation}
\begin{aligned}
q_1^{br}(p_\mathrm{eq}=0) < 1 & \iff X(0,\hat{\rho}_1,N) < 2|Y(0,\hat{\rho}_1,c_1,\lambda,N)| \\& \iff 2\lambda N^2 \hat{\rho}_1^2 + [N(c_1+1) - 2\lambda N^2] \hat{\rho}_1 - (c_1+1) < 0\ .
\end{aligned}
\end{equation}
Given that $2\lambda N^2 > 0$, this condition is satisfied when $\hat{\rho}_1$ lies between the two roots of this convex parabola. These are given by
\begin{equation}
\begin{aligned}
\hat{\rho}_1= \frac{2\lambda N - (c_1+1) \pm \sqrt{[2\lambda N - (c_1+1)]^2 + 8\lambda(c_1+1)}}{4\lambda N}\ .
\end{aligned}
\end{equation}
One can verify that the smaller root is negative. Letting $\hrho_1^+$ denote the larger root, we can then write
\begin{equation}
\begin{aligned}
q_1^{br}(p_\mathrm{eq}=0) < 1 \iff \hat{\rho}_1 \in [0, \hat{\rho}_1^+)\ ,
\end{aligned}
\end{equation}
with
\begin{equation}
\begin{aligned}
\hat{\rho}_1^+= \frac{2\lambda N - (c_1+1) + \sqrt{[2\lambda N - (c_1+1)]^2 + 8\lambda(c_1+1)}}{4\lambda N}\ .
\end{aligned}
\end{equation}
We now verify that $\hrho_1^+$ lies strictly between $\frac{1}{N}$ and 1:
\begin{equation}
\begin{aligned}
\hat{\rho}_1^+ > \frac{1}{N} & \iff 2\lambda(N-2) - (c_1+1) + \sqrt{[2\lambda N - (c_1+1)]^2 + 8\lambda(c_1+1)} > 0 \\ & \iff \sqrt{[2\lambda N - (c_1+1)]^2 + 8\lambda(c_1+1)} > (c_1+1) - 2\lambda(N-2)\ ,
\end{aligned}
\end{equation}
\begin{equation}
\begin{aligned}
\hat{\rho}_1^+ < 1 &\iff -2\lambda N - (c_1+1) + \sqrt{[2\lambda N - (c_1+1)]^2 + 8\lambda(c_1+1)} < 0\\
&\iff 2\lambda N + (c_1+1) > \sqrt{[2\lambda N - (c_1+1)]^2 + 8\lambda(c_1+1)}\\
&\iff 4\lambda N(c_1+1) > -4\lambda N(c_1+1) + 8\lambda(c_1+1)\\
&\iff N > 1\ .
\end{aligned}
\end{equation}
Hence, when $\hrho \in [0,\frac{1}{N}]$, we can conclude:
\begin{equation}
\begin{aligned}
q_1^{br}(p_\mathrm{eq}=0) =0 \iff \hat{\rho}_1 \in [0,\frac{1}{N}]\ ,
\end{aligned}
\end{equation}
\begin{equation}
\begin{aligned}
q_1^{br}(p_\mathrm{eq}=0) \in (0,1) \iff \hat{\rho}_1 \in (\frac{1}{N}, \hat{\rho}_1^+)\ ,
\end{aligned}
\end{equation}
\begin{equation}
\begin{aligned}
q_1^{br}(p_\mathrm{eq}=0) =1 \iff \hat{\rho}_1 \in [\hat{\rho}_1^+,1]\ .
\end{aligned}
\end{equation}
This confirms once again the possibility of mixed optimal strategies for P1.
\subsubsection{$\hrho \geq 1-\frac{1}{N} \iff p_\mathrm{eq} \to 1$}
The limit for $p_\mathrm{eq} \to 1$ of \Eqref{eq:payoff_seller_before_lambda} yields
\begin{equation}
\begin{aligned}
\lim_{p_\mathrm{eq} \to 1} \U_1(\Vec{p},\hrho_1,c_1;\lambda)&= Y(1,\hat{\rho}_1,c_1,\lambda,N) (q_1-1)^2 + X(1,\hat{\rho}_1,N) (q_1-1) \\&= -\frac{\lambda}{c_1+1} \hat{\rho}_1(1-\hat{\rho}_1) N^2 (q_1-1)^2 + [N\hat{\rho}_1 + (1-N)] (q_1-1)\ ,
\end{aligned}
\end{equation}
where we have defined:
\begin{equation}
\begin{aligned}
 Y(1,\hat{\rho}_1,c_1,\lambda,N) \coloneqq \lim_{p_\mathrm{eq}\to 1} Y(p_\mathrm{eq},\hat{\rho}_1,c_1,\lambda,N)=-\frac{\lambda}{c_1+1} \hat{\rho}_1(1-\hat{\rho}_1) N^2\ ,
\end{aligned}
\end{equation}
\begin{equation}
\begin{aligned}
 X(1,\hat{\rho}_1,N) \coloneqq \lim_{p_\mathrm{eq}\to 1} X(p_\mathrm{eq},\hat{\rho}_1,N)=N\hat{\rho}_1 + (1-N)\ .
\end{aligned}
\end{equation}
Since $Y(1,\hat{\rho}_1,c_1,\lambda,N)$ is strictly negative, the utility function remains a concave parabola in $q_1$, allowing us to proceed as in the previous subsection. Following the same approach, we now identify the conditions under which P1's optimal strategy is mixed or pure. In this case, \Eqref{eq:q1_br_seller_pre} simplifies to:
\begin{equation}
\begin{aligned}
q_1^{br}(p_\mathrm{eq}=1) = \min\{\max\{0,-\frac{X(1,\hat{\rho}_1,N)}{2Y(1,\hat{\rho}_1,c_1,\lambda,N)}+1\}, 1\}\ ,
\end{aligned}
\end{equation}
from which we deduce:
\begin{equation}
\begin{aligned}
q_1^{br}(p_\mathrm{eq}=1) < 1 \iff \frac{X(1,\hat{\rho}_1,N)}{2|Y(1,\hat{\rho}_1,c_1,\lambda,N)|} + 1 < 1 \iff X(1,\hat{\rho}_1,N) < 0 \iff \hat{\rho}_1 < 1 - \frac{1}{N}\ ,
\end{aligned}
\end{equation}
\begin{equation}
\begin{aligned}
q_1^{br}(p_\mathrm{eq}=1) = 1 \iff  X(1,\hat{\rho}_1,N) \geq 0 \iff \hat{\rho}_1 \ge 1 - \frac{1}{N}\ ,
\end{aligned}
\end{equation}
\begin{equation}
\begin{aligned}
q_1^{br}(p_\mathrm{eq}=1) > 0 & \iff \frac{X(1,\hat{\rho}_1,N)}{2|Y(1,\hat{\rho}_1,c_1,\lambda,N)|} > -1  \iff X(1,\hat{\rho}_1,N) + 2|Y(1,\hat{\rho}_1,c_1,\lambda,N)| > 0 \\ & \iff 2\lambda N^2 \hat{\rho}_1^2 - [2\lambda N^2 + N(c_1+1)]\hat{\rho}_1 + (N-1)(c_1+1) < 0\ .
\end{aligned}
\end{equation}
Given that $2\lambda N^2 > 0$, this condition is satisfied when $\hat{\rho}_1$ lies between the two roots of this convex parabola. These are given by:
\begin{equation}
\begin{aligned}
\hat{\rho}_1^{\uparrow} = \frac{(c_1+1) + 2\lambda N + \sqrt{[2\lambda N - (c_1+1)]^2 + 8\lambda(c_1+1)}}{4\lambda N}\ ,
\end{aligned}
\end{equation}
\begin{equation}
\begin{aligned}
\hat{\rho}_1^{\downarrow} = \frac{(c_1+1) + 2\lambda N - \sqrt{[2\lambda N - (c_1+1)]^2 + 8\lambda(c_1+1)}}{4\lambda N}\ .
\end{aligned}
\end{equation}
We prove that the larger root is bigger than one, i.e., 
\begin{equation}
\begin{aligned}
\hat{\rho}_1^{\uparrow} > 1 &\iff (c_1+1) - 2\lambda N + \sqrt{[2\lambda N - (c_1+1)]^2 + 8\lambda(c_1+1)} > 0 \\& \iff
\sqrt{[2\lambda N - (c_1+1)]^2 + 8\lambda(c_1+1)} >  2\lambda N - (c_1+1)\ ,
\end{aligned}
\end{equation}
and that the smaller root lies strictly between 0 and $1-\frac{1}{N}$
\begin{equation}
\begin{aligned}
\hat{\rho}_1^{\downarrow} > 0 &\iff (c_1+1) + 2\lambda N - \sqrt{[2\lambda N - (c_1+1)]^2 + 8\lambda(c_1+1)} > 0\\
&\iff 4\lambda N(c_1+1) > -4\lambda N(c_1+1) + 8\lambda(c_1+1)\\
&\iff N > 1\ ,
\end{aligned}
\end{equation}
\begin{equation}
\begin{aligned}
\hat{\rho}_1^{\downarrow} < 1 - \frac{1}{N} &\iff 4\lambda - 2\lambda N + (c_1+1) - \sqrt{[2\lambda N - (c_1+1)]^2 + 8\lambda(c_1+1)} < 0\\
&\iff \sqrt{[2\lambda N - (c_1+1)]^2 + 8\lambda(c_1+1)} > 2\lambda N - (c_1+1) - 4\lambda\ .
\end{aligned}
\end{equation}
Hence, when $\hrho \in [1-\frac{1}{N},1]$, we can conclude:
\begin{equation}
\begin{aligned}
q_1^{br}(p_\mathrm{eq}=1) = 0 \iff \hat{\rho}_1 \in [ 0,\hat{\rho}_1^{\downarrow}]
\end{aligned}
\end{equation}
\begin{equation}
\begin{aligned}
q_1^{br}(p_\mathrm{eq}=1) \in (0,1) \iff \hat{\rho}_1 \in (\hat{\rho}_1^{\downarrow}, 1 - \frac{1}{N})
\end{aligned}
\end{equation}
\begin{equation}
\begin{aligned}
q_1^{br}(p_\mathrm{eq}=1) = 1 \iff \hat{\rho}_1 \in [ 1 - \frac{1}{N},1]
\end{aligned}
\end{equation}
Confirming once again the possibility of mixed optimal strategies for P1.
\subsection{Calculation of $q_{\mathrm{eq}}$}\label{B3_buyer_post_lambda}
In this section, we compute the symmetric Nash equilibrium that emerges between the seller and the buyer, assuming the transaction takes place. The setup and assumptions are the same as in section \ref{section:A3_buyer_after}, and the buyer’s utility function in the volatility-neutral case is given by \Eqref{eq:payoff_buyer_after_app}, which, being linear in $\hrho_2$, can be rewritten as
\begin{equation}
\begin{aligned}
 \U_2(\Vec{p},\hrho_2)
=S^{(2)}(q_1, q_2, p_\mathrm{eq}, N) \hat{\rho}_2 + P^{(2)}(q_1, q_2, p_\mathrm{eq}, N)\ ,
\end{aligned}
\end{equation}
where the strategy profile considered is $\Vec{p}=(q_1,q_2,p_\mathrm{eq},...,p_\mathrm{eq})$  and $p_\mathrm{eq}$ denotes the equilibrium strategy of the less informed players, as derived in section~\ref{B1_uninformed}, which is known to P2. For volatility-averse players, the utility function is modified by introducing a penalty term proportional to the variance of the posterior distribution, resulting in the following expression (see \Eqref{eq:variance_rho}):
\begin{equation}\label{eq:payoff_buyer_after_lambda}
\begin{aligned}
 \U_2(\Vec{p},\hrho_2,c_2;\lambda)
&=S^{(2)}(q_1, q_2, p_\mathrm{eq}, N) \hat{\rho}_2 + P^{(2)}(q_1, q_2, p_\mathrm{eq}, N) - \lambda  \frac{\hat{\rho}_2(1-\hat{\rho}_2)}{c_2 + 1} [S^{(2)}(q_1, q_2, p_\mathrm{eq}, N)]^2 \\
 &=Y^{(2)}(q_1, \hat{\rho}_2, p_\mathrm{eq}, \lambda, c_2, N) q_2^2 + X^{(2)}(q_1, \hat{\rho}_2, p_\mathrm{eq}, \lambda, c_2, N) q_2 + R^{(2)}(q_1, \hat{\rho}_2, p_\mathrm{eq}, \lambda, c_2, N)\ .
\end{aligned}
\end{equation}
Although the explicit expressions of these coefficients are algebraically complex and not particularly informative for our analysis, the key property is that the leading coefficient, $Y^{(2)}(q_1, \hat{\rho}_2, p_\mathrm{eq}, \lambda, c_2, N)$, remains strictly negative for all $p_\mathrm{eq} \in (0,1)$, including in the limits as $p_\mathrm{eq} \to 0$ and $p_\mathrm{eq} \to 1$. This coefficient is given by
\begin{equation}
\begin{aligned}
Y^{(2)}(q_1, \hat{\rho}_2, p_\mathrm{eq}, \lambda, c_2, N) \coloneqq-\dfrac{\lambda (1-\hat{\rho}_2) \hat{\rho}_2 }{(1+c_2)(N-1)^2(1-p_\mathrm{eq})^4 p_\mathrm{eq}^4} \gamma^2(q_1, p_\mathrm{eq}, N)\ ,
\label{Y^2_2}
\end{aligned}
\end{equation}
with the auxiliary function $\gamma(q_1, p_\mathrm{eq}, N)$ defined by:
\begin{equation}
\begin{aligned}
\gamma(q_1, p_\mathrm{eq}, N) \coloneqq(-1+(1-p_\mathrm{eq})^N+p_\mathrm{eq}^N)(p_\mathrm{eq}^2+q_1-2p_\mathrm{eq}q_1)+N(p_\mathrm{eq}(1-p_\mathrm{eq}+(1-p_\mathrm{eq})^N(-1+q_1))+(-1+p_\mathrm{eq})p_\mathrm{eq}^N q_1)\ .
\end{aligned}
\end{equation}
The limits of the coefficient as the symmetric strategy $p_\mathrm{eq}$ of the less informed players approaches the pure strategies are given by:
\begin{equation}
\begin{aligned}
\lim_{p_\mathrm{eq}\to 0} Y^{(2)}(q_1, \hat{\rho}_2, p_\mathrm{eq}, \lambda, c_2, N) =-\frac{ \lambda N^2 (1-\hrho_2) \hrho_2 }{4 (1+c_2) (N-1)^2} (N (q_1-2)-3 q_1+2)^2\ ,
\end{aligned}
\end{equation}
\begin{equation}
\begin{aligned}
\lim_{p_\mathrm{eq}\to 1} Y^{(2)}(q_1, \hat{\rho}_2, p_\mathrm{eq}, \lambda, c_2, N) = - \frac{\lambda N^2 (1-\hrho_2) \hrho_2 }{4 (1+c_2) (N-1)^2}(N q_1+N-3 q_1+1)^2\ .
\end{aligned}
\end{equation}
Given that $Y^{(2)}$ is always strictly negative, the resulting utility function is concave in $q_2$ (we restrict ourselves again to the case $\hrho_2 \in (0,1)$, where the volatility correction is nonzero), enabling us to compute the buyer’s best response following the same procedure adopted in earlier sections:
\begin{equation}
\begin{aligned}
q_2^{br} = \min\left\{\max\left\{0, -\frac{X^{(2)}(q_1, \hat{\rho}_2, p_\mathrm{eq}, \lambda, c_2, N)}{2Y^{(2)}(q_1, \hat{\rho}_2, p_\mathrm{eq}, \lambda, c_2, N)}\right\}, 1\right\}.
\end{aligned}
\end{equation}
In this setting, the seller evaluates her utility in the same way as the buyer. That is, their utility function shares the same functional form as in \Eqref{eq:payoff_buyer_after_lambda}; the only difference lies in the belief parameters $(\hat{\rho}_1, c_1)$ used in the computation. Consequently, the expressions for the coefficients remain structurally identical. The condition for a Nash equilibrium between buyer and seller is therefore given by:
\begin{equation}
\begin{aligned}
\begin{cases}
q_2^{br} = \min\left\{\max\left\{0, -\dfrac{X^{(2)}(q_1^{br}, \hat{\rho}_2, p_\mathrm{eq}, \lambda, c_2, N)}{2Y^{(2)}(q_1^{br}, \hat{\rho}_2, p_\mathrm{eq}, \lambda, c_2, N)}\right\}, 1\right\}\ ,\\[1em]
q_1^{br} = \min\left\{\max\left\{0, -\dfrac{X(q_2^{br}, \hat{\rho}_1, p_\mathrm{eq}, \lambda, c_1, N)}{2Y(q_2^{br}, \hat{\rho}_1, p_\mathrm{eq}, \lambda, c_1, N)}\right\}, 1\right\}\ .
\end{cases}
\end{aligned}
\end{equation}
The strategy $p_\mathrm{eq}$ of the remaining $N-2$ players is common knowledge, as it is uniquely determined by public information and given by the results of section \ref{B1_uninformed}. 
Note, however, that the buyer does not know the exact values of the seller’s belief parameters $(\hat{\rho}_1, c_1)$, since she does not have access to the seller’s full information string. As already discussed in section \ref{section:A3_buyer_after}, the buyer assumes that P1 will act based on $(\hat{\rho}_2, c_2)$.
This implies that the Nash equilibrium condition, from the point of view of P2, should be:
\begin{equation}
\begin{aligned}
\begin{cases}
q_2^{br} = \min\left\{\max\left\{0, -\dfrac{X^{(2)}(q_1^{br}, \hat{\rho}_2, p_\mathrm{eq}, \lambda, c_2, N)}{2Y^{(2)}(q_1^{br}, \hat{\rho}_2, p_\mathrm{eq}, \lambda, c_2, N)}\right\}, 1\right\}\ ,\\[1em]
q_1^{br} = \min\left\{\max\left\{0, -\dfrac{X(q_2^{br}, \hat{\rho}_2, p_\mathrm{eq}, \lambda, c_2, N)}{2Y(q_2^{br}, \hat{\rho}_2, p_\mathrm{eq}, \lambda, c_2, N)}\right\}, 1\right\}\ .
\end{cases}
\end{aligned}
\end{equation}
With this assumption, a possible solution for a Nash equilibrium could be symmetric, i.e. with $q_2^{br} =q_1^{br}=q_\mathrm{eq}$, where:
\begin{equation}
\begin{aligned}
q_\mathrm{eq} = \min\left\{\max\left\{0, -\frac{X^{(2)}(q^{eq}, \hat{\rho}_2, p_\mathrm{eq}, \lambda, c_2, N)}{2Y^{(2)}(q^{eq}, \hat{\rho}_2, p_\mathrm{eq}, \lambda, c_2, N)}\right\}, 1\right\}\ .
\label{eq:buyer_post_labda}
\end{aligned}
\end{equation}
The equilibrium equation above is challenging to solve analytically due to the number of parameters involved. Nonetheless, we solved it numerically and confirmed the solution's uniqueness across all tested parameter values. 
For completeness, we provide a Wolfram Mathematica notebook in which the interested reader can find the exact expressions of all coefficients\cite{github}.
\subsection{Calculation of $(q_1)_\mathrm{eq}^\mathrm{post}$}
In this section, we compute the equilibrium strategy $(q_1)_\mathrm{eq}^\mathrm{post}$ of the volatility-averse seller after the transaction. Since P1 retains access to the data she sold, she knows the values $c_2$ and $\hat{\rho}_2$ that P2 will estimate, and can thus anticipate her strategy as the solution of \Eqref{eq:buyer_post_labda}. Also the strategy $p_\mathrm{eq}$ of the remaining $N-2$ players is common knowledge and determined by public information alone (see section \ref{B1_uninformed}).
The utility computation for P1 follows the same structure as in section \ref{B3_buyer_post_lambda}, with the same strategy profile $\Vec{p}=(q_1,q_2,p_\mathrm{eq},...,p_\mathrm{eq})$ and buyer and seller roles interchanged:
\begin{equation}\label{eq:payoff_seller_after_lambda}
\begin{aligned}
 \U_1(\Vec{p},\hrho_1,c_1;\lambda)&=Y^{(2)}(q_2, \hat{\rho}_1, p_\mathrm{eq}, \lambda, c_1, N) q_1^2 + X^{(2)}(q_2, \hat{\rho}_1, p_\mathrm{eq}, \lambda, c_1, N) q_1 + R^{(2)}(q_2, \hat{\rho}_1, p_\mathrm{eq}, \lambda, c_1, N)\ .
\end{aligned}
\end{equation}
Here, $Y^{(2)}(q_2, \hat{\rho}_1, p_\mathrm{eq}, \lambda, c_1, N)$ retains the same functional form as in \Eqref{Y^2_2}, remains strictly negative, and ensures concavity of the utility function. It follows that P1’s best response and equilibrium strategy after the transaction now is:
\begin{equation}
\begin{aligned}
(q_1)_\text{eq}^\text{post} = \min\left\{\max\left\{0, -\frac{X^{(2)}(q_\mathrm{eq}, \hat{\rho}_1, p_\mathrm{eq}, \lambda, c_1, N)}{2Y^{(2)}(q_\mathrm{eq}, \hat{\rho}_1, p_\mathrm{eq}, \lambda, c_1, N)}\right\}, 1\right\}\ , 
\end{aligned}
\end{equation}
where $q_\mathrm{eq}$ is the strategy played by the buyer after the transaction (solution of \Eqref{eq:buyer_post_labda}), and $p_\mathrm{eq}$ denotes the symmetric equilibrium strategy of the less informed players  (derived in section \ref{B1_uninformed}). This equation can be efficiently solved numerically for any given set of parameters.
\section{The information structure and the hierarchy of beliefs}\label{sec:Appendix_C}

This Appendix's goal is to state precisely what each player in our game believes, and to show that, thanks to the informational assumptions {\bf (A1)}--{\bf (A2)} and {\bf (A5)}, this apparently unmanageable object collapses onto a handful of numbers. Appendix~\ref{sec:Appendix_D} then explains how players turn those beliefs into strategies. The two questions are genuinely distinct: here we describe \emph{what players think}, there \emph{what they do about it}.

\subsection{What has to be specified, and why one number per player is not enough}\label{C_motivation}

Consider the post-transaction configuration, and fix for concreteness $\hrho=0.5$ (the belief formed from the public string $\vec s_0$), $\hrho_2=0.65$ (the buyer's belief after the transaction) and $\hrho_1=0.8$ (the seller's belief). A player's optimal bet depends on the bets of the others, which in turn depend on \emph{their} beliefs; so specifying the game requires more than the list $(\hrho_1,\hrho_2,\hrho,\dots,\hrho)$. It requires, for instance, all of the following:

\begin{itemize}
    \item what player $3$ believes the coin bias is: $0.5$;
    \item what player $3$ believes P1 believes about the coin: $0.5$ as well, since by {\bf (A5)} player $3$ cannot conceive of a better-informed participant;
    \item what P1 believes P2 believes about the coin: $0.65$, since P1 knows exactly which substring she sold;
    \item what P1 believes P2 believes \emph{about P1}: $0.65$ (P1 knows that the buyer regards her as an equal).
\end{itemize}

The last item is a crucial mechanism of the model. P1 has an edge on P2 precisely because she knows that P2 is playing against an imagined opponent who is weaker than the real one. A description that stopped at ``what $i$ believes about $j$'' would not allow this to happen.

Continuing in the same way (what P1 believes P2 believes player $3$ believes, and so on) produces an infinite tower. This object is standard in the theory of games with incomplete information and is called the players' \emph{hierarchy of beliefs} \cite{harsanyi1967games,mertens1985formulation}. It should not be confused with the \emph{information hierarchy} of our model, which is simply the statement that players can be ranked from best to worst informed. The first is a feature of every Bayesian game; the second is a modelling assumption we impose.

\subsection{The universal belief space}\label{C_UBS}

The natural home of the tower described above is the \emph{universal belief space} of Mertens and Zamir \cite{mertens1985formulation} (see also \cite{brandenburger1993hierarchies} and the review \cite{zamir2013bayesian}). Let $S$ denote the set of \emph{states of nature}, i.e.\ everything that is payoff-relevant and not under the players' control; in our game $S=\{0,1\}$, the outcome $w$ of the next toss, since the payoff Eq.~(6) of the main text depends on nothing else. Writing $\Delta(A)$ for the set of probability distributions over $A$, one defines inductively the spaces of hierarchies truncated at level $k$,
\begin{equation}\label{eq:MZ_construction}
    X_1=\Delta(S)\ ,\qquad X_{k+1}=X_k\times\Delta\!\left(S\times X_k^{\,N-1}\right)\ ,\qquad k\ge 1\ .
\end{equation}
In words: a first-order belief is a probability distribution over the state of nature; a second-order belief is a joint distribution over the state of nature \emph{and} the first-order beliefs of the other $N-1$ players; and so on. Not every element of $X_k$ is meaningful: a player's higher-order belief must be compatible with her lower-order ones (its marginal on $S$ must reproduce her first-order belief), and she must consider only such compatible beliefs possible for others. Retaining only the elements that satisfy this \emph{coherence} requirement at every level and taking the limit $k\to\infty$ yields the \emph{universal type space} $T$, which satisfies the self-referential relation
\begin{equation}\label{eq:MZ_fixedpoint}
    T\;\cong\;\Delta\!\left(S\times T^{\,N-1}\right)\ .
\end{equation}
A \emph{type} of a player is thus a joint probability distribution over the state of nature and the types of everybody else (notice the circularity). A \emph{state of the world} is then a specification of the state of nature together with one type per player,
\begin{equation}\label{eq:MZ_omega}
    \omega=\bigl(s(\omega);\,t_1(\omega),\dots,t_N(\omega)\bigr)\ \in\ \Omega \coloneqq S\times T^{N}\ ,
\end{equation}
and $\Omega$, the universal belief space, contains, by construction, every conceivable informational situation of $N$ players facing $S$.

The universal belief space is the correct general answer, and it is useless for computation: already the second level is a space of probability distributions on a continuum, the third a space of probability distributions on \emph{that}, and so on. Any tractable model must therefore single out a small subset of $\Omega$ and argue that it is the relevant one. There are two classical ways of doing so; we now recall the standard one, explain why it is unavailable here, and then present ours.

\subsection{The standard route: types with a common prior, and why we cannot take it}\label{C_harsanyi}

Harsanyi's solution \cite{harsanyi1967games,harsanyi1968games,harsanyi1968games3} is to postulate a finite set of \emph{types} for each player together with a \emph{common prior} $p$ over type profiles, from which each player's beliefs are obtained by conditioning on her own type. The entire infinite hierarchy is then generated by $p$, the game becomes an ordinary game preceded by a chance move that draws the type profile, and Nash's theorem applies.

The price of this construction is the assumption of \emph{consistency}: all differences in beliefs must originate from differences in information, never from differences in the underlying model of the world. Consistency has strong and well-studied consequences. Players with a common prior cannot ``agree to disagree'' \cite{aumann1976agreeing}, and purely speculative trade between them is impossible \cite{milgrom1982information}: if it is common knowledge that a bet is being accepted, then someone must expect to lose, and would decline.

Our Assumption {\bf (A5)} is incompatible with this. It states not that the least-informed players hold a coarse or imprecise view of P1, but that they hold a \emph{false} one, and hold it with certainty: they assign probability one to a configuration of the game, ``everybody has only $\vec s_0$'', which is not the true one. Belief structures of this kind are legitimate elements of the universal belief space $\Omega$; they are what Zamir \cite{zamir2013bayesian} calls \emph{highly inconsistent} belief subspaces, in which the true state of the world lies outside the set of states some player considers possible. They cannot, however, be generated by a common prior, and consequently our game admits no representation as a Harsanyi game with types, nor an equivalent extensive form with an initial chance move.

This is a deliberate model choice. It is precisely the failure of consistency that keeps the market non-trivial: under a common prior the least-informed players would anticipate being exploited by whoever holds more data and would price that expectation into their participation, whereas here they perceive a fair, symmetric game with zero expected gain. The same mechanism (unawareness of better-informed counterparties as a way of escaping no-trade theorems) has been studied in the literature on games with unawareness \cite{heifetz2013unawareness}, to which we return in Section~\ref{C_comparison}.

\subsection{Our route, step 1: dogmatic higher-order beliefs}\label{C_step1}

The first simplification is to let players be \emph{Bayesian about the coin but dogmatic about each other}. Concretely, player $i$'s belief about the next outcome remains a genuine probability distribution: the Bernoulli distribution of parameter $\hrho_i$ obtained from the posterior Eq.~(7) of the main text. Instead, her beliefs about the other players' beliefs, and about their beliefs about beliefs, are point masses. For example player $3$ assigns probability 1 to the statement ``P1 has access only to $\vec s_0$''.

This strong assumption considerably simplifies the structure of the universal belief space. If a level-$r$ belief is a point mass, it is specified by a single number rather than by a measure over measures, and the whole hierarchy becomes a family of real tensors. We define the \emph{$r$-th order belief tensor} $T^{(r)}\in[0,1]^{\overbrace{\scriptstyle N \times N \times \dots \times N}^{r \text{ times}}}$ by
\begin{equation}\label{eq:belief_tensor_def}
     i_1\text{ believes }i_2\text{ believes}\;\dots\;i_{r-1}\text{ believes that }i_r\text{'s belief about the coin is } T^{(r)}_{i_1,i_2,\dots,i_r}\ .
\end{equation}
For example:
\begin{itemize}
    \item $T^{(1)}_i=\hrho_i$, the belief of player $i$ herself;
    \item $T^{(2)}_{ij}$ is the value of $\hrho$ that player $i$ believes player $j$ attributes to the coin -- the quantity denoted $\hat\rho^{(i)}_j$ in the main text;
    \item $T^{(3)}_{ijk}$ is what $i$ believes $j$ believes $k$ believes.
\end{itemize}
Two elementary consistency requirements are worth noting. A player knows her own belief, so repeating an index changes nothing,
\begin{equation}\label{eq:idempotence}
    T^{(r+1)}_{\dots,\,j,\,j}=T^{(r)}_{\dots,\,j}\ ;
\end{equation}
and, since point masses have point marginals, no further ``coherence'' condition of the kind required in \Eqref{eq:MZ_construction} arises. Only chains in which no two consecutive indices coincide therefore carry independent information.

The sequence $\{T^{(r)}\}_{r\in\N}$ specifies the belief hierarchy completely, and it is now a countable family of numbers rather than a tower of measure spaces. It is, however, still infinite. A second simplification is needed.
\begin{center}\rule{0.85\linewidth}{0.4pt}\end{center}
\noindent\textit{Mathematical digression (may be skipped on a first reading): building the tensor order by order.}

\medskip
\noindent
The tensor \eqref{eq:belief_tensor_def} is not a new object: it is what the general construction \eqref{eq:MZ_construction} reduces to once the dogmatic restriction is imposed. Here, we carry out the reduction explicitly, showing where each entry of $T^{(r)}$ comes from and how many of those entries are genuinely independent. Write $\mathcal X_k\subseteq X_k$ for the set of order-$k$ hierarchies that are both coherent and dogmatic, i.e.\ in which every belief of order $\geq 2$ is a point mass. Computing $\mathcal X_k$ requires only two elementary rules, applied in this order. In the displays below, a bracket $[\,\cdot\,]$ carries as a subscript the constraints still to be imposed, and the label above each equals sign records which rule has just been used; every step is an equality of sets, up to the canonical identifications stated in the two rules.

\paragraph{Rule 1 (dogmaticity: erase an outer $\Delta$).} A point mass on a space $Y$ is the same thing as a point of $Y$. Hence a joint law on $S\times Y$ whose marginal on $Y$ is a point mass is concentrated on $S\times\{y\}$ for a single $y$, and is therefore the pair (its $S$-marginal, that point):
\begin{equation}\label{eq:rule1}
    \Bigl[\,\Delta\bigl(S\times Y\bigr)\Bigr]_{\text{dogmatic}}\;=\;\Delta(S)\times Y\ .
\end{equation}
The $\Delta$ in front of $Y$ simply disappears. Two things happen at once: the measure over the opponents' beliefs is replaced by a point, and the belief about the coin decouples from the belief about the opponents.

\paragraph{Rule 2 (consistency: erase a duplicated factor).} A player's order-$(k{+}1)$ description must agree with her order-$k$ description wherever the two overlap; in particular both assign the same marginal to $S$. Coherence therefore restricts a duplicated pair of factors to its diagonal, which is a single copy:
\begin{equation}\label{eq:rule2}
    \Bigl[\,\underbrace{\Delta(S)}_{\text{order }k}\times\underbrace{\Delta(S)}_{\text{order }k+1}\Bigr]_{\text{coherent}}\;=\;\Delta(S)\ .
\end{equation}

\paragraph{Order 1.} $\mathcal X_1=\Delta(S)$ is my own belief about the coin, the entry $T^{(1)}_i$. Nothing to simplify.

\paragraph{Order 2.} Substituting $\mathcal X_1=\Delta(S)$ into \eqref{eq:MZ_construction},
\begin{equation}\label{eq:order2}
\begin{aligned}
\mathcal X_2 \;&=\; \Bigl[\,\underbrace{\Delta(S)}_{\mathcal X_1}\times\Delta\bigl(S\times\Delta(S)^{N-1}\bigr)\Bigr]_{\text{dogmatic, coherent}}\\[4pt]
&\overset{\text{\scriptsize dogmaticity}}{=}\;\; \Bigl[\,\Delta(S)\times\Delta(S)\times\Delta(S)^{N-1}\Bigr]_{\text{coherent}}\\[4pt]
&\overset{\text{\scriptsize coherence}}{=}\;\; \Delta(S)\times\Delta(S)^{N-1}\ .
\end{aligned}
\end{equation}
The surviving factors are exactly my own belief $T^{(1)}_i$ and the $N-1$ beliefs $T^{(2)}_{ij}$, one for each opponent $j\neq i$.

\paragraph{Order 3.} Now $\mathcal X_2=\Delta(S)\times\Delta(S)^{N-1}$, so that $\mathcal X_2^{\,N-1}=\Delta(S)^{N-1}\times\Delta(S)^{(N-1)^2}$, and
\begin{equation}\label{eq:order3}
\begin{aligned}
\mathcal X_3 \;&=\; \Bigl[\,\underbrace{\Delta(S)\times\Delta(S)^{N-1}}_{\mathcal X_2}\times\Delta\Bigl(S\times\bigl(\Delta(S)\times\Delta(S)^{N-1}\bigr)^{N-1}\Bigr)\Bigr]_{\text{dogmatic, coherent}}\\[4pt]
&\overset{\text{\scriptsize dogmaticity}}{=}\;\; \Bigl[\,\underbrace{\Delta(S)\times\Delta(S)^{N-1}}_{\mathcal X_2}\times\Delta(S)\times\underbrace{\Delta(S)^{N-1}\times\Delta(S)^{(N-1)^2}}_{\mathcal X_2^{\,N-1}}\Bigr]_{\text{coherent}}\\[4pt]
&\overset{\text{\scriptsize coherence}}{=}\;\; \Delta(S)\times\Delta(S)^{N-1}\times\Delta(S)^{(N-1)^2}\ .
\end{aligned}
\end{equation}
Note what coherence has done at this order: it has removed the \emph{entire} leading $\mathcal X_2$, not just one factor. Its $\Delta(S)$ duplicates the newly produced $\Delta(S)$, and its $\Delta(S)^{N-1}$ duplicates the first-order beliefs sitting inside $\mathcal X_2^{\,N-1}$ --- for each opponent $j$, ``what I think $j$ believes about the coin'' appears both in my own order-2 description and as the leading component of the order-2 hierarchy I attribute to $j$.

\paragraph{General order.} The same two steps apply verbatim at every level and always erase the leading factor, so that
\begin{equation}\label{eq:dogmatic_recursion}
    \mathcal X_{k+1}\;=\;\Delta(S)\times \mathcal X_k^{\,N-1}\ ,
    \qquad\text{hence}\qquad
    \mathcal X_k\;=\;\prod_{j=1}^{k}\Delta(S)^{(N-1)^{j-1}}\ .
\end{equation}
Compare this with \eqref{eq:MZ_construction}. In the general construction, coherence is a non-trivial closed constraint that has to be imposed by selecting a subset of $X_k$, and the factor $X_k$ must be carried along because a hierarchy is \emph{not} determined by its deepest level. Under the dogmatic restriction coherence is automatically accounted for, and the hierarchy \emph{is} determined by its deepest level. Note also that the order of the two rules matters: without dogmaticity, $\Delta(S\times X_k^{N-1})$ does not factorize and there is no duplicated $\Delta(S)$ for coherence to remove.

\paragraph{Counting the entries.} Each copy of $\Delta(S)$ in \eqref{eq:dogmatic_recursion} is one entry of the tensor. Unfolding the recursion, the factor $\Delta(S)^{(N-1)^{r-1}}$ at order $r$ is indexed by the chains $(i,i_2,\dots,i_r)$ in which each index differs from the one before it: $i_2\neq i$ because a player's belief is about the \emph{others}, then $i_3\neq i_2$ for the same reason one level down, and so on. There are $(N-1)^{r-1}$ such chains for a given holder $i$, hence $N(N-1)^{r-1}$ in total.

This resolves an apparent mismatch. Written as in \eqref{eq:belief_tensor_def}, $T^{(r)}$ has $N^r$ entries, but only $N(N-1)^{r-1}$ of them are free parameters. The remaining ones are the chains with a repeated adjacent index, and they are \emph{determined}, by the convention \eqref{eq:idempotence} --- which is nothing but the statement that a player knows her own belief:
\begin{center}
\begin{tabular}{cccc}
\hline
order $r$ & free entries & entries fixed by \eqref{eq:idempotence} & total \\
\hline
$1$ & $N$ & $0$ & $N$ \\
$2$ & $N(N-1)$ & $N$ & $N^2$ \\
$3$ & $N(N-1)^2$ & $N(2N-1)$ & $N^3$ \\
$r$ & $N(N-1)^{r-1}$ & $N^r-N(N-1)^{r-1}$ & $N^r$ \\
\hline
\end{tabular}
\end{center}
At order $2$, for instance, the free entries are the off-diagonal ones $T^{(2)}_{ij}$ with $j\neq i$, while the $N$ diagonal entries are forced to $T^{(2)}_{ii}=T^{(1)}_i$.

\paragraph{What is left to do.} Dogmaticity and coherence alone therefore leave $N(N-1)^{r-1}$ free numbers at every order $r$; this is a countable but still infinite set of numbers. It is the hierarchical assumption of Section~\ref{C_step2} that collapses all of them onto the $N$ numbers $(\hrho_1,\dots,\hrho_N)$, through the rule \eqref{eq:collapse_info_hier}. Finally, since every $\mathcal X_k$ obtained above consists of coherent hierarchies, the limiting object is a genuine subset of the universal type space of \cite{mertens1985formulation}, and not merely a convenient bookkeeping device.

\begin{center}\rule{0.85\linewidth}{0.4pt}\end{center}
\subsection{Our route, step 2: the information hierarchy and the $\max$ rule}\label{C_step2}

We now use the ordering of the players. Three assumptions are involved; the first two are the content of {\bf (A1)}--{\bf (A2)} and {\bf (A5)}, the third makes explicit a requirement that is implicit in the notion of a hierarchy.

\begin{itemize}
    \item {\bf (H1)} \emph{Nested information.} Players are labelled so that if $i<j$, then whatever player $j$ knows, player $i$ knows too. In our game this holds because all datasets are nested consecutive strings, $\vec s_0\subseteq\vec s_2\subseteq\vec s_1$. In particular a better-informed player can reconstruct the dataset, and hence the belief, of a worse-informed one: $T^{(2)}_{ij}=\hrho_j$ for $j>i$.
    \item {\bf (H2)} \emph{Upward blindness.} Player $i$ is certain that every player $j<i$ holds exactly the same information as she does, and therefore the same belief: $T^{(2)}_{ij}=\hrho_i$ for $j<i$. This is Assumption {\bf (A5)} applied along the whole hierarchy: a player cannot imagine an informational advantage she does not possess.
    \item {\bf (H3)} \emph{Self-similarity.} Rules {\bf (H1)} and {\bf (H2)}, with the same ordering, hold not only in the real game but also inside the game that each player imagines, and inside the game that each imagined player imagines, and so on.
\end{itemize}

Together with \Eqref{eq:idempotence}, {\bf (H1)} and {\bf (H2)} give the second-order tensor at once,
\begin{equation}\label{eq:T2_maxrule}
    T^{(2)}_{ij}=\hrho_{\max(i,j)}\ ,
\end{equation}
and {\bf (H3)} propagates the same rule to every order:
\begin{equation}
\label{eq:collapse_info_hier}
    \boxed{\;T^{(r)}_{I}=\hrho_{\max_{a=1,\dots,r} I_a}\;}\qquad\text{for any chain } I=(i_1,\dots,i_r)\ .
\end{equation}
In words: \emph{along any chain of nested beliefs, the belief that emerges is the one held by the least informed player appearing anywhere in the chain.}

The derivation is a repeated application of {\bf (H2)}. Take $I=(3,1,2)$, that is: what player $3$ believes player $1$ believes player $2$ believes. Player $1$ is better informed than player $3$, so by {\bf (H2)} player $3$ cannot tell her apart from himself, and the chain is equivalent to $I'=(3,3,2)$. Player $2$ is better informed than player $3$ as well, and by {\bf (H3)} the same reasoning is at work inside player $3$'s picture of the game, so the imagined believer again replaces player $2$ by a copy of herself and $I''=(3,3,3)$. Hence $T^{(3)}_{3,1,2}=\hrho_3$.

The general case works in the same way. Scan the chain from left to right, keeping track of the running maximum, i.e.\ of the least informed player met so far. Each new index is either larger than the running maximum (a player less informed than everyone before him, whose belief the current believer knows exactly by {\bf (H1)}, so that nothing is replaced and the running maximum simply updates) or smaller (a player better informed, whom the current believer cannot distinguish from himself by {\bf (H2)}, so that the index is replaced by the running maximum). Either way the chain $I$ becomes the equivalent chain $I'$ with $I'_a=\max(I_a,I'_{a-1})$, whose last entry is $\max_a I_a$, and this yields \Eqref{eq:collapse_info_hier}. Note that {\bf (H1)} and {\bf (H2)} are statements about second-order beliefs only, so they only allow to carry out  the first step of the index replacement: it is {\bf (H3)} that allows every subsequent step, each of which applies {\bf (H1),(H2)} inside somebody's imagined world. Without it, $T^{(2)}$ would be pinned down but all higher orders would remain free.

One consequence of \Eqref{eq:collapse_info_hier} is worth noting explicitly: the belief tensor is \emph{symmetric}. Its right-hand side depends on the chain $I$ only through the \emph{set} of players it contains, not through their order or multiplicity, so that under hierarchical information ``P1 believes P2 believes P3 believes'' and ``P3 believes P1 believes P2 believes'' evaluate to the same number.

\subsection{What the collapse buys: $N$ views of the world}\label{C_worlds}

Equation~\eqref{eq:collapse_info_hier} says that the infinite belief hierarchy is entirely determined by the ordering of the players together with the $N$ first-order beliefs $(\hrho_1,\dots,\hrho_N)$. Equivalently, and more concretely, the whole model can be pictured as a chain of $N$ ``worlds''. Let $\omega_k$ denote the configuration in which
\begin{equation}\label{eq:worlds}
    \text{players }1,\dots,k \text{ hold the belief } \hrho_k\ ,\qquad \text{players } k+1,\dots,N \text{ hold their true beliefs } \hrho_{k+1},\dots,\hrho_N\ .
\end{equation}
Then $\omega_1$ is the real world, and \Eqref{eq:collapse_info_hier} says that a player who is at level $k$ believes she is in world $\omega_k$, while believing that any player $j>k$ believes herself to be in world $\omega_j$. Beliefs run one way only: everyone looks downwards along the chain $\omega_1\to\omega_2\to\dots\to\omega_N$ and never upwards.

For the post-transaction configuration of our model, with $\hrho_3=\dots=\hrho_N=\hrho$, the chain has only three distinct links:

\begin{center}
\begin{tabular}{lll}
\hline
World & Beliefs of $(\text{P1},\text{P2},\text{P3},\dots,\text{P}N)$ & Who believes they are here \\
\hline
$\omega_1$ (real) & $(\hrho_1,\hrho_2,\hrho,\dots,\hrho)$ & P1 \\
$\omega_2$ & $(\hrho_2,\hrho_2,\hrho,\dots,\hrho)$ & P2, and P1's model of P2 \\
$\omega_3=\dots=\omega_N$ & $(\hrho,\hrho,\hrho,\dots,\hrho)$ & players $3,\dots,N$, and everyone's model of them \\
\hline
\end{tabular}
\end{center}

\noindent
Before the transaction $\hrho_2=\hrho$, the middle row merges with the last one, and only two worlds remain. In the language of \cite{mertens1985formulation,zamir2013bayesian}, the set $\{\omega_1,\dots,\omega_N\}$ is a \emph{belief subspace} of the universal belief space $\Omega$: a set of states of the world that is closed under the players' beliefs, so that no player, and no player imagined by a player, ever needs to consider a state outside it. Its distinctive feature is that the true state $\omega_1$ does not belong to the set of states considered possible by players $2,\dots,N$; this is the signature of Assumption {\bf (A5)}.

\subsection{Relation to other constructions}\label{C_comparison}

It may help to situate the construction relative to several familiar ones.

\begin{itemize}
    \item \emph{Nested information} \cite{jacobovic2024bayesian}. The ordering of players by informativeness, and the resulting collapse of the belief hierarchy, are shared with the theory of Bayesian games with nested information. There, however, the nesting is a nesting of information fields on a common probability space, so that beliefs are consistent in Harsanyi's sense and the less-informed players \emph{know} that they are less informed, holding correct but coarse beliefs about the others. Assumption {\bf (A5)} places our model outside that class; in exchange, our collapse is exact and finite rather than approximate.
    \item \emph{Games with unawareness} \cite{heifetz2006interactive,heifetz2013unawareness,halpern2014extensive}. This is the closest match to {\bf (A5)}: a player is simply unable to conceive of certain features of the game -- here, the existence of better-informed opponents. Such models are built from a lattice of state spaces of increasing expressiveness; the chain $\omega_1\to\dots\to\omega_N$ of Section~\ref{C_worlds} is precisely such a lattice in the special case where it is totally ordered. Unawareness is also known to permit speculative trade that no-trade theorems otherwise exclude, which is the role it plays in our market.
    \item \emph{Cognitive hierarchies and level-$k$ reasoning} \cite{stahl1995players,camerer2004cognitive}. The assumption that a player cannot imagine anyone above herself is structurally the same as the one used in cognitive hierarchy models, where a level-$k$ player believes all others reason at lower levels. The difference is what the hierarchy is over: depth of strategic reasoning there, quantity of data here. Our hierarchy is also deterministic, whereas cognitive hierarchy models place a distribution over levels.
    \item \emph{Cursed equilibrium and analogy-based expectations} \cite{eyster2005cursed,jehiel2008revisiting}. These describe agents who fail to appreciate the information content of their opponents' behaviour -- the same observable symptom displayed by our least-informed players. Formally they differ: both retain a common prior and a commonly known type space, and the error concerns the \emph{correlation} between opponents' actions and their types rather than the players' picture of who is informed at all.
    \item Finally, the qualifier ``subjective'' in \emph{subjective Nash equilibrium} (Appendix~\ref{sec:Appendix_D}) refers to the player-dependence of the perceived game and should not be confused with the subjective equilibrium of \cite{kalai1993subjective}, where players' conjectures, though possibly wrong, are required to be confirmed by observed play in a repeated interaction.
\end{itemize}

\section{The equilibrium concept}\label{sec:Appendix_D}

Appendix~\ref{sec:Appendix_C} described what the players believe; it remains to say what they do. Assumption {\bf (A4)} of the main text fixes the principle: each player plays her part of a Nash equilibrium of the game she believes she is in. For this to mean anything, the game a player believes she is in must be an \emph{ordinary} game defined by a definite set of participants, with definite strategies and payoffs, in which the concept of Nash equilibrium applies. In this appendix, we show that this is the case.

\subsection{What a single world contains}\label{D_oneworld}

Fix one world $\omega_k$ and read off from Section~\ref{C_worlds} who is in it. There are two groups.

The first consists of the $k$ participants $1,\dots,k$. All of them hold the belief $\hrho_k$, and none of them can be told apart from the others: in $\omega_2$, for instance, there are two participants with belief $\hrho_2$, namely the buyer and the seller as the buyer pictures her. Since each believes the others to reason exactly as she does, and each is playing against them, no attribution of strategies to this group is self-consistent unless the $k$ strategies are mutually optimal, that is, unless they form a Nash equilibrium among themselves.

The second group consists of the participants $k+1,\dots,N$. These are perceived correctly: their beliefs are $\hrho_{k+1},\dots,\hrho_N$, and, by Appendix~\ref{sec:Appendix_C}, each of them is in turn in her own world $\omega_{k+1},\dots,\omega_N$. Nothing about them is decided in $\omega_k$: whatever they do, they do because of the world they are in, and $\omega_k$ merely records it.

So a world does present an ordinary game, provided the behaviour of the second group is already known. Writing $p^\star_j(k)$ for the strategy of participant $j$ in $\omega_k$, we may state this as follows.

\begin{definition}[The game of a world]\label{def:Gamma}
Given strategies $p^\star_{k+1},\dots,p^\star_N$ for the participants $k+1,\dots,N$, let $\Gamma_k$ be the ordinary $k$-player game in which participants $1,\dots,k$ choose $p^\star_1(k),\dots,p^\star_k(k)$ in $[0,1]$, each maximizing the utility $\U_j(\,\cdot\,,\hrho_k)$ of Eq.~(9) (or Eq.~(17) in the volatility-averse case) evaluated at the full profile $\bigl(p^\star_1(k),\dots,p^\star_k(k),p^\star_{k+1},\dots,p^\star_N\bigr)$.
\end{definition}

\subsection{Putting the worlds together}\label{D_def}

Definition~\ref{def:Gamma} presupposes the strategies of the less-informed participants, so to solve for $\Gamma_k$'s equilibrium one needs to know $p^\star_{k+1},\dots,p^\star_N$. This implies that worlds must be solved starting from the most ignorant player and working our way upwards. At $k=N$ there is no second group at all and $\Gamma_N$ is a symmetric $N$-player game in which everybody holds $\hrho$; each subsequent $\Gamma_k$ is then well posed once $\Gamma_{k+1},\dots,\Gamma_N$ have been solved.

\begin{definition}[Equilibrium]\label{def:equilibrium}
An \emph{equilibrium} is a family of profiles $\bigl(p^\star_1(k),\dots,p^\star_k(k)\bigr)$, one for each $k=N,N-1,\dots,1$, such that each is a Nash equilibrium of the game $\Gamma_k$ of Definition~\ref{def:Gamma} built on the strategies $p^\star_j:=p^\star_j(j)$ of the participants $j>k$. Player $i$, who by Appendix~\ref{sec:Appendix_C} is in the world $\omega_i$, plays the strategy $p^\star_i(i)$ assigned to her there.
\end{definition}

\noindent
Each $\Gamma_k$ has finitely many players and compact strategy sets, so an equilibrium always exists. When a $\Gamma_k$ has several Nash equilibria we select the symmetric one, its players being indistinguishable. The whole strategic content of the model is thus $N$ ordinary Nash equilibria computed in sequence; the rest is the information structure of Appendix~\ref{sec:Appendix_C}.

Definition~\ref{def:equilibrium} is Eqs.~(12) and (13) of the main text. There $\vec p^{\,\star}(k)$ denotes the full $N$-vector of strategies attributed to the participants of $\omega_k$: its first $k$ entries are those of $\Gamma_k$, and Eq.~(12), $p^\star_j(k)=p^\star_j(j)$ for $j>k$, is the statement that the remaining ones are imported from the worlds of the less-informed participants. The strategies actually played, $\vec p^{\,\star}=\bigl(p^\star_1(1),\dots,p^\star_N(N)\bigr)$, are Eq.~(16).

\subsection{The pricing game}\label{D_worked}

In our model $\hrho_3=\dots=\hrho_N=\hrho$, so only three distinct worlds occur after a transaction and the descent has three steps:

\begin{center}
\begin{tabular}{llll}
\hline
World & Participants of $\Gamma_k$, with belief & Strategies imported & Nash equilibrium of $\Gamma_k$ \\
\hline
$\omega_N=\dots=\omega_3$ & all $N$, \ $\hrho$ & --- & $p_\mathrm{eq}$ for all, \Eqref{eq:eq_condition_uninformed_app} \\
$\omega_2$ & P1 and P2, \ $\hrho_2$ & $p_\mathrm{eq}$ for $j\geq3$ & $q_\mathrm{eq}$ for both, \Eqref{eq:qeq_buyer_after} \\
$\omega_1$ & P1, \ $\hrho_1$ & $q_\mathrm{eq}$ for P2, $p_\mathrm{eq}$ for $j\geq3$ & $(q_1)^{\mathrm{post}}_\mathrm{eq}$ \\
\hline
\end{tabular}
\end{center}

\noindent
Each player plays her strategy in her own world, so $\vec p^{\,\star}=\bigl((q_1)^{\mathrm{post}}_\mathrm{eq},q_\mathrm{eq},p_\mathrm{eq},\dots,p_\mathrm{eq}\bigr)=\vec p^{\,\text{post}}$, which is Eq.~(3) of the main text. Before a transaction $\hrho_2=\hrho$, the middle row merges with the one above it, and the same descent returns $\vec p^{\,\text{pre}}$, Eq.~(2). The four quantities derived in Appendices~\ref{sec:Appendix_A} and \ref{sec:Appendix_B}, and the order in which Section V\,D of the main text computes them, are exactly these three steps.

The two entries for P1 are worth comparing. In her own world she plays $(q_1)^{\mathrm{post}}_\mathrm{eq}$, whereas in $\omega_2$ (the world the buyer is in) she plays $q_\mathrm{eq}$.

\subsection{Two remarks}\label{D_status}

The profile $\vec p^{\,\star}$ is not a Nash equilibrium of any single game played by all $N$ participants, instead, its components are optimal in $\Gamma_1,\dots,\Gamma_N$ respectively, and these are different games. Similarly, $\vec p^{\,\star}$ is not a Bayesian Nash equilibrium in Harsanyi's sense since no common prior is available here. It is the form that equilibrium takes on a belief structure of the kind constructed in Appendix~\ref{sec:Appendix_C} \cite{mertens1985formulation,zamir2013bayesian}.

Because participants evaluate the same strategy profile with different beliefs, any utility appearing in the analysis has to be attributed to a specific evaluator. Throughout this work we adopt the point of view of P1: the bounds of Eqs.~(4) and (5) of the main text are computed with her belief $\hrho_1$, so that $\Psi_{\min}$ is the loss P1 computes for herself and $\Psi_{\max}$ the gain P1 computes for P2. Her information set being the finest in the game, these are the most accurate assessments available to any participant, and $[\Psi_{\min},\Psi_{\max}]$ is the range of prices that the best-informed player identifies as mutually beneficial.

\bibliography{apssamp}

\end{document}